\documentclass{ejm}
\usepackage{amssymb}
\usepackage{amsmath}
\usepackage{amsfonts}
\usepackage{array}
\usepackage{graphicx}
\usepackage{enumerate}
\usepackage{color}
\usepackage{epsfig}
\usepackage{subfig}
\usepackage{siunitx}
\newcommand*{\rom}[1]{\uppercase\expandafter{\romannumeral #1\relax}}
\newcommand*{\vecb}[1]{\boldsymbol{#1}}

\newcommand{\bsub}{\begin{subequations}}
\newcommand{\esub}{\end{subequations}$\!$}

\renewcommand{\theequation}{\arabic{section}.\arabic{equation}}
\renewcommand{\thesection}{\arabic{section}}

\newtheorem{result}{Principal Result}[section]
\newtheorem{remark}{Remark}[section]

\numberwithin{equation}{section}

\def\ni{\noindent}
\def\proof{{\ni \bf \underline{Proof:} }}
\def\endproof{\hfill$\blacksquare$\vspace{6pt}}

\renewcommand{\arraystretch}{1.5}

\newcommand{\R}{{\mathbb{R}}}

\title[A 2-D Model of Dynamically Active Compartments Coupled by Bulk
  Diffusion] {An Asymptotic Analysis of a 2-D Model of Dynamically Active
  Compartments Coupled by Bulk Diffusion}

\author[J. Gou, M. J. Ward]{%
  J.\ns G\ls O\ls U, \ns  M.\ns J. \ns W\ls A\ls R\ls D}
\affiliation{Department of Mathematics, University of British Columbia,
 Vancouver, British Columbia, V6T 1Z2, Canada,}

\date{\today}
\pubyear{2004}
\volume{000}
\pagerange{\pageref{firstpage}--\pageref{lastpage}}

\begin{document}

\label{firstpage}
\maketitle

\baselineskip=12pt

\begin{abstract}

A class of coupled cell-bulk ODE-PDE models is formulated and analyzed
in a two-dimensional domain, which is relevant to studying quorum
sensing behavior on thin substrates. In this model, spatially
segregated dynamically active signaling cells of a common small radius
$\epsilon\ll 1$ are coupled through a passive bulk diffusion
field. For this coupled system, the method of matched asymptotic
expansions is used to construct steady-state solutions and to
formulate a spectral problem that characterizes the linear stability
properties of the steady-state solutions, with the aim of predicting
whether temporal oscillations can be triggered by the cell-bulk
coupling. Phase diagrams in parameter space where such collective
oscillations can occur, as obtained from our linear stability
analysis, are illustrated for two specific choices of the
intracellular kinetics.  In the limit of very large bulk diffusion, it
is shown that solutions to the ODE-PDE cell-bulk system can be
approximated by a finite-dimensional dynamical system. This limiting
system is studied both analytically, using a linear stability
analysis, and globally, using numerical bifurcation software. For one
illustrative example of the theory it is shown that when the number of
cells exceeds some critical number, i.e. when a {\em quorum} is
attained, the passive bulk diffusion field can trigger oscillations
that would otherwise not occur without the coupling.  Moreover, for
two specific models for the intracellular dynamics, we show that there
are rather wide regions in parameter space where these triggered
oscillations are synchronous in nature. Unless the bulk diffusivity is
asymptotically large, it is shown that a clustered spatial
configuration of cells inside the domain leads to larger regions in
parameter space where synchronous collective oscillations between the
small cells can occur. Finally, the linear stability analysis for
these cell-bulk models is shown to be qualitatively rather similar to
the linear stability analysis of localized spot patterns for
activator-inhibitor reaction-diffusion systems in the limit of
long-range inhibition and short-range activation.
\end{abstract}


\baselineskip=16pt

\setcounter{equation}{0}
\setcounter{section}{0}
\section{Introduction}\label{sec:form}

In multicellular organisms ranging from cellular amoebae to the human
body, it is essential for cells to communicate with each other. One
common mechanism to initiate communication between cells that are not
in close contact is for cells to secrete diffusible signaling
molecules into the extracellular space between the spatially
segregated units. Examples of this kind of signaling range from
colonies of the amoebae Dictyostelium discoideum, which release cAMP
into the medium where it diffuses and acts on each separate colony
(cf.~\cite{Goldbeter1990}), to some endocrine neurons that secrete a
hormone to the extracellular medium where it influences the secretion
of this hormone from a pool of such neurons
(cf.~\cite{Krasmanovic_etal2003}, \cite{Li2008}), to the effect of
catalysts in surface science \cite{riecke}, and to quorum sensing
behavior for various applications (cf.~\cite{CLL}, \cite{Muller1},
\cite{Muller2}). In many of these systems, the individual cells or
localized units can exhibit sustained temporal oscillations. In this
way, signaling through a diffusive chemical can often trigger
synchronous oscillations among all the units.

In this paper we provide a theoretical investigation of the mechanism
through which this kind of synchronization occurs for a class of
coupled cell-bulk ODE-PDE models in bounded two-dimensional
domains. Our class of models consists of $m$ small cells with
multi-component intracellular dynamics that are coupled together by a
diffusion field that undergoes constant bulk decay. We assume that the
cells can release a specific signaling molecule into the bulk region
exterior to the cells, and that this secretion is regulated by both
the extracellular concentration of the molecule together with its
number density inside the cells. Our aim is to characterize conditions
for which the release of the signaling molecule leads to the
triggering of some collective synchronous oscillatory behavior among
the localized cells. Our modeling framework is closely related to the
study of quorum sensing behavior in bacteria done in \cite{Muller1}
and \cite{Muller2} through the formulation and analysis of similar
coupled cell-bulk models in $\R^3$. For this 3-D case, in
\cite{Muller1} and \cite{Muller2} steady-state solutions were
constructed and large-scale dynamics studied in the case where the
signaling compartments have small radius of order ${\mathcal
  O}(\epsilon)$. However, due to the rapid ${1/r}$ decay of the
free-space Green's function for the Laplacian in 3-D, it was shown in
\cite{Muller1} and \cite{Muller2} that the release of the signaling
molecule leads to only a rather weak communication between the cells
of the same ${\mathcal O}(\epsilon)$ order of the cell radius. As a
result, small cells in 3-D are primarily influenced by their own
signal, and hence no robust mechanism to trigger collective
synchronous oscillations in the cells due to Hopf bifurcations was
observed in \cite{Muller1} and \cite{Muller2}. We emphasize that the
models of \cite{Muller1} and \cite{Muller2} are based on postulating a
diffusive coupling mechanism between distinct, spatially segregated,
dynamically active sites. Other approaches for studying quorum sensing
behavior, such as in \cite{NSSGM2014}, are based on reaction-diffusion
(RD) systems, which adopt a homogenization theory approach to treat
large populations or colonies of individual cells as a continuum
density, rather than as discrete units as in \cite{Muller1} and
\cite{Muller2}.

\begin{figure}[htbp]
\begin{center}
\includegraphics[width=0.65\textwidth,height=5.0cm]{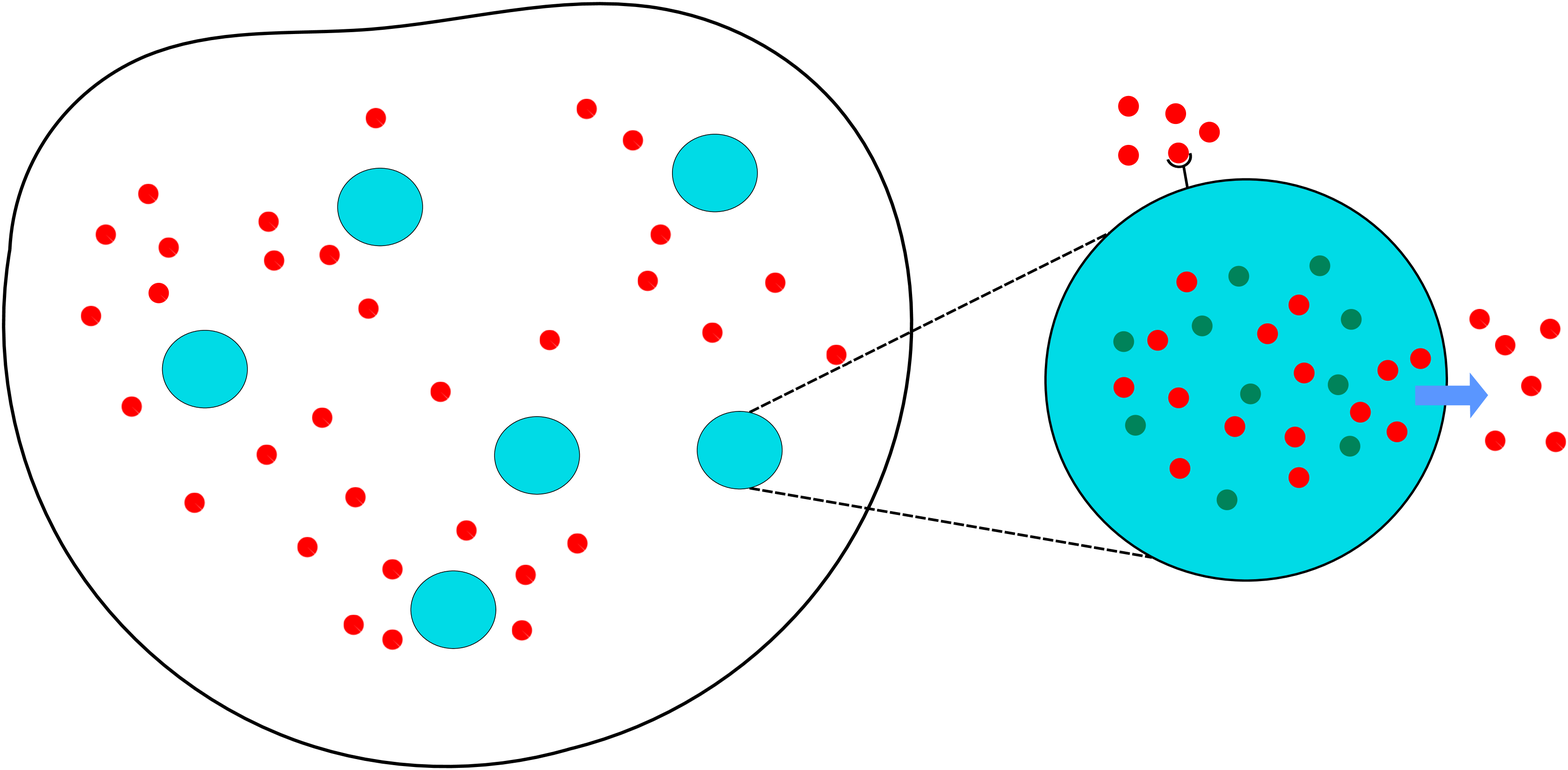}
\caption{Schematic diagram showing the intracellular reactions and
  external bulk diffusion of the signal. The small shaded regions are
  the signaling compartments or cells.}\label{fig:schem}
\end{center}
\end{figure}

Before discussing our main results, we first formulate and
non-dimensionalize our coupled cell-bulk model assuming that there is
only one signaling compartment $\Omega_0$ inside the two-dimensional
domain $\Omega$. The corresponding dimensionless model for $m>1$ small
signaling cells is given in (\ref{mainbd}).  Fig.~\ref{fig:schem}
shows a schematic plot of the geometry for $m$ small cells. We assume
that the cell can release a specific signaling molecule into the bulk
region exterior to the cell, and that this secretion is regulated by
both the extracellular concentration of the molecule together with its
number density inside the cell. If ${\mathcal U}({\vecb X},T)$
represents the concentration of the signaling molecule in the bulk
region $\Omega\backslash\Omega_0$, then its spatial-temporal evolution
in this region is assumed to be governed by the PDE model
\bsub\label{f:mainbd}
\begin{equation}\label{f:mainU}
\begin{aligned}
 {\mathcal U}_T &= D_B\Delta_{\vecb X} {\mathcal U} - k_B {\mathcal U}\,,
 \qquad \vecb X \in \Omega\backslash\Omega_{0}\,; \qquad
\partial_{n_{\vecb X}} {\mathcal U}= 0\,,\qquad \vecb X \in\partial \Omega\,,\\
 D_B \partial_{n_{\vecb X}} {\mathcal U}&= \beta_1 {\mathcal U} -\beta_2 {\mu}^1
 \,,  \;\;\quad \vecb X\in\partial \Omega_{0}\,, \\
\end{aligned}
\end{equation}
where, for simplicity, we assume that the signaling compartment
$\Omega_0\in\Omega$ is a disk of radius $\sigma$ centered at some
$\vecb X_0\in \Omega$. Inside the cell we assume that there are $n$
interacting species $\vecb \mu \equiv (\mu^1,\ldots,\mu^n)^T$ whose
dynamics are governed by $n$-ODEs, with a source term representing the
exchange of material across the cell membrane $\partial\Omega_0$, of
the form
\begin{equation}\label{f:mainuj}
\frac{d \vecb \mu}{dT} = k_R \mu_c \vecb F\left({{\vecb \mu}/\mu_c}\right)
+ \vecb e_1 \int_{\partial \Omega_{0}}\left( \beta_1 {\mathcal U} -
\beta_2 \mu^1\right) \, dS_{\vecb X} \,,
\end{equation}
\esub where $\vecb e_1\equiv (1,0,\dots,0)^T$. Here $\vecb \mu$ is the
total amount of the $n$ species inside the cell, while $k_R>0$ is the
reaction rate for the dimensionless intracellular dynamics ${\vecb
  F}(\vecb u)$. The scalar $\mu_c>0$ is a typical value for $\vecb
\mu$.

In this coupled cell-bulk model, $D_B>0$ is the diffusion coefficient
for the bulk process, $k_B$ is the rate at which the signaling
molecule is degraded in the bulk, while $\beta_1>0$ and $\beta_2>0$
are the dimensional influx (eflux) constants modeling the permeability
of the cell wall. In addition, $\partial_{n_{\vecb X}}$ denotes either
the outer normal derivative of $\Omega$, or the outer normal to
$\Omega_{0}$ (which points inside the bulk region).  The flux
$\beta_1{\mathcal U}-\beta_2\mu^1$ on the cell membrane models the
influx of the signaling molecule into the extracellular bulk region,
which depends on both the external bulk concentration ${\mathcal
  U}({\vecb X},T)$ at the cell membrane $\partial\Omega_0$ as well as
on the intracellular concentration $\mu^1$ within the cell. We assume
that only one of the intracellular species, $\mu^1$, is capable of
being transported across the cell membrane $\partial\Omega_0$ into the
bulk. We remark that a related class of models was formulated and
analyzed in \cite{bm1} and \cite{bm2} in their study of the initiation
of the biological cell cycle, where the dynamically active compartment
is the nucleus of the biological cell.

Next, we introduce our scaling assumption that the radius $\sigma$ of
the cell is small compared to the radius of the domain, so that
$\epsilon\equiv {\sigma/L}\ll 1$, where $L$ is the length-scale of
$\Omega$. However, in order that the signaling compartment has a
non-negligible effect on the bulk process, we need to assume that
$\beta_1$ and $\beta_2$ are both ${\mathcal O}(\epsilon^{-1})\gg 1$ as
$\epsilon\to 0$. In this way, in Appendix \ref{app:A} we show that
(\ref{f:mainbd}) reduces to the dimensionless coupled system
\bsub\label{f4:mainbd}
\begin{equation}\label{f4:mainU}
\begin{aligned}
  \tau { U}_t &= D \Delta_{\vecb x} {U} - {U}\,, \qquad \vecb x \in
  {\Omega}\backslash{\Omega}_{0}\,; \qquad
  \partial_{n_{\vecb x}} {U} = 0\,,\qquad \vecb x \in\partial
  {\Omega}\,,\\ \epsilon D \partial_{n_{\vecb x}} {U}&= d_1 { U}
  - d_2 u^1 \,, \;\;\quad \vecb x\in\partial {\Omega}_{0}\,, \\
\end{aligned}
\end{equation}
where ${\Omega}_0$ is a disk of radius $\epsilon\ll 1$ centered
at some $x_0\in {\Omega}$. The bulk process is coupled to the
intracellular dynamics by
\begin{equation}\label{f4:mainuj}
  \frac{d \vecb u}{dt} = \vecb F\left(\vecb u\right)
+ \frac{\vecb e_1}{\tau \epsilon} \int_{\partial {\Omega}_{0}}
  \left( d_1 {U} - d_2 u^1\right) \,  dS_{\vecb x} \,.
\end{equation}
\esub
The four ${\mathcal O}(1)$ dimensionless parameters in
(\ref{f4:mainbd}) are $\tau$, $D$, $d_1$, and $d_2$, defined by
\begin{equation}\label{form:par}
  \tau \equiv \frac{k_R}{k_B}  \,, \qquad  D \equiv \frac{D_B}{k_B L^2} \,,
\qquad \beta_1 \equiv \left(k_B L\right) \frac{d_1}{\epsilon} \,, \qquad
    \beta_2 \equiv \left( \frac{k_B}{L} \right) \frac{d_2}{\epsilon} \,.
\end{equation}
We remark that the limit $\tau\ll 1$ ($\tau\gg 1$) corresponds to when
the intracellular dynamics is very slow (fast) with respect to the
time-scale of degradation of the signaling molecule in the bulk. The
limit $D\gg 1$ corresponds to when the bulk diffusion length
$\sqrt{D_B/k_B}$ is large compared to the length-scale $L$ of the confining
domain. 

A related class of coupled cell-bulk models involving two bulk
diffusing species, and where the cells, centered at specific spatial
sites, are modeled solely by nonlinear flux boundary conditions, have
been used to model cellular signal cascades (cf.~\cite{levy1},
\cite{levy2}), and the effect of catalyst particles on chemically
active substrates (cf.~\cite{Peirce}, \cite{Glass}). In contrast to
these nonlinear flux-based models, in the coupled cell-bulk models of
\cite{Muller1} and \cite{Muller2}, and the one considered herein, the
cells are not quasi-static but are, instead, dynamically active
units. Our main goal for (\ref{f4:mainbd}) and (\ref{mainbd}) is to
determine conditions that lead to the triggering of synchronized
oscillations between the dynamically active cells.  Related 1-D
cell-bulk models, where the cells are dynamically active units at the
ends of a 1-D spatial domain, have been analyzed in
\cite{Gomez-Marin2007}--\cite{glnw}.

Our analysis of the 2-D coupled cell-bulk model (\ref{f4:mainbd}), and
its multi-cell counterpart (\ref{mainbd}), which extends the 3-D
modeling paradigm of \cite{Muller1} and \cite{Muller2}, has the
potential of providing a theoretical framework to model quorum sensing
behavior in experiments performed in petri dishes, where cells live on
a thin substrate. In contrast to the assumption of only one active
intracellular component used in \cite{Muller1} and \cite{Muller2}, in
our study we will allow for $m$ small spatially segregated cells with
multi-component intracellular dynamics. We will show for our 2-D case
that the communication between small cells through the diffusive
medium is much stronger than in 3-D, and leads in certain parameter
regimes to the triggering of synchronous oscillations, which otherwise
would not be present in the absence of any cell-bulk coupling. In
addition, when $D={\mathcal O}(1)$, we show that the spatial
configuration of small cells in the domain is an important factor in
triggering collective synchronous temporal instabilities in the cells.

The outline of this paper is as follows. In \S \ref{sec:bd} we use the
method of matched asymptotic expansions to construct steady-state
solutions to our 2-D multi-cell-bulk model (\ref{mainbd}), and we
derive a globally coupled eigenvalue problem whose spectrum
characterizes the stability properties of the steady-state. In our 2-D
analysis, the interaction between the cells is of order
$\nu\equiv-1/\log\epsilon$, where $\epsilon\ll 1$ is the assumed
common radius of the small circular cells. In the distinguished limit
where the bulk diffusion coefficient $D$ is of the asymptotic order
$D=\mathcal{O}(\nu^{-1})$, in \S \ref{sec:largeD} we show that the
leading order approximate steady-state solution and the associated
linear stability problem are both independent of the spatial
configurations of cells and the shape of the domain. In this regime,
we then show that the steady-state solution can be destabilized by
either a synchronous perturbation in the cells or by $m-1$ possible
asynchronous modes of instability. In \S \ref{sec:largeD}
leading-order-in-$\nu$ limiting spectral problems when $D={D_0/\nu}$,
with $\nu\ll 1$, for both these classes of instabilities are
derived. In \S \ref{sec:examples}, we illustrate our theory for
various intracellular dynamics. When there is only a single
dynamically active intracellular component, we show that no triggered
oscillations can occur. For two specific intracellular reaction
kinetics involving two local species, modeled either by Sel'kov or
Fitzhugh-Nagumo (FN) dynamics, in \S \ref{sec:examples} we perform
detailed analysis to obtain Hopf bifurcation boundaries, corresponding
to the onset of either synchronous or asynchronous oscillations, in
various parameter planes. In addition to this detailed stability
analysis for the $D=\mathcal{O}(\nu^{-1})$ regime, in \S
\ref{sec:odes} we show for the case of one cell that when $D\gg
\mathcal{O}(\nu^{-1})$ the coupled cell-bulk model is effectively
well-mixed and its solutions can be well-approximated by a
finite-dimensional system of nonlinear ODEs. The analytical and
numerical study of these limiting ODEs in \S \ref{sec:odes} reveals
that their steady-states can be destabilized through a Hopf
bifurcation. Numerical bifurcation software is then used to show the
existence of globally stable time-periodic solution branches that are
intrinsically due to the cell-bulk coupling. For the $D={\mathcal
  O}(1)$ regime, where the spatial configuration of the cells in the
domain is an important factor, in \S \ref{sec:finite_d} we perform a
detailed stability analysis for a ring-shaped pattern of cells that is
concentric within the unit disk. For this simple spatial configuration
of cells, phase diagrams in the $\tau$ versus $D$ parameter space, for
various ring radii, characterizing the existence of either synchronous
or asynchronous oscillatory instabilities, are obtained for the case
of Sel'kov intracellular dynamics. These phase diagrams show that
triggered synchronous oscillations can occur when cells become more
spatially clustered. In \S \ref{sec:finite_d} we also provide a clear
example of quorom sensing behavior, characterized by the triggering of
collective dynamics only when the number of cells exceeds a critical
threshold. Finally, in \S \ref{sec:disc} we briefly summarize our main
results and discuss some open directions.

Our analysis of synchronous and asynchronous instabilities for
(\ref{mainbd}) in the $D={\mathcal O}(\nu^{-1})$ regime, where the
stability thresholds are to, to leading-order, independent of the
spatial configuration of cells, has some similarities with the
stability analysis of \cite{wei1}, \cite{wei2}, \cite{rozada}, and
\cite{chen} (see also the references therein) for localized spot
solutions to various activator-inhibitor RD systems with short range
activation and long-range inhibition. In this RD context, when the
inhibitor diffusivity is of the order ${\mathcal O}(\nu^{-1})$,
localized spot patterns can be destabilized by either synchronous or
asynchronous perturbations, with the stability thresholds being, to
leading-order in $\nu$, independent of the spatial configuration of
the spots in the domain. The qualitative reason for this similarity
between the coupled cell-bulk and localized spot problems is
intuitively rather clear. In the RD context, the inhibitor diffusion
field is the long-range ``bulk'' diffusion field, which mediates the
interaction between the ``dynamically active units'', consisting of
$m$ spatially segregated localized regions of high activator
concentration, each of which is is self-activating. In this RD
context, asynchronous instabilities lead to asymmetric spot patterns,
while synchronous oscillatory instabilities lead to collective
temporal oscillations in the amplitudes of the localized spots
(cf.~\cite{wei1}, \cite{wei2}, \cite{rozada}, and \cite{chen}). A more
detailed discussion of this analogy is given in Remark \ref{remark}.

Finally, we remark that the asymptotic framework for the construction
of steady-state solutions to the cell-bulk model (\ref{mainbd}), and
the analysis of their linear stability properties, relies heavily on
the methodology of strong localized perturbation theory
(cf.~\cite{ward}). Related problems where such techniques are used
include \cite{ktcw}, \cite{pill}, \cite{levy1}, and \cite{levy2}.

\setcounter{equation}{0}
\setcounter{section}{1}
\section{Analysis of the Dimensionless 2-D Cell-Bulk System}\label{sec:bd}

We first generalize the one-cell model of \S \ref{sec:form} by
formulating a class of dimensionless coupled cell-bulk dynamics that
contains $m$ small, disjoint, cells or compartments that are scattered
inside the bounded two-dimensional domain $\Omega$. We assume that
each cell is a small disk of a common radius $\epsilon\ll 1$ that
shrinks to a point $\vecb x_j\in \Omega$ as $\epsilon\to 0$, and that
are well-separated in the sense that dist$(\vecb x_i, \vecb x_j)=O(1)$ for
$i\neq j$ and dist$(\vecb x_j, \partial \Omega)=O(1)$ for
$j=1,\ldots,m$, as $\epsilon\to 0$.

As motivated by the dimensional reasoning provided in \S
\ref{sec:form}, if $U({\vecb x},t)$ is the dimensionless
concentration of the signaling molecule in the bulk region between the
cells, then in this region it satisfies the dimensionless PDE 
\bsub\label{mainbd}
\begin{equation}\label{mainU}
\begin{aligned}
\tau U_t &= D\Delta U -U\,,\qquad \vecb x\in \Omega\backslash
\cup_{j=1}^m\Omega_{\epsilon_j}\,; \qquad
\partial_nU = 0\,, \qquad \vecb x\in\partial \Omega\,,\\
\epsilon D \partial_{n_j} U&=d_1U-d_2u^1_j\,,  \;\;\quad \vecb x\in\partial 
\Omega_{\epsilon_j}\,, \quad j=1,\ldots,m\,.
\end{aligned}
\end{equation}
Here $D>0$ is the effective diffusivity of the bulk, $d_1>0$ and
$d_2>0$ are the dimensionless influx (eflux) constants modeling the
permeability of the cell membrane, $\partial_n$ denotes the outer normal
derivative of $\Omega$, and $\partial_{n_j}$ denotes the outer normal
to each $\Omega_{\epsilon_j}$, which points inside the bulk
region. The signaling cell, or compartment,
$\Omega_{\epsilon_j}$ is assumed to lie entirely within $\Omega$.  The
flux $d_1U-d_2u^1_j$ on each cell membrane models the influx of the
signaling molecule into the extracellular bulk region, which depends
on both the external bulk concentration $U({\vecb x},t)$ at the cell
membrane $\partial\Omega_{\epsilon_j}$ as well as on the amount
$u_j^1$ of one of the intracellular species within the $j$-th cell.  We
suppose that inside each of the $m$ cells there are $n$ interacting
species, with intracellular dynamics
\begin{equation}\label{mainuj}
\frac{d \vecb u_j}{dt} = \vecb F_j(\vecb u_j)+ \frac{\vecb e_1}{\epsilon
\tau}  \int_{\partial \Omega_{\epsilon_j}}( d_1 U -d_2 u^1_j)\, ds \,,
\end{equation}
\esub where $\vecb e_1\equiv (1,0,\dots,0)^T$. Here $\vecb u_j =
(u^1_j,\dots,u^n_j)^T$ is the mass of the $n$ species inside the
$j$-th cell and $\vecb F_j(\vecb u_j)$ is the vector nonlinearity
modeling the reaction dynamics within the $j$-th cell. The integration
in (\ref{mainuj}) is over the boundary $\partial\Omega_{\epsilon_j}$
of the compartment. Since its perimeter has length $|\partial
\Omega_{\epsilon_j}|={\mathcal O}(\epsilon)$, this source term for the
ODE in (\ref{mainuj}) is ${\mathcal O}(1)$ as $\epsilon\to 0$.
The dimensionless parameters $D$, $\tau$, $d_1$, and $d_2$, are related
to their dimensional counterparts by (\ref{form:par}). The qualitative
interpretation of the limits $\tau\ll 1$, $\tau\gg 1$, and $D\gg 1$, were
discussed following (\ref{form:par}).

\subsection{The Steady-State Solution for the $m$ Cells System}
We construct a steady-state solution to (\ref{mainbd}) under the
assumption that the cells are well-separated in the sense described
preceding (\ref{mainbd}). In \S \ref{sec:stability} we will then
formulate the linear stability problem for this steady-state solution.

Since in an ${\mathcal O}(\epsilon)$ neighborhood near each cell the
solution $U$ has a sharp spatial gradient, we use the method of
matched asymptotic expansions to construct the steady-state solution
to (\ref{mainbd}). In the inner region near the $j$-th cell, we
introduce the local variables $U_j$ and $\vecb y$, defined by $\vecb y
= \epsilon^{-1}(\vecb x-\vecb x_j)$ and $U_j(\vecb y) = U({\vecb x_j}
+ \epsilon {\vecb y}) \,,$ so that (\ref{mainU}) transforms to
\begin{equation}\label{ssy}
D\Delta_{\vecb y} U_j -\epsilon^2 U_j= 0\,,\qquad |\vecb y|>1\,; \qquad
D\partial_{n_j} U_j = d_1 U_j-d_2 u^1_j\,, \qquad |\vecb y|=1\,.
\end{equation}
We look for a radially symmetric solution to (\ref{ssy}) in the form
$U_j=U_j(\rho)$, where $\rho\equiv |{\vecb y}|$ and $\Delta_{\vecb
  y}=\partial_{\rho\rho}+\rho^{-1}\partial_\rho$ denotes the radially
symmetric part of the Laplacian. Therefore, to leading order, we have that
$U_j(\rho)$ satisfies
\begin{equation}\label{main:ssrho}
\partial_{\rho\rho}U_j+\rho^{-1}\partial_\rho U_j = 0\,,\qquad 1<\rho<\infty\,;
\qquad D\frac{\partial U_j}{\partial \rho} = d_1 U_j-d_2 u^1_j\,, \qquad 
\rho =1\,.
\end{equation}
The solution to (\ref{main:ssrho}) in terms of a constant
$S_j$, referred to as the {\em source strength} of the $j$-th cell, is
\begin{equation}\label{chijs}
U_j = S_j \log\rho +\chi_j\,, \qquad \chi_j = 
\frac{1}{d_1}(DS_j+d_2u^1_j)\,,\qquad j = 1, \dots, m\,.
\end{equation} 
The constant $S_j$ will be determined below upon matching the inner
solutions to the outer solution.

From the steady-state of the intracellular dynamics (\ref{mainuj})
inside each cell, we find that the source strength $S_j$ and the
steady-state solution $\vecb u_j$ satisfy the nonlinear algebraic
system 
\begin{equation}\label{bduj}
\vecb F_j(\vecb u_j)+ \frac{2\pi D}{ \tau} S_j \vecb e_1=0\,.
\end{equation}
In principle, we can determine $u^1_j$ in terms of the unknown $S_j$
as $u^1_j = u^1_j(S_j)$. The other values $u_2^j,\dots, u_n^j$ also
depend on $S_j$. Next, in terms of $u_j^1$, we will derive a system of
algebraic equations for $S_1,\ldots,S_m$, which is coupled to
(\ref{bduj}).

Upon matching the far-field behavior of the inner solution
(\ref{chijs}) to the outer solution, we obtain the outer problem
\begin{equation}\label{bdouterU}
\begin{aligned}
 \Delta U-\varphi_0^2 U &=0\,, \qquad  \vecb x\in\Omega\backslash
\lbrace{\vecb x_1,\ldots,\vecb x_m\rbrace}\,; \qquad \partial_n U=0\,, 
 \qquad \vecb x\in\partial \Omega\,,\\
 U &\sim S_j\log|\vecb x-\vecb x_j|+\frac{S_j}{\nu}+\chi_j\,, \quad 
 \text{as }\;\vecb x\rightarrow \vecb x_j\,, \quad j=1,\ldots,m\,,
\end{aligned}
\end{equation}
where we have defined $\varphi_0$ and $\nu\ll 1$ by
\begin{equation}
 \varphi_0\equiv1/\sqrt{D}\,,\qquad \nu\equiv {-1/\log\epsilon}\,.
\end{equation}
We remark that the singularity condition in (\ref{bdouterU}) is
derived by matching the outer solution for $U$ to the far-field
behavior of the inner solution (\ref{chijs}). We then introduce the
reduced-wave Green's function $G(\vecb x;\vecb x_j)$ satisfying
\bsub \label{bdgreenss_all}
\begin{equation}\label{bdgreenss}
\Delta G-\varphi_0^2 G = -\delta(\vecb x-\vecb x_j)\,,\qquad \vecb x\in 
\Omega \,; \qquad \partial_n G = 0\,, \qquad \vecb x\in \partial \Omega\,.
\end{equation}
As $\vecb x\rightarrow \vecb x_j$, this Green's function has the local behavior
\begin{equation}\label{bdgreenss_reg}
G(\vecb x;\vecb x_j)\sim -\frac{1}{2\pi}\log|\vecb x - \vecb x_j|+R_j+o(1)\,, 
\qquad \mbox{as} \quad \vecb x\rightarrow \vecb x_j\,,
\end{equation}
\esub
where $R_j=R_j(\vecb x_j)$ is called the regular part of $G(\vecb
x;\vecb x_j)$ at $\vecb x=\vecb x_j$. In terms of $G(\vecb x;\vecb
x_j)$, the solution to (\ref{bdouterU}) is
\begin{equation}\label{bdUso}
U(\vecb x)=-2\pi\sum_{i=1}^{m}S_iG(\vecb x, \vecb x_i)\,.
\end{equation}
By expanding $U$ as $\vecb x\rightarrow \vecb x_j$, and equating the
resulting expression with the required singularity behavior in
(\ref{bdouterU}), we obtain the following algebraic system
for $\vecb S=(S_1,\ldots,S_m)^T$, which we write in matrix form as
\begin{equation}\label{bdsysS2m}
\left(1+\frac{D\nu}{d_1}\right)\vecb S+2\pi\nu {\mathcal G}\vecb S=-
\frac{d_2}{d_1}\nu \vecb u^1\,.
\end{equation}
Here the Green's matrix ${\mathcal G}$, with matrix entries
$({\mathcal G})_{ij}$, and the vector $\vecb u^1$, whose $j$-th element
is the first local species in the $j$-th cell, are given by
\begin{equation*}
({\mathcal G})_{ii} = R_{i} \,, \qquad ({\mathcal G})_{ij} = 
G(\vecb x_i;\vecb x_j) \equiv G_{ij} \,, \quad i\neq j\,; \qquad \vecb u^1\equiv
\left(u_1^1,\ldots,u_m^1\right)^T \,.
\end{equation*}
Since $G_{ji}=G_{ij}$, by the reciprocity of the Green's function,
${\mathcal G}$ is a symmetric matrix.

Together with (\ref{bduj}), (\ref{bdsysS2m}) provides an approximate
steady-state solution for $\vecb u$, which is coupled to the source
strengths $\vecb S$. It is rather intractable analytically to write
general conditions on the nonlinear kinetics to ensure the existence
of a solution to the coupled algebraic system (\ref{bduj}) and
(\ref{bdsysS2m}). As such, in \S \ref{sec:examples} below we will
analyze in detail some specific choices for the nonlinear kinetics. We
remark that even if we make the assumption that the nonlinear kinetics
in the cells are identical, so that $\vecb F_j=\vecb F$ for
$j=1,\ldots,m$, we still have that $S_j$ and $\vecb u^1$ depend on $j$
through the Green's interaction matrix ${\mathcal G}$, which depends
on the spatial configuration $\lbrace{\vecb x_1,\ldots,\vecb
  x_m\rbrace}$ of the cells within $\Omega$.

In summary, after solving the nonlinear algebraic system (\ref{bduj})
and (\ref{bdsysS2m}), the approximate steady-state solution for $U$ is
given by (\ref{bdUso}) in the outer region, defined at ${\mathcal
  O}(1)$ distances from the cells, and (\ref{chijs}) in the
neighborhood of each cell.  This approximate steady-state solution is
accurate to all orders in $\nu$, since our analysis has effectively
summed an infinite order logarithmic expansion in powers of $\nu$ for
the steady-state solution. Related 2-D problems where infinite
logarithmic expansions occur for various specific applications were
analyzed in \cite{ktcw} and \cite{pill} (see also the references therein).

\subsection{Formulation of the Linear Stability Problem}\label{sec:stability}
Next, we consider the linear stability of the steady-state solution
constructed in the previous subsection. We perturb this steady-state
solution, denoted here by $U_e(\vecb x)$ in the bulk region and
$\vecb u_{e,j}$ in the $j$-th cell as $U=U_e + e^{\lambda
  t}\eta({\vecb x})$ and $\vecb u_j = \vecb u_{e,j}+ e^{\lambda
  t}\vecb \phi_j$. Upon substituting this perturbation into
(\ref{mainbd}), we obtain in the bulk region that
\bsub\label{bdeigall}
\begin{equation}\label{bdeigeta}
\begin{aligned}
\tau \lambda \eta &=D\Delta \eta-\eta\,,\qquad 
 \vecb x\in\Omega\backslash\cup_{j=1}^m\Omega_{\epsilon_j}\,; \qquad
\partial_n \eta = 0\,,\qquad \vecb x\in\partial \Omega\,,\\
\epsilon D\partial_{n_j}\eta &= d_1\eta-d_2\vecb \phi_j^1\,,
\qquad \vecb x\in\partial \Omega_{\epsilon_j}\,, \qquad j=1,\ldots,m\,.
\end{aligned}
\end{equation}
Within the $j$-th cell the linearized problem is
\begin{equation}\label{bdeigphij}
\lambda \phi_j=J_j\phi_j+ \frac{\vecb e_1}{\epsilon  \tau} 
 \int_{\partial\Omega_{\epsilon_j}}\left(d_1\eta-d_2\vecb \phi_j^1\right) \, ds\,,
\end{equation}
\esub 
where $J_j$ denotes the Jacobian matrix of the nonlinear
kinetics $\vecb F_j$ evaluated at $\vecb u_{e,j}$.  We now study
(\ref{bdeigall}) in the limit $\epsilon\to 0$ using the method of
matched asymptotic expansions. The analysis will provide a limiting
globally coupled eigenvalue problem for $\lambda$, from which we can
investigate possible instabilities.

In the inner region near the $j$-th cell, we introduce the local
variables $\vecb y=\epsilon^{-1}(\vecb x-\vecb x_j)$, with
$\rho=|{\vecb y}|$, and let $\eta_j(\vecb y)=\eta(\vecb x_j + \epsilon
\vecb y)$. We will look for the radially symmetric
eigenfunction $\eta_j$ in the inner variable $\rho$. Then, from
(\ref{bdeigeta}), upon neglecting higher order algebraic terms in $\epsilon$,
the leading order inner problem becomes
\begin{equation}\label{bdeig}
\partial_{\rho\rho}\eta_j+\rho^{-1}\partial_\rho\eta_j= 0\,, \qquad 
1<\rho<\infty\,; \qquad
D\frac{\partial \eta_j}{\partial \rho} = d_1\eta_j-d_2\phi_j^1\,,
\qquad \rho=1\,,
\end{equation}
which has the solution
\begin{equation}\label{bdBj}
\eta_j=c_j\log \rho+B_j\,, \qquad B_j=\frac{1}{d_1}(Dc_j+d_2\phi_j^1)\,,
\end{equation}
where $c_j$ is an unknown constant to be determined.  Then, upon
substituting (\ref{bdBj}) into (\ref{bdeigphij}), we obtain that
\begin{equation}\label{bdeig1}
(J_j-\lambda I)\vecb \phi_j+ \frac{2\pi D}{ \tau} c_j\vecb
  e_1=0\,, \qquad j=1,\ldots, m\,.
\end{equation}

In the outer region, defined at ${\mathcal O}(1)$ distances from the cells,
the outer problem for the eigenfunction $\eta(\vecb x)$ is
\begin{equation}\label{bdeigout}
\begin{aligned}
 &\Delta \eta-\frac{(1+\tau\lambda)}{D}\eta =0, \qquad   \vecb x\in 
\Omega\backslash\lbrace{\vecb x_1,\ldots,\vecb x_m\rbrace} \,; \qquad
 \partial_n \eta =0,\qquad \vecb x\in \partial \Omega\,,\\
 &\eta \sim c_j\log|\vecb x-\vecb x_j|+\frac{c_j}{\nu}+B_j, \qquad 
 \mbox{as} \quad \vecb x\rightarrow \vecb x_j\,, \quad j=1,\ldots, m\,,
\end{aligned}
\end{equation}
where $\nu\equiv -{1/\log\epsilon}$. We remark that the singularity
condition in (\ref{bdeigout}) as $\vecb x\rightarrow \vecb x_j$ is
derived by matching the outer solution for $\eta$ to the far-field
behavior of the inner solution (\ref{bdBj}). To solve
(\ref{bdeigout}), we introduce the eigenvalue-dependent Green's
function $G_{\lambda}(\vecb x; \vecb x_j)$, which satisfies
\begin{equation}\label{bdgneig}
\begin{aligned}
&\Delta G_{\lambda}-\varphi_{\lambda}^2G_{\lambda}=-\delta(\vecb x-\vecb x_j), 
\qquad \vecb x\in\Omega\,; \qquad \partial_n G_\lambda = 0\,,\qquad
 \vecb x\in\partial \Omega\,,\\
&G_\lambda(\vecb x;\vecb x_j) \sim -
\frac{1}{2\pi}\log|\vecb x-\vecb x_j|+R_{\lambda,j}+ o(1)\,, \qquad 
\mbox{as} \quad \vecb x\rightarrow\vecb x_j\,,
\end{aligned}
\end{equation}
where $R_{\lambda, j}\equiv R_{\lambda}(\vecb x_j)$ is the regular part of
$G_\lambda$ at $\vecb=\vecb x_j$. Here we have defined $\varphi_\lambda$ by
\begin{equation}
\varphi_\lambda\equiv\sqrt{\frac{1+\tau\lambda}{D}}\,.
\end{equation}
We must choose the principal branch of $\varphi_\lambda$, which
ensures that $\varphi_\lambda$ is analytic in $\mbox{Re}(\lambda)>0$.
For the case of an asymptotically large domain $\Omega$, this choice
for the branch cut, for which $\mbox{Re}(\theta_\lambda)>0$, also
ensures that $G_\lambda$ decays as $|\vecb x-\vecb x_j|\to \infty$.

In terms of $G_\lambda(\vecb x; \vecb x_j)$, we can represent the
outer solution $\eta(\vecb x)$ satisfying (\ref{bdeigout}), as
\begin{equation}\label{eig:out}
\eta(\vecb x) = -2\pi \sum_{i=1}^mc_iG_\lambda(\vecb x,\vecb x_i)\,.
\end{equation}
By matching the singularity condition at $\vecb x\rightarrow\vecb
x_j$, we obtain a system of equations for $c_j$ as
\begin{equation}\label{bdcj1}
 c_j + \nu B_j=-2\pi\nu\left(c_j R_{\lambda_j}+\sum_{i\neq j}^{m} c_i 
G_{\lambda, ij} \right) \,, \qquad j=1,\ldots,m \,,
\end{equation}
where $G_{\lambda, ij }\equiv G_\lambda(\vecb x_j;\vecb x_i)$.
Upon recalling that $B_j = \frac{1}{d_1}(Dc_j+d_2\phi_j^1)$ from (\ref{bdBj}),
we can rewrite (\ref{bdcj1}) in matrix form in
terms of $\vecb c =(c_1,\ldots,c_m)^T$ as 
\begin{equation}\label{bdeig2m}
\left(1+\frac{D\nu}{d_1}\right)\vecb c+\frac{d_2}{d_1}\nu\vecb\phi^1+2\pi \nu
 {\mathcal G}_\lambda \vecb c=0\,.
\end{equation}
Here we have defined the symmetric Green's matrix ${\mathcal
  G}_\lambda$, with matrix entries $({\mathcal G})_{\lambda,ij}$, and
the vector $\vecb \phi^1$ by
\begin{equation}\label{gcep:green_red}
({\mathcal G})_{\lambda,ii} = R_{\lambda,i} \,, \qquad ({\mathcal G})_{\lambda, ij} = 
G_{\lambda}(\vecb x_i;\vecb x_j) \equiv G_{\lambda, ij} \,, \quad i\neq j\,; \qquad 
\vecb \phi^1\equiv \left(\phi_1^1,\ldots,\phi_m^1\right)^T \,.
\end{equation}
The $j$-th entry of the vector $\phi^1=(\phi^1_1,\cdots,\phi^1_m)^T$
is simply the first element in the eigenvector for the $j$-th cell.
Together with (\ref{bdeig1}), the system (\ref{bdeig2m}) will yield an
eigenvalue problem for $\lambda$ with eigenvector $\vecb c$.

Next, we calculate $\phi_1$ in terms of $\vecb c$ from (\ref{bdeig1})
and then substitute the resulting expression into (\ref{bdeig2m}). If
$\lambda$ is not an eigenvalue of $J_j$, (\ref{bdeig1}) yields that
$\vecb \phi_j = 2\pi D\tau^{-1} (\lambda I-J_j)^{-1} c_j\vecb
e_1$. Upon taking the dot product with the $n$-vector $\vecb
e_1=(1,0,\ldots,0)^T$, we get $\phi_j^1 = 2\pi D \tau^{-1} c_j {\vecb
  e_1}^T (\lambda I-J_j)^{-1} {\vecb e_1}$, which yields in vector form that
\bsub \label{kdef}
\begin{equation}\label{kdef:1}
  \phi^1 = \frac{2\pi D}{ \tau} {\mathcal K} \vecb c \,,
\end{equation}
where ${\mathcal K}={\mathcal K}(\lambda)$ is the $m\times m$ diagonal
matrix with diagonal entries
\begin{equation}\label{kdef:2}
     {\mathcal K}_j = {\vecb e_1}^T (\lambda I-J_j)^{-1} {\vecb e_1} =
     \frac{1}{\det(\lambda I - J_j)} \vecb e_1 M_j^T \vecb e_1 = 
   \frac{M_{j,11}}{\det(\lambda I - J_j)} \,. 
\end{equation}
\esub
Here $M_j$ is the $n\times n$ matrix of cofactors of the matrix
$\lambda I-J_j$, with $M_{j,11}$ denoting the matrix entry in the first row
and first column of $M_{j}$, given explicitly by
\begin{equation}\label{kdef:m11}
M_{j,11}=M_{j,11}(\lambda) 
 \equiv \det\left(
\begin{array}{ccc}
\lambda - 
\frac{\partial {F}_j^2}{\partial u_2}\Big|_{\vecb{u}=\vecb{u}_{e,j}},
&\cdots, &-\frac{\partial {F}_j^2}{\partial u_n}
\Big|_{\vecb{u}=\vecb{u}_{e,j}}\\
\cdots,&\cdots,&\cdots\\
-\frac{\partial {F}_j^n}{\partial u_2}\Big|_{\vecb{u}=\vecb{u}_{e,j}}
 ,&\cdots, &\lambda 
-\frac{\partial {F}_j^n}{\partial u_n}\Big|_{\vecb{u}=\vecb{u}_{e,j}}
\end{array}
\right) \,.
\end{equation}
Here ${F}_j^2,\ldots,{F}_j^n$ denote the components of
the vector $\vecb F_j \equiv (F_{j}^1,\ldots,F_j^n)^T$, characterizing
the intracellular kinetics.

Next, upon substituting (\ref{kdef:1}) into (\ref{bdeig2m}), we obtain the
homogeneous $m\times m$ linear system
\bsub \label{gcep:full}
\begin{equation}
   {\mathcal M} \vecb c = \vecb 0 \,, 
\end{equation}
where the $m\times m$ matrix ${\mathcal M}={\mathcal M}(\lambda)$ is
defined by
\begin{equation}\label{gcep:mdef}
\mathcal{M}\equiv \left(1+\frac{D\nu}{d_1}\right)I+2\pi \nu \frac{d_2}
        {d_1 { \tau}} D {\mathcal K} + 2\pi \nu {\mathcal
          G}_\lambda\,.
\end{equation}
\esub In (\ref{gcep:mdef}), the diagonal matrix ${\mathcal K}$ has
diagonal entries (\ref{kdef:2}), and ${\mathcal G}_\lambda$ is the
Green's interaction matrix defined in (\ref{gcep:green_red}), which
depends on $\lambda$ as well as on the spatial configuration
$\lbrace{\vecb x_1,\ldots,\vecb x_m \rbrace}$ of the centers of the
small cells within $\Omega$.

We refer to (\ref{gcep:full}) as the globally coupled eigenvalue
problem (GCEP). In the limit $\epsilon\to 0$, we conclude that $\lambda$ is
a discrete eigenvalue of the linearized problem (\ref{bdeigall}) if
and only if $\lambda$ is a root of the transcendental equation
\begin{equation}\label{bdMdet}
\det\mathcal{M}=0\,.
\end{equation}
To determine the region of stability, we must seek conditions to
ensure that all such eigenvalues satisfy $\mbox{Re}(\lambda)<0$. The
corresponding eigenvector $\vecb c$ of (\ref{gcep:full}) gives the
spatial information for the eigenfunction in the bulk via
(\ref{eig:out}).

We now make some remarks on the form of the GCEP. We first observe
from (\ref{gcep:mdef}) that when $D={\mathcal O}(1)$, then to
leading-order in $\nu\ll 1$, we have that ${\mathcal M}\sim I +
{\mathcal O}(\nu)$. As such, when $D={\mathcal O}(1)$, we conclude
that to leading order in $\nu$ there are no discrete eigenvalues of
the linearized problem with $\lambda={\mathcal O}(1)$, and hence no
${\mathcal O}(1)$ time-scale instabilities. However, since
$\nu={-1/\log\epsilon}$ is not very small unless $\epsilon$ is
extremely small, this prediction of no instability in the $D={\mathcal
  O}(1)$ regime may be somewhat misleading at small finite
$\epsilon$. In \S \ref{sec:finite_d} we determine the roots of
(\ref{bdMdet}) numerically, without first assuming that $\nu\ll 1$,
for a ring-shaped pattern of cells within the unit disk $\Omega$, for
which the Green's matrix is cyclic.  In the next section we will
consider the distinguished limit $D={\mathcal O}(\nu^{-1})\gg 1$ for
(\ref{gcep:mdef}) where the linearized stability problem becomes
highly tractable analytically.

\setcounter{equation}{0}
\setcounter{section}{2}
\section{The Distinguished Limit of $D={\mathcal O}(\nu^{-1})\gg 1$}
\label{sec:largeD}

In the previous section, we constructed the steady-state solution for
the coupled cell-bulk system (\ref{mainbd}) in the limit $\epsilon\to
0$ and we derived the spectral problem that characterizes the linear
stability of this solution. In this section, we consider the
distinguished limit where the signaling molecule in the bulk diffuses
rapidly, so that $D\gg 1$. More specifically, we will consider the
distinguished limit where $D={\mathcal O}(\nu^{-1})$, and hence for
some $D_0={\mathcal O}(1)$, we set
\begin{equation}\label{bdD0}
D={D_0/\nu} \,.
\end{equation}

For $D={\mathcal O}(\nu^{-1})$, we determine a leading order
approximation for the steady-state solution and the associated
spectral problem. To do so, we first approximate the reduced-wave
Green's function for large $D$ by writing (\ref{bdgreenss}) as
\begin{equation}\label{bdgreenssD}
\Delta G-\frac{\nu}{D_0}G=-\delta(\vecb x-\vecb x_j)\,,\qquad \vecb 
x\in\Omega\,; \qquad \partial_n G=0, \qquad \vecb x\in\partial\Omega\,.
\end{equation}
This problem has no solution when $\nu=0$. Therefore, we expand 
$G=G(\vecb x; \vecb x_j)$ for $D={D_0/\nu} \gg 1$ as
\begin{equation}\label{bdgexp}
G=\frac{1}{\nu}G_{-1}+G_0+\nu G_1+\ldots \,.
\end{equation}
Upon substituting (\ref{bdgexp}) into (\ref{bdgreenssD}), we equate
powers of $\nu$ to obtain a sequence of problems for $G_i$ for
$i=-1,0,1$. This leads to the following two-term expansion for 
$G(\vecb x;\vecb x_j)$ and its regular part $R_j$ in the limit
$D={D_0/\nu} \gg 1$:
\begin{equation}\label{bdgexp1}
G(\vecb x; \vecb x_j)=\frac{D_0}{\nu |{\Omega}|}+
G_{0}(\vecb x; \vecb x_j)+\cdots\,, \qquad R_j =\frac{D_0}{\nu |{\Omega}|}+
R_{0,j} + \cdots \,.
\end{equation}
Here $G_{0}(\vecb x; \vecb x_j)$, with regular part $R_{0j}$, is the
Neumann Green's function defined as the unique solution to
\begin{equation}\label{g0:neum}
\begin{aligned}
\Delta G_{0}&=\frac{1}{|{\Omega}|}-\delta(\vecb x-\vecb x_j)\,, \qquad 
\vecb x\in\Omega\,; \qquad \partial_n G_0=0\,,\quad\vecb x\in 
\partial\Omega\,; \qquad \int_{\Omega} G_0\: d\vecb x =0\,,\\
G_0(\vecb x;\vecb x_j) &\sim -\frac{1}{2\pi}\log|{\vecb x-\vecb x_j}|+
 R_{0,j}, \qquad \vecb x\rightarrow\vecb x_j\,.
\end{aligned}
\end{equation}

We then substitute the expansion (\ref{bdgexp1}) and $D={D_0/\nu}$ into
the nonlinear algebraic system (\ref{bduj}) and (\ref{bdsysS2m}), which
characterizes the steady-state solution, to obtain that
\begin{equation}\label{bdsysS2m1}
\left(1+\frac{D_0}{d_1}\right)\vecb S+\frac{2\pi m D_0}{|\Omega|} 
  {\mathcal E} \vecb S + 2\pi \nu {\mathcal G}_0 \vecb S 
  =-\frac{d_2}{d_1}\nu \vecb u^1\,; \qquad \vecb F_j(\vecb u_j) + 
  \frac{2\pi D_0}{\tau \nu} S_j \vecb e_1 = 0\,, \quad j=1,\ldots, m \,,
\end{equation}
where the $m\times m$ matrices ${\mathcal E}$ and the Neumann Green's 
matrix ${\mathcal G}_0$, with entries $({\mathcal G}_0)_{ij}$, are
defined by
\begin{equation}\label{bdEJ}
{\mathcal E} \equiv \frac{1}{m} \vecb e \vecb e^T \,; \qquad
\left({\mathcal G}_0\right)_{ij} = G_{0}(\vecb x_i;\vecb x_j)\equiv G_{0,ij} \,,
 \quad i\neq j \,, \qquad
\left({\mathcal G}_0\right)_{ii} = R_{0,i} \,.
\end{equation}
Here $\vecb e$ is the $m$-vector $\vecb e\equiv (1,\ldots,1)^T$.  The
leading-order solution to (\ref{bdsysS2m1}) when $\nu\ll 1$ has the
form
\begin{equation}\label{dlarge:s}
   \vecb S = \nu \vecb S_0 + {\mathcal O}(\nu^2) \,, \qquad
    \vecb u_j = \vecb u_{j0} + {\mathcal O}(\nu) \,.
\end{equation}
From (\ref{bdsysS2m1}) we conclude that $\vecb S_0$ and $\vecb u_{j0}$
satisfy the limiting leading-order nonlinear algebraic system
\begin{equation}\label{Dlarge:1ss}
\left(1+\frac{D_0}{d_1}\right)\vecb S_0+\frac{2\pi m D_0}{|\Omega|} 
  {\mathcal E} \vecb S_0 
  =-\frac{d_2}{d_1} \vecb u_0^1\,; \qquad \vecb F_j(\vecb u_{0j}) + 
  \frac{2\pi D_0}{ \tau} S_{0j} \vecb e_1 = 0\,, \quad j=1,\ldots, m \,.
\end{equation}
Since this leading order system does not involve the Neumann Green's
matrix ${\mathcal G}_0$, we conclude that $\vecb S_0$ is independent of the 
spatial configuration of the cells within $\Omega$.

For the special case where the kinetics $\vecb F_j$ is
identical for each cell, so that $\vecb F_j=\vecb F$ for
$j=1,\ldots,m$, we look for a solution to (\ref{Dlarge:1ss}) with
identical source strengths, so that $S_{0j}$ and 
$\vecb u_{0j}=\vecb u_{0}$ are independent of $j$. Therefore, we write
\begin{equation}
    \vecb S_0 = S_{0c} \vecb e \,, \label{Dlarge:scom}
\end{equation}
where $S_{0c}$ is the common source strength.  From
(\ref{Dlarge:1ss}), where we use ${\mathcal E}\vecb e=\vecb e$, this
yields that $S_{0c}$ and $\vecb u_0$ satisfy the $m+1$ dimensional
nonlinear algebraic system
\begin{equation}\label{Dlarge:ss}
\left(1+\frac{D_0}{d_1} + \frac{2\pi m D_0}{|\Omega|} \right) S_{0c}
  = -\frac{d_2}{d_1} u_0^1\,, \qquad \vecb F(\vecb u_{0}) + 
  \frac{2\pi D_0}{ \tau} S_{0c} \vecb e_1 = 0\,,
\end{equation}
where $u_0^1$ is the first component of $\vecb u_0$.  This simple
limiting system will be studied in detail in the next section for
various choices of the nonlinear intracellular kinetics $\vecb F(\vecb
u_0)$.

Next, we will simplify the GCEP, given by (\ref{gcep:full}), when
$D={D_0/\nu}\gg 1$, and under the assumption that the reaction
kinetics are the same in each cell. In the same way as was derived in
(\ref{bdgreenssD})--(\ref{bdgexp1}), we let $D={D_0/\nu}\gg 1$ and
approximate the $\lambda$-dependent reduced Green's function
$G_\lambda(\vecb x;\vecb x_j)$, which satisfies (\ref{bdgneig}). For
$\tau={\mathcal O}(1)$, we calculate, in place of (\ref{bdgexp1}),
that the two-term expansion in terms of the Neumann Green's function
$G_0$ is
\begin{equation*}
G_\lambda(\vecb x;\vecb x_j) = \frac{D_0}{\nu (1+\tau\lambda) |{\Omega}|}
 +G_{0}(\vecb x;\vecb x_j)+ {\mathcal O}(\nu)\,, \qquad
R_{\lambda,j} = \frac{D_0}{\nu (1+\tau\lambda)|{\Omega}|}
 +R_{0,j} + {\mathcal O}(\nu)\,.
\end{equation*}
Therefore, for $D={D_0/\nu}\gg 1$ and $\tau={\mathcal O}(1)$, we have in
terms of ${\mathcal E}$ and the Neumann Green's matrix ${\mathcal G}_0$ of
(\ref{bdEJ}), that
\begin{equation}\label{dlarge:glam}
 {\mathcal G}_\lambda=\frac{mD_0}{\nu(1+\tau\lambda)|{\Omega}|}
  {\mathcal E} + {\mathcal G}_0 + {\mathcal O}(\nu) \,.
\end{equation}

We substitute (\ref{dlarge:glam}) into (\ref{gcep:mdef}), and set
$D={D_0/\nu}$.  In (\ref{gcep:mdef}), we calculate to leading order in
$\nu$ that the matrix ${\mathcal   K}(\lambda)$, defined in (\ref{kdef:2}), 
reduces to 
\begin{equation}\label{klambda:2t}
   {\mathcal K} \sim \frac{M_{11}}{\det(\lambda I - J)} + {\mathcal O}(\nu)\,,
\end{equation}
where $J$ is the Jacobian of $\vecb F$ evaluated at the solution
$\vecb u_0$ to the limiting problem (\ref{Dlarge:ss}), and $M_{11}$ is
the cofactor of $\lambda I - J$ associated with its first row and
first column. The ${\mathcal O}(\nu)$ correction in ${\mathcal
  K}(\lambda)$ arises from the higher order terms in the Jacobian
resulting from the solution to the full system (\ref{bdsysS2m1}). 

In this way, the matrix ${\mathcal M}$ in (\ref{gcep:mdef}), which is
associated with the GCEP, reduces to leading order to
\bsub \label{mdef:dlarge}
\begin{equation}\label{gcep:mdef_large}
    {\mathcal M} = a(\lambda) I + b(\lambda) {\mathcal E} + {\mathcal O}(\nu)
  \,,
\end{equation}
where $a(\lambda)$ and $b(\lambda)$ are defined by
\begin{equation} \label{mdef:dlarge_ab}
  a(\lambda) = 1 + \frac{D_0}{d_1} + \frac{2\pi d_2}{d_1 { \tau}} 
  \frac{D_0 M_{11}}
  {\det(\lambda I-J)} \,, \qquad b(\lambda) = 
  \frac{2\pi m D_0}{(1+\tau\lambda)|\Omega|} \,. 
\end{equation}
\esub We remark that the ${\mathcal O}(\nu)$ correction terms in
(\ref{gcep:mdef_large}) arises from both $2\pi\nu {\mathcal G}_0$,
which depends on the spatial configuration of the cells within
$\Omega$, and the ${\mathcal O}(\nu)$ term in ${\mathcal K}$ as
written in (\ref{klambda:2t}).

Therefore, when $D={D_0/\nu}$, it follows from (\ref{gcep:mdef_large})
and the criterion (\ref{bdMdet}) of the GCEP that $\lambda$ is a
discrete eigenvalue of the linearization if and only if there exists a
nontrivial solution $\vecb c$ to
\begin{equation}\label{dlarge:key}
     \left(  a(\lambda) I + b(\lambda) {\mathcal E} \right) \vecb c = 
  \vecb 0\,.
\end{equation}
Any such eigenvalue with $\mbox{Re}(\lambda)>0$ leads to a linear instability
of the steady-state solution when $D={\mathcal O}(\nu^{-1})$.

We now derive explicit stability criteria from (\ref{dlarge:key}) by
using the key properties that ${\mathcal E}\vecb e=\vecb e$ and
${\mathcal E} \vecb q_j=0$ for $j=2,\ldots,m$, where $\vecb q_j$ for
$j=2,\ldots,m$ are an orthogonal basis of the $m-1$ dimensional
perpendicular subspace to $\vecb e$, i.e $\vecb q_j^T\vecb e=0$.
We obtain that $\lambda$ is a discrete eigenvalue for 
the {\em synchronous mode}, corresponding to $\vecb c=\vecb e$, whenever
$\lambda$ satisfies
\begin{equation}\label{dlarge:sync}
   a(\lambda)+b(\lambda) \equiv 
   1 + \frac{D_0}{d_1} + \frac{2\pi d_2}{d_1 { \tau}} \frac{D_0 M_{11}}
  {\det(\lambda I-J)} + \frac{2\pi m D_0}{(1+\tau\lambda)|\Omega|} =0 \,.
\end{equation}
This expression can be conveniently written as
\begin{equation}\label{dlarge:sync_new}
   \frac{M_{11}}{\det(\lambda I-J)} = -\frac{{ \tau}}{2\pi d_2} \left( 
  \frac{\kappa_1 \tau \lambda + \kappa_2}{\tau\lambda + 1} \right) \,, \qquad
 \mbox{where} \qquad 
   \kappa_1 \equiv \frac{d_1}{D_0} + 1 \,, \quad \kappa_2 \equiv \kappa_1 + 
  \frac{2m \pi d_1}{|\Omega|} \,. 
\end{equation}
In contrast, $\lambda$ is a discrete eigenvalue for 
the {\em asynchronous or competition modes}, corresponding to 
$\vecb c=\vecb q_j$ for $j=2,\ldots,m$, whenever $\lambda$ satisfies
$a(\lambda)=0$. This yields for any $m\geq 2$ that
\begin{equation}\label{dlarge:async}
   \frac{M_{11}}{\det(\lambda I-J)} = - \frac{{ \tau}}{2\pi
     d_2} \left( \frac{d_1}{D_0} + 1 \right) \,.
\end{equation}

Any discrete eigenvalue for either of the two modes that
satisfies $\mbox{Re}(\lambda)>0$ leads to an instability.  If all such
eigenvalues satisfy $\mbox{Re}(\lambda)<0$, then the steady-state
solution for the regime $D={D_0/\nu}$ is linearly stable on an
${\mathcal O}(1)$ time-scale.

\begin{remark}\label{remark} The spectral problems (\ref{dlarge:sync_new}) and
(\ref{dlarge:async}) have a remarkably similar form to the spectral
  problem characterizing the linear stability of localized spot
  solutions to the Gierer-Meinhardt (GM) RD system
\begin{equation}\label{gm}
  a_t=\epsilon^2 \Delta a - a + {a^2/h} \,, \qquad \tau h_t=D \Delta h-
h + \epsilon^{-2} a^2 \,,
\end{equation}
posed in a bounded two-dimensional domain $\Omega$ with $\partial_n
a=\partial_n h=0$ on $\partial\Omega$. For $\epsilon\to 0$, and for
$D={D_0/\nu}$ with $\nu={-1/\log\epsilon}\ll 1$, the linear stability
of an $m$-spot solution, with eigenvalue parameter $\lambda$,
is characterized by the roots of a transcendental equation of the
form (see \cite{wei1})
\begin{equation}\label{gm:stab}
     \frac{a+b \tau\lambda}{c+d \tau \lambda} = {\mathcal F}(\lambda) \,,
  \qquad {\mathcal F}(\lambda)\equiv \frac{\int_{0}^{\infty} w\left(L_0-\lambda
  \right)^{-1} w^2 \rho \, d\rho}{\int_{0}^{\infty} w^2\rho\, d\rho} \,,
\end{equation}
where $w(\rho)>0$ is the radially symmetric ground-state solution of
$\Delta w-w+w^2=0$ with $w\to 0$ as $\rho\to\infty$, and $L_0$ is the
local operator $L_0\Phi\equiv \Delta \Phi - \Phi + 2w\Phi$, where
$L_0^{-1}$ is restricted to radially symmetric functions. In direct
analogy to our cell-bulk problem, it was shown in \cite{wei1} that
there can be either synchronous or asynchronous instabilities of the
amplitudes of the localized spots. For synchronous instabilities, the
coefficients in (\ref{gm:stab}) are $a=b=1$, $c=2$, and
$d={2/(1+\mu)}$, where $\mu\equiv {2\pi m D_0/|\Omega|}$, while for
asynchronous instabilities we have $a={(1+\mu)/2}$, and
$b=c=d=0$. Since $L_0$ has a unique eigenpair in $H^{1}$ with positive
eigenvalue $\sigma_0>0$, the function ${\mathcal F}(\lambda)$ in
(\ref{gm:stab}) is analytic in $\mbox{Re}(\lambda)>0$, with the
exception of a simple pole at $\lambda=\sigma_0>0$. In this way,
(\ref{gm:stab}) is remarkably similar to our cell-bulk spectral
problems (\ref{dlarge:sync_new}) and (\ref{dlarge:async}) when there
is only a single intracellular species that is self-activating in the
sense that $J\equiv F_{u}(u_e)>0$ so that ${M_{11}/\det(\lambda
  I-J)}={1/(\lambda-F_u(u_e))}$.  Spectral problems similar to
(\ref{gm:stab}) also occur for the Schnakenberg RD
system \cite{wei2}, the Brusselator \cite{rozada}, and the Gray-Scott
model \cite{chen}. For the GM model, it can be shown that there is a
unique Hopf bifurcation value of $\tau$ for the synchronous mode
whenever $\mu>1$. However, for our cell-bulk model, below in \S
\ref{one:ode} we prove that Hopf bifurcations are impossible for the
synchronous mode whenever there is only one intracellular species.
\end{remark}

\setcounter{equation}{0}
\setcounter{section}{3}
\section{Examples of the Theory: Finite Domain With $D={\mathcal O}(\nu^{-1})$}
\label{sec:examples}

In this section we will study the leading-order steady-state problem
(\ref{Dlarge:ss}), and its associated spectral problem
(\ref{dlarge:sync_new}) and (\ref{dlarge:async}), for various special
cases and choices of the reaction kinetics $\vecb F$. We investigate
the stability properties of the steady-state (\ref{Dlarge:ss}) as
parameters are varying, and in particular, find conditions for Hopf
bifurcations to occur.

\subsection{Example 1: $m$ Cells; One Local Component}\label{one:ode}
To illustrate our theory, we first consider a system such that the
local dynamics inside each cell consists of only a single component
with arbitrary scalar kinetics $F$.  For this case, the steady-state
problem for $\vecb u_0=u_0$ and $S$, given by (\ref{Dlarge:ss}),
reduces to two algebraic equations.  We will study the stability
problem for both the synchronous and asynchronous modes.  We show that
for any $F(u)$, the steady-state can never be destablilized by a Hopf
bifurcation.

For the one-component case, we calculate $M_{11}=1$ and $\det(\lambda
I-J)=\lambda-F_u^e$, where $F_u^e$ is defined as the derivative of
$F(u)$ evaluated at the steady-state $u_0$. From (\ref{dlarge:sync_new}),
the spectral problem for the synchronous mode reduces to 
\bsub \label{d0:one}
\begin{equation}\label{mcell1ode:sync}
\lambda^2-\lambda p_1 + p_2 =0 \,, \qquad p_1 \equiv
F_u^e-\frac{\gamma}{ \tau} -\frac{\zeta}{\tau} \,, \qquad p_2 \equiv
\frac{1}{\tau}\left(\frac{\gamma}{ \tau}-\zeta F_u^e\right) \,,
\end{equation}
where
\begin{equation}\label{alphabeta}
\gamma\equiv\frac{2\pi d_2 D_0}{d_1+D_0}>0\,, \qquad
\zeta\equiv1+\frac{2\pi md_1D_0}{|\Omega| (d_1+D_0)}>1\,.
\end{equation}
\esub The following result characterizes the stability properties for
the synchronous mode:

\begin{result}\label{d0:theorem_1}
There can be no Hopf bifurcations associated with the synchronous mode.
Moreover, suppose that
\begin{equation}\label{d0:stab_one}
  F_{u}^e < \frac{\gamma}{\zeta  \tau} = \frac{2\pi
    d_2}{ \tau} \left[ 1+ \frac{d_1}{D_0} + \frac{2\pi m
      d_1}{|\Omega|} \right]^{-1} \,.
\end{equation}
Then, we have $\mbox{Re}(\lambda)<0$, and so the steady-state is
linearly stable to synchronous perturbations. If $F_{u}^{e} >
{\gamma/\left(\zeta \tau\right)}$, the linearization has
exactly one positive eigenvalue.
\end{result}

\begin{proof}
For a Hopf bifurcation to occur we need $p_1=0$ and $p_2>0$. Upon
setting $p_1=0$, we get $F_u^e={\left(\gamma+ \zeta\right)/\tau} >0$.
Upon substituting this expression into the formula for
$p_2$ in (\ref{mcell1ode:sync}) we get
\begin{equation}
p_2 = \frac{1}{ \tau}\left( \frac{\gamma}{ \tau}
 -\zeta F_u^e\right)=\frac{1}{{\tau^2}} \left( \gamma(1-\zeta)-
\zeta^2\right)<0\,,
\end{equation}
since $\gamma>0$ and $\zeta>1$. Therefore, there can be no Hopf
bifurcation for the synchronous mode.

Next, to establish the stability threshold, we note that the
steady-state solution is stable to synchronous perturbations if and
only if $p_1<0$ and $p_2>0$. From (\ref{mcell1ode:sync}), we have that
$p_1<0$ and $p_2>0$ when
\begin{equation}\label{equal}
   \tau F_u^e < \zeta + \gamma \,, \qquad \tau F_u^e < {\gamma/\zeta}\,,
\end{equation}
respectively, which implies that we must have $\tau F_u^e
<\min(\zeta+\gamma,{\gamma/\zeta})$.  Since $\zeta>1$, the two
inequalities in (\ref{equal}) hold simultaneously only when $\tau
F_u^e <{\gamma/\zeta}$. This yields that
$\mbox{Re}(\lambda)<0$ when (\ref{d0:stab_one}) holds. Finally, when
$F_{u}^{e} >{\gamma/\left(\zeta \tau\right)}$, then $p_2<0$,
and so there is a unique positive eigenvalue. \end{proof}

This result shows that the effect of cell-bulk coupling is that the
steady-state of the coupled system can be linearly stable even when
the reaction kinetics is self-activating in the sense that
$F_{u}^{e}>0$. We observe that the stability threshold
${\gamma/\zeta}$ is a monotone increasing function of $D_0$, with
${\gamma/\zeta}\to 0$ as $D_0\to 0$ and ${\gamma/\zeta}$ tending to a
limiting value as $D_0\to \infty$. This shows that as $D_0$ is
decreased, corresponding to when the cells are effectively more
isolated from each other, there is a smaller range of $F_u^e>0$ where
stability can still be achieved.

Next, we will consider the spectral problem for the asynchronous mode.
From (\ref{dlarge:async}), we get
\begin{equation}
\frac{1}{\lambda-F_u^e}=-\frac{{\tau}}{\gamma}\,,
\end{equation}
where $\gamma$ is defined in (\ref{alphabeta}). Therefore,
$\lambda=F_u^e-{\gamma/{\tau}}$, and so $\lambda$ is real and
no Hopf bifurcation can occur. This asynchronous mode is stable if
$F_u^e<{\gamma/ \tau}$. Since $\zeta>1$, we observe, upon
comparing this threshold with that for the synchronous mode in
(\ref{d0:stab_one}), that the stability criterion for the synchronous
mode is the more restrictive of the two stability thresholds.

In summary, we conclude that a Hopf bifurcation is impossible for
(\ref{mainbd}) in the parameter regime $D={D_0/\nu}$ when there is only
one dynamically active species inside each of $m$ small cells.
Moreover, if $F_{u}^e < {\gamma/(\zeta \tau)}$, where
$\gamma$ and $\zeta$ are defined in (\ref{alphabeta}), then the
steady-state solution is linearly stable to both the synchronous and
asynchronous modes.

\subsection{Example 2: $m$ Cells; Two Local Components}

Next, we assume that there are two dynamically active local species
inside each of $m$ distinct cells. For ease of notation, we write the
intracellular variable as $\vecb u=(v,w)^T$ and the local kinetics as
$\vecb F(v,w)=(F(v,w),G(v,w))^T$.  In this way, the steady-state
problem (\ref{Dlarge:ss}) becomes
\begin{equation}\label{mcell2ode:ss}
\left(1+\frac{D_0}{d_1} + \frac{2\pi m D_0}{|\Omega|} \right) S_{0c} =
-\frac{d_2}{d_1} v_e\,, \qquad F(v_e,w_e)+\frac{2\pi D_0}{
  \tau} S_{0c}= 0\,, \qquad G(v_e,w_e)=0\,.
\end{equation}
Given specific forms for $F$ and $G$, we can solve the steady-state
problem (\ref{mcell2ode:ss}) either analytically or numerically.

To analyze the stability problem, we first calculate the cofactor
$M_{11}$ as $M_{11}=\lambda-G_w^e$ and $\det(\lambda
I-J)=\lambda^2-\mbox{tr}(J)\lambda+\det(J)$, where $\mbox{tr}(J)$ and
$\det(J)$ are the trace and determinant of the Jacobian of $\vecb F$,
given by
\begin{equation}
 \mbox{tr}(J)=F_v^e+G_w^e\,,\qquad \det(J)=F_v^eG_w^e-F_w^eG_v^e\,.
\end{equation}
Here $F_i^e$, $G_i^e$ are partial derivatives of $F$, $G$ with respect
to $i$, with $i\in(v,w)$, evaluated at the solution to
(\ref{mcell2ode:ss}).

Next, we analyze the stability of the steady-state solution with respect to
either synchronous or asynchronous perturbations. For the synchronous mode, 
we obtain, after some algebra,  that (\ref{dlarge:sync_new}) can be
reduced to the study of the cubic
\bsub \label{d0:cubic}
\begin{equation}\label{mcell2ode:H3}
\mathcal{H}(\lambda)\equiv\lambda^3+\lambda^2 p_1+\lambda p_2+p_3=0\,.
\end{equation}
where $p_1$, $p_2$, and $p_3$, are defined in terms of
$\gamma$ and $\zeta$, as given in (\ref{alphabeta}),  by
\begin{equation}\label{mcell2ode:p123}
\begin{aligned}
p_1\equiv \frac{\gamma}{ \tau}
+\frac{\zeta}{\tau}-\mbox{tr}(J)\,,\qquad p_2\equiv
\det(J)-\frac{\gamma}{ \tau}
G_w^e+\frac{1}{\tau}(\frac{\gamma}{ \tau}-\zeta
\mbox{tr}(J))\,,\qquad p_3\equiv \frac{1}{\tau}(\zeta \det(J)-
 \frac{\gamma}{ \tau} G_w^e)\,.
\end{aligned}
\end{equation}
\esub

To determine whether there is any eigenvalue in
$\mbox{Re}(\lambda)>0$, and to detect any Hopf bifurcation boundary in
parameter space, we use the well-known \textit{Routh-Hurwitz
  criterion} for a cubic function.  It is well known that all three
roots to the cubic satisfy $\mbox{Re}(\lambda)<0$ if and only if the
following three conditions on the coefficients hold:
\begin{equation}
p_1>0\,, \qquad p_3>0\,,\qquad p_1p_2>p_3\,.
\end{equation}
To find the Hopf bifurcation boundary, we need only consider a special
cubic equation which has roots $\lambda_1=a<0$ and $\lambda_{2,3}=\pm
i\omega$, for which
$(\lambda-a)(\lambda-i\omega)(\lambda+i\omega)=\lambda^3-
a\lambda^2+\omega^2\lambda-a\omega^2=0\,.$ Comparing this expression
with (\ref{mcell2ode:H3}), and using the \textit{Routh-Hurwitz
  criterion}, we conclude that on any Hopf bifurcation boundary the
parameters must satisfy
\begin{equation}
p_1>0\,,\qquad p_3>0\,, \qquad p_1p_2=p_3\,. \label{mcell2ode:syn}
\end{equation} 
We will return to this criterion below when we study two specific
models for the local kinetics $(F,G)$.

Next, we consider the spectral problem for the asynchronous mode. Upon
substituting the expressions of $M_{11}$ and $\det(\lambda I-J)$ into
(\ref{dlarge:async}) and reorganizing the resulting expression,
(\ref{dlarge:async}) becomes the quadratic equation 
\begin{equation}
\lambda^2-\lambda q_1 + q_2 = 0 \,, \qquad \mbox{where} \qquad q_1
\equiv \mbox{tr}(J)- \frac{\gamma}{ \tau} \,, \qquad q_2 \equiv
\det(J)- \frac{\gamma}{ \tau} G_w^e \,.
\end{equation}
For a Hopf bifurcation to occur, we require that $q_1=0$ and $q_2>0$, which
yields that
\begin{equation}\label{mcell2ode:asyn_all}
\frac{\gamma}{ \tau} =\mbox{tr}(J)=F_v^e+G_w^e\,, \qquad \mbox{provided that}
 \qquad  \det(J)-\frac{\gamma}{ \tau} G_w^e=-G_v^eF_w^e-(G_w^e)^2>0\,.
\end{equation}
Finally, we conclude that $\mbox{Re}(\lambda)<0$ for the
asynchronous modes if and only if
\begin{equation}
 \mbox{tr}(J)<{\gamma/ \tau} \,, \qquad \mbox{and} \qquad 
 \det(J) -\frac{\gamma}{ \tau} G_w^e>0
  \,.
\end{equation}

To write the stability problem for the asynchronous mode in terms of
$D_0$, we use (\ref{alphabeta}) for $\gamma$ in terms of $D_0$ to
obtain from the conditions (\ref{mcell2ode:asyn_all}) that the Hopf
bifurcation threshold satisfies the transcendental equation
\begin{equation}\label{mcell:d0_async}
D_0=\frac{\tau d_1 \mbox{tr}(J)}{2\pi d_2-\tau \mbox{tr}(J)}\,,  \qquad
  \mbox{provided that} \qquad 
 D_0\left( \frac{2\pi d_2}{\tau} G_w^e-\det (J)\right)<d_1\det (J)\,.
\end{equation}
We observe that in this formulation, both $\mbox{tr}(J)$ and $\det(J)$
depend on the local kinetics and the steady-state solution, with the
latter depending on $D_0$ and $\tau$.  In the next two subsections we
study in some detail two specific choices for the local kinetics, and
we calculate phase diagrams where oscillatory instabilities can occur.

\subsubsection{Local Kinetics Described by the Sel'kov model}\label{sec:ex:selkov}

We first consider the Sel'kov model, used in simple models of glycolysis, 
where $F(v,w)$ and $G(v,w)$ are given in terms of parameters
$\alpha>0$, $\mu>0$, and $\epsilon_0>0$ by
\begin{equation}\label{2dsel:fg}
F(v,w)=\alpha w+wv^2-v\,, \qquad G(v,w) =\epsilon_0\left
(\mu-(\alpha w+wv^2)\right) \,.
\end{equation}
First, we determine the approximate steady-state solution $v_e$ and
$w_e$ by substituting (\ref{2dsel:fg}) into (\ref{mcell2ode:ss}). This
yields that
\begin{equation}\label{2dsel:ss_sol}
 v_e=\frac{\mu}{\left[1+ {2\pi D_0\beta/{ \tau}}\right]}\,, \qquad
 w_e=\frac{\mu}{\alpha+v_e^2}\,, \qquad S_{0c}= -\beta v_e\,, \qquad
 \mbox{where} \qquad \beta \equiv \frac{d_2}{d_1+D_0+2\pi md_1D_0/|\Omega|}\,.
\end{equation}
As needed below, the partial derivatives of $F$ and $G$ evaluated at
the steady-state solution are
\bsub \label{2dsel:jac}
\begin{equation}\label{2dsel:fgp}
F_v^e = 2v_ew_e-1\,, \qquad F_w^e=\alpha+v_e^2\,, \qquad
G_v^e=-2\epsilon_0 v_ew_e\,, \qquad G_w^e=-\epsilon_0(\alpha+v_e^2)\,,
\end{equation}
which yields
\begin{equation}
  \mbox{det}(J) = \epsilon_0\left(\alpha+ v_e^2\right)=-G_w^e >0 \,, \qquad
  \mbox{tr}(J)= 2 v_e w_e -1  -\epsilon_0 \left(\alpha + v_e^2\right) \,.
\end{equation}
\esub

\begin{figure}[htbp]
\begin{center}
\includegraphics[width=0.45\textwidth,height=5.0cm]{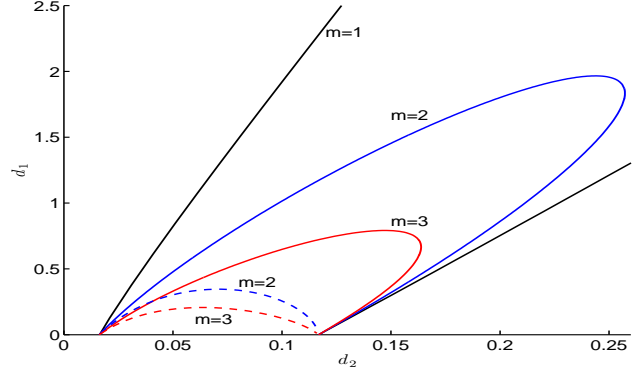}
\caption{Hopf bifurcation boundaries for the synchronous (solid curve)
  and asynchronous (dashed curve) modes for the Sel'kov model for
  different numbers $m$ of cells in the $d_1$ versus $d_2$ parameter
  plane. The synchronous mode for $m=1$ is unstable between the two
  black lines. For $m=2$ and $m=3$ the synchronous mode is
  unstable in the horseshoe-shaped region bounded by the blue and red
  solid curves, respectively. Inside the dotted regions for $m=2$ and
  $m=3$ the asynchronous mode is unstable. For the asynchronous mode,
  the boundary of these regions is given parametrically by
  (\ref{2dcell:a_par}). The parameters used are $\mu=2$, $\alpha=0.9$,
  $\epsilon_0=0.15$, $\tau=1$, $D_0=1$, and $|\Omega|=10$.}
\label{d1d2sel}
\end{center}
\end{figure}

To study possible synchronous oscillations of the $m$ cells, we
compute the Hopf bifurcation boundaries as given in
(\ref{mcell2ode:syn}), where we use (\ref{2dsel:jac}). For the
parameter set $\tau=1$, $D_0=1$, $|\Omega|=10$, $\mu=2$, $\alpha=0.9$, and
$\epsilon_0=0.15$, we obtain the Hopf bifurcation boundary in the
$d_1$ versus $d_2$ parameter plane as shown by the solid curves in
Fig.~\ref{d1d2sel} for $m=1,2,3$.

Next, to obtain instability thresholds corresponding to the asynchronous
mode, we substitute (\ref{2dsel:jac}) into (\ref{mcell2ode:asyn_all})
to obtain that the Hopf bifurcation boundary is given by
\begin{equation}
\gamma=\tau \mbox{tr}(J) \equiv \tau
  \left[2v_ew_e-1-\epsilon(\alpha+v_e^2) \right]\,,
\label{2dsel:asy_1}
\end{equation}
provided that $\mbox{det}(J)-\mbox{tr}(J) G_{w}^e>0$. This latter
condition can be written, using (\ref{2dsel:jac}), as
$\epsilon_0(\alpha+v_e^2)\left(1+\mbox{tr}(J)\right)>0$, and so is
satisfied provided that $\mbox{tr}(J)>-1$. Since $\gamma>0$ from
(\ref{alphabeta}), (\ref{2dsel:asy_1}) implies that we must have
$\mbox{tr}(J)>0$, which guarantees that $\mbox{det}(J)-\mbox{tr}(J)
G_{w}^e>0$ always holds at a Hopf bifurcation. In this way, and by
substituting (\ref{2dsel:ss_sol}) for $w_e$ into (\ref{2dsel:asy_1}),
we obtain that the asynchronous mode has a pure imaginary pair of
complex conjugate eigenvalues when
\begin{equation}\label{2dsel:hopf_async}
\gamma=  \tau \left[\frac{2 v_e \mu}{\alpha + v_e^2} -1 -
  \epsilon_0 \left(\alpha + v_e^2 \right) \right] \,, \qquad \mbox{where}
  \qquad v_e = \frac{\mu}{\left[1+ {2\pi D_0\beta/ \tau}\right]}\,.
\end{equation}
Here $\gamma$ and $\beta$, depending on $d_1$, $d_2$, $m$,
$|\Omega|$, and $D_0$, are defined in (\ref{alphabeta}) and
(\ref{2dsel:ss_sol}), respectively. By using these expressions for
$\gamma$ and $\beta$, we can readily determine a parametric form for
the Hopf bifurcation boundary in the $d_1$ versus $d_2$ plane as the
solution to a linear algebraic system for $d_1$ and $d_2$ in terms of
the parameter $v_e$ with $0<v_e<\mu$, given by
\bsub \label{2dcell:a_par}
\begin{equation}
   d_1 = \frac{D_0(a_{12}-a_{22})}{a_{11}a_{22}-a_{21}a_{12}}\,, \qquad
   d_2 = \frac{D_0(a_{21}-a_{11})}{a_{11}a_{22}-a_{21}a_{12}}\,,
 \label{2dcell:a_par_1}
\end{equation}
where $a_{11}$, $a_{12}$, $a_{22}$, and $a_{21}$, are defined in terms of the
parameterization $v_e$ by
\begin{equation}
  a_{11}\equiv  1 + \frac{2\pi m D_0}{|\Omega|} \,, \qquad
  a_{12}\equiv -\frac{1}{\beta(v_e)} \,, \qquad a_{21}\equiv 1 \,, \qquad
  a_{22}\equiv -\frac{2\pi D_0}{\gamma(v_e)} \,,
\end{equation}
where
\begin{equation}
\gamma(v_e)\equiv \tau \left[\frac{2 v_e \mu}{\alpha + v_e^2}
  -1 - \epsilon_0 \left(\alpha + v_e^2 \right)\right]\,, \qquad
  \beta(v_e)\equiv  \tau \frac{(\mu-v_e)}{2\pi D_0 v_e} \,.
\end{equation}
\esub

By varying $v_e$, with $0<v_e<\mu$, and retaining only the portion of the
curve for which $d_1>0$ and $d_2>0$, we obtain a parametric form for
the Hopf bifurcation boundary for the asynchronous mode in the 
$d_1$ versus $d_2$ parameter plane. For $m=2$ and $m=3$, these are
the dashed curves shown in Fig.~\ref{d1d2sel}.

We now discuss qualitative aspects of the Hopf bifurcation boundaries
for both synchronous and asynchronous modes for various choices of $m$
as seen in Fig.~\ref{d1d2sel}. For $m=1$, we need only consider the
synchronous instability. The Hopf bifurcation boundary is given by the
two black lines, and the region with unstable oscillatory dynamics is
located between these two lines. For $m=2$, inside the region bounded
by the blue solid curve, the synchronous mode is unstable and under
the blue dashed curve, the asynchronous mode is unstable. Similar
notation applies to the case with $m=3$, where the Hopf bifurcation
boundaries for synchronous/asynchronous mode are denoted by red
solid/dashed curves.

One key qualitative feature we can observe from Fig.~\ref{d1d2sel}, for
the parameter set used, is that the oscillatory region for a larger
value of $m$ lies completely within the unstable region for smaller
$m$ for both the synchronous and asynchronous modes. This suggests
that if a coupled system with $m_1$ cells is unstable to synchronous
perturbations, then a system with $m_2<m_1$ cells will also be
unstable to such perturbations. However, if a two-cell system is
unstable, it is possible that a system with three cells, with the same
parameter set, can be stable.  Finally, we observe qualitatively that
the Hopf bifurcation boundary of the asynchronous mode always lies
between that of the synchronous mode. This suggests that as we vary
$d_1$ and $d_2$ from a stable parameter region into an unstable
parameter region, we will always first trigger a synchronous
oscillatory instability rather than an asynchronous instability. It is
an open problem to show that these qualitative observations still hold
for a wide range of other parameter sets.

\begin{figure}[htbp]
\begin{center}
\includegraphics[width=0.45\textwidth,height=5.0cm]{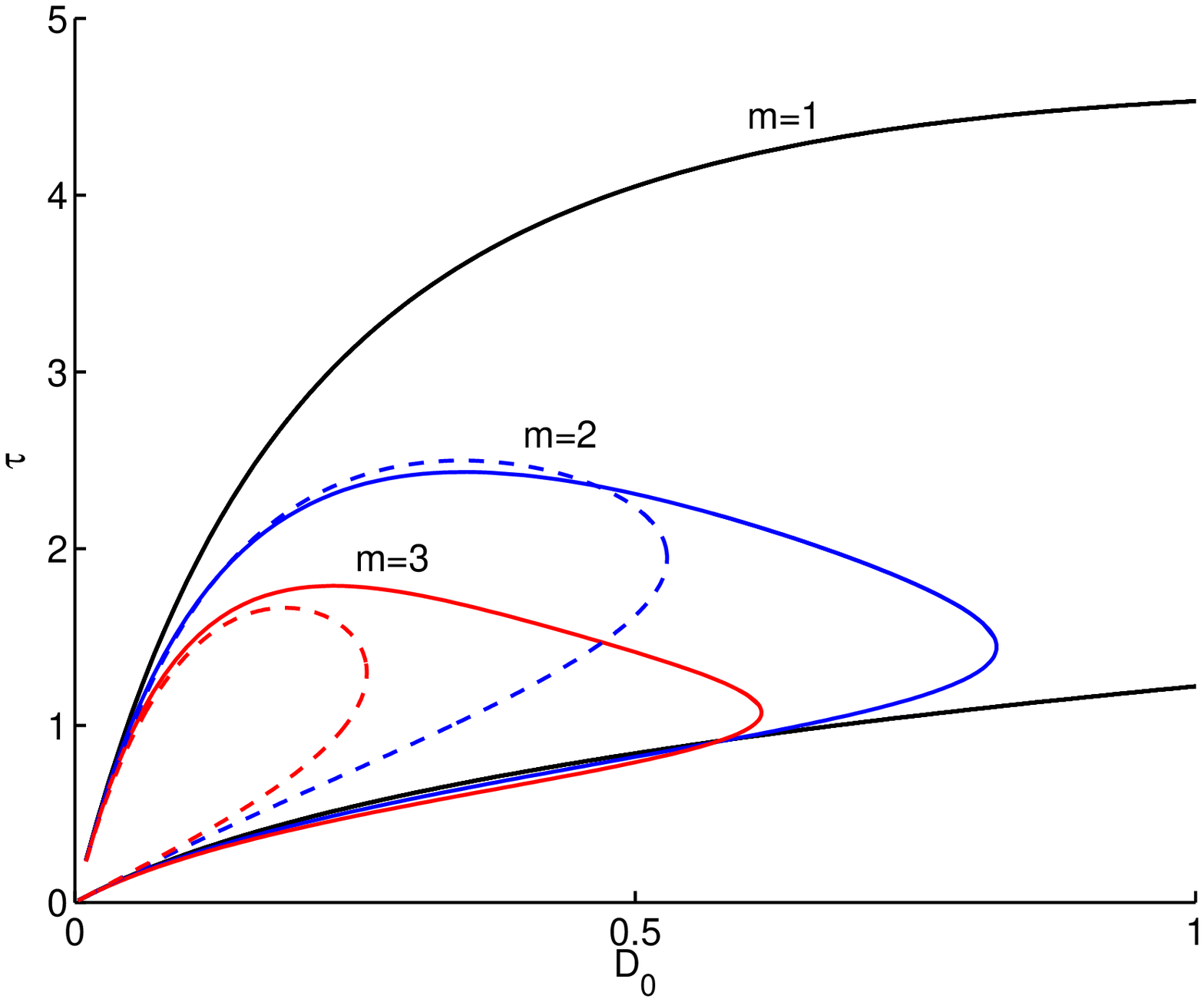}
\includegraphics[width=0.45\textwidth,height=5.0cm]{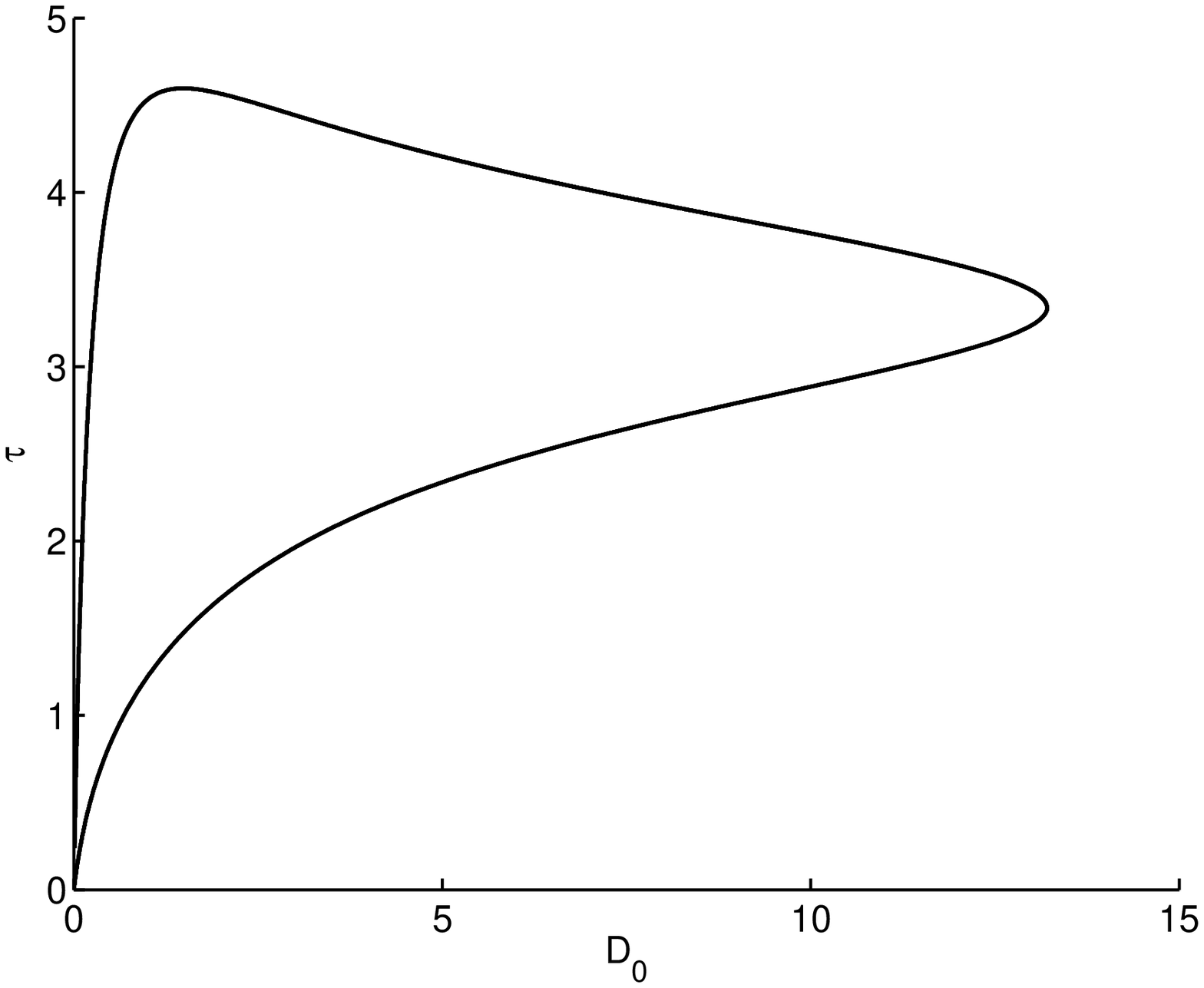}
\caption{Left panel: Hopf bifurcation boundaries for the synchronous
  (solid curves) and asynchronous (dashed curves) modes for the
  Sel'kov model with different numbers $m$ of cells in the $\tau$
  versus $D_0$ plane. The synchronous mode for $m=1$ is unstable only
  inside the region bounded by the two black solid curves. Similarly,
  in the lobe formed by the blue solid and red solid curves the
  synchronous mode is unstable for $m=2$ (blue) and $m=3$ (red),
  respectively. In the region enclosed by the blue (red) dashed curve,
  the asynchronous mode is unstable for $m=2$ ($m=3$). Right panel:
  Hopf bifurcation boundaries for the synchronous mode with $m=1$
  shown in a larger region of the $\tau$ versus $D_0$ plane, showing
  that the instability lobe is a bounded region. Parameters used are
  $\mu=2$, $\alpha=0.9$, $\epsilon_0=0.15$, $d_1=0.5$, $d_2=0.2$ and
  $|\Omega|=10$.}\label{tauDsel}
\end{center}
\end{figure}

  Next, we show the region where oscillatory instabilities can
  occur in the $\tau$ versus $D_0$ parameter plane for the synchronous
  and asynchronous modes. We fix the Sel'kov parameter values as
  $\mu=2$, $\alpha=0.9$, and $\epsilon_0=0.15$, so that the uncoupled
  intracelluar kinetics has a stable steady-state. We then take
  $d_1=0.5$, $d_2=0.2$, and $|\Omega|=10$. For this parameter set, we
  solve the Hopf bifurcation conditions (\ref{mcell2ode:syn}) by a
  root finder. In this way, in the left panel of Fig.~\ref{tauDsel} we
  plot the Hopf bifurcation boundaries for the synchronous mode in the
  $\tau$ versus $D_0$ plane for $m=1,2,3$. Similarly, upon using
  (\ref{mcell:d0_async}), in the left panel of Fig.~\ref{tauDsel} we
  also plot the Hopf bifurcation boundaries for the asynchronous
  mode. In the right panel of Fig.~\ref{tauDsel}, where we plot in a
  larger region of the $\tau$ versus $D$ plane, we show that the
  instability lobe for the $m=1$ case is indeed closed. We observe for
  $m=2$ and $m=3$ that, for this parameter set, the lobes of
  instability for the asynchronous mode are almost entirely contained
  within the lobes of instability for the synchronous mode.

\begin{figure}[htbp]
\begin{center}
\includegraphics[width=0.45\textwidth,height=5.0cm]{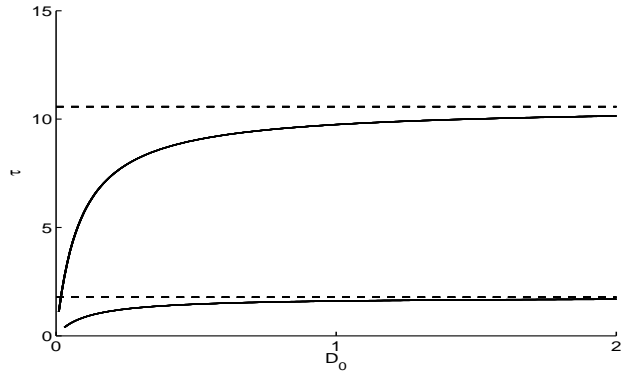}
\caption{Hopf bifurcation boundaries for the synchronous mode for the
  Sel'kov model with $m=1$ in the $\tau$ versus $D_0$ plane when
  $d_1=0.1$ and $d_2=0.2$. The other parameters are the same as in
  Fig.~\ref{tauDsel}. Inside the region bounded by the two black solid
  curves, which were computed using (\ref{mcell2ode:syn}), the
  synchronous mode is unstable. The instability region is no longer a
  closed lobe as in Fig.~\ref{tauDsel}. The dashed lines represent the
  two Hopf bifurcation points that are obtained by using the
  bifurcation software \cite{xpp} on the steady-state problem for the
  ODE system (\ref{d:odes}). As $D_0$ increases, the Hopf bifurcation
  thresholds in $\tau$ gradually approach those obtained from the large
  $D$ approximation. }\label{tauDsel_2}
\end{center}
\end{figure}

  Finally, we consider the effect of changing $d_1$ and $d_2$ to
  $d_1=0.1$ and $d_2=0.2$, while fixing the Sel'kov parameters as
  $\mu=2$, $\alpha=0.9$, and $\epsilon_0=0.15$, and keeping
  $|\Omega|=10$. In Fig.~\ref{tauDsel_2} we plot the Hopf bifurcation
  curve for the synchronous mode when $m=1$, computed using
  (\ref{mcell2ode:syn}), in the $\tau$ versus $D_0$ plane. We observe
  that there is no longer any closed lobe of instability. In this
  figure we also show the two Hopf bifurcation values that correspond
  to taking the limit $D_0\gg 1$. These latter values are Hopf
  bifurcation points for the linearization of the ODE system
  (\ref{d:odes}) around its steady-state value. This ODE system
  (\ref{d:odes}), derived in \S \ref{sec:odes}, describes large-scale
  cell-bulk dynamics in the regime $D\gg {\mathcal O}(\nu^{-1})$.

\subsubsection{Local Kinetics Described by a Fitzhugh-Nagumo Type System}\label{sec:ex:fn}

Next, for the cell kinetics we take the Fitzhugh-Nagumo (FN) nonlinearities,
taken from \cite{Gomez-Marin2007}, given by
\begin{equation}\label{2dsf:fg}
F(v,w)=\epsilon_0 (wz-v)\,, \qquad G(v,w) =w-q(w-2)^3+4-v\,,
\end{equation}
where the parameters satisfy $\epsilon_0>0$, $q>0$, and $z>1$. 

Upon substituting (\ref{2dsf:fg}) into (\ref{mcell2ode:ss}) we
calculate that the steady-state solution $w_e>0$ is given by the unique
real positive root of the cubic ${\mathcal C}(w)=0$ given by
\begin{equation}
 {\mathcal C}(w)\equiv qw^3-6qw^2+w\left(12q-1+\Lambda\right)-(8q+4)\,,\qquad
 \mbox{where} \qquad
\Lambda\equiv\frac{\epsilon_0 z}{\left[\epsilon_0 + {2\pi
      D_0\beta/ \tau} \right]}\,, \label{2dfitz:cubic}
\end{equation}
where $\beta$ is defined in (\ref{2dsel:ss_sol}). The uniqueness of the
positive root of this cubic for any $\Lambda>0$ was proved previously
in \S 2 of \cite{gou_ward}. In terms of the solution $w_e>0$ to the
cubic equation, we calculate $v_e=\Lambda w_e$ and $S_{0c}=-\beta \Lambda w_e$.

As needed below, we first calculate the partial derivatives of $F$ and $G$
at the steady-state solution as
\bsub \label{2dsf:jac}
\begin{equation}\label{2dsf:fgp}
F_v^e = -\epsilon_0\,, \qquad F_w^e=\epsilon_0 z\,, \qquad
G_v^e=-1\,, \qquad G_w^e=1 - 3q(w_e-2)^2 \,,
\end{equation}
which yields
\begin{equation}
  \mbox{det}(J) = \epsilon_0\left[ z-1 + 3q(w_e-2)^2\right]>0 \,, \qquad
  \mbox{tr}(J)= -\epsilon_0 + 1 - 3q(w_e-2)^2 \,.
\end{equation}
\esub

To determine conditions for which the synchronous mode has a Hopf
bifurcation we first substitute (\ref{2dsf:fgp}) into
(\ref{mcell2ode:p123}). The Hopf bifurcation boundary is then found by
numerically computing the roots of (\ref{mcell2ode:syn}).
Similarly, to study instabilities associated with the asynchronous
oscillatory mode we substitute (\ref{2dsf:fgp}) into
(\ref{mcell2ode:asyn_all}) to obtain that 
\bsub \label{2dfitz:async}
\begin{equation}
\gamma =  \tau \left[-\epsilon_0 +1-3q(w_e-2)^2\right]\,,
\end{equation}
which yields the Hopf bifurcation boundary, provided that
\begin{equation}\label{2dsf:asy1}
\det(J)-\frac{\gamma}{ \tau} G_w^e=-G_v^eF_w^e-(G_w^e)^2 =
-G_v^eF_w^e-(G_w^e)^2=\epsilon_0 z- \left[1-3q(w_e-2)^2\right]^2>0\,,
\end{equation}
\esub where $w_e$ is the positive root of the cubic
(\ref{2dfitz:cubic}). As was done for the Sel'kov model in \S
\ref{sec:ex:selkov}, the Hopf bifurcation boundary for the
asynchronous mode in the $d_1$ versus $d_2$ parameter plane can be
parametrized as in (\ref{2dcell:a_par_1}) where $a_{11}$, $a_{12}$,
$a_{22}$, and $a_{21}$, are now defined in terms of the parameter
$w_e>0$ by \bsub
\begin{equation}
  a_{11}\equiv  1 + \frac{2\pi m D_0}{|\Omega|} \,, \qquad
  a_{12}\equiv -\frac{1}{\beta(w_e)} \,, \qquad a_{21}\equiv 1 \,, \qquad
  a_{22}\equiv -\frac{2\pi D_0}{\gamma(w_e)} \,,
\end{equation}
where
\begin{equation}
  \beta(w_e) \equiv  \frac{\epsilon_0\tau}{2\pi D_0}
    \left( \frac{z}{\Lambda(w_e)} -1 \right) \,, \quad \mbox{with}
    \quad \Lambda(w_e)\equiv -\frac{q(w_e-2)^3}{w_e} + 1 +
    \frac{4}{w_e} \,; \quad \gamma(w_e) \equiv  \tau
    \left[-\epsilon_0 + 1 - 3q(w_e-2)^2 \right]\,.
\end{equation}
\esub

\begin{figure}[htbp]
\begin{center}
\includegraphics[width=0.75\textwidth,height=5.0cm]{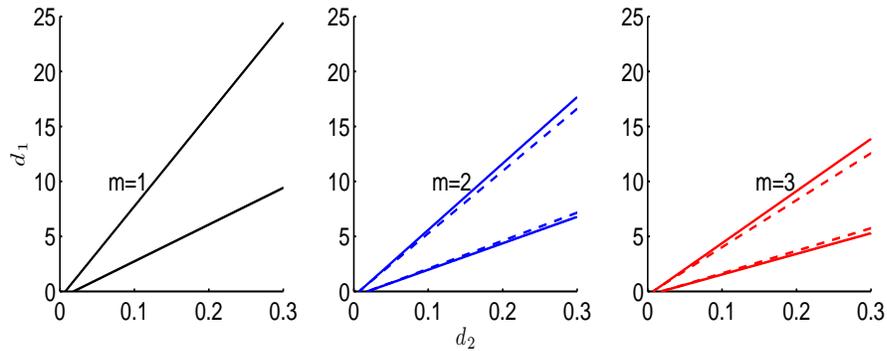}
\caption{Hopf bifurcation boundaries for the synchronous (solid curve)
  and asynchronous (dashed curve) modes for the FN system 
  (\ref{2dsf:fg}) with various number $m$ of cells in the $d_1$ versus
  $d_2$ parameter plane. Between the solid lines the synchronous mode
  is unstable, whereas between the dashed lines the asynchronous mode
  is unstable. Notice that the region of instability for the asynchronous
  mode is contained within the instability region for the synchronous mode.
  Parameters used are $z=3.5$, $q=5$, $\epsilon_0=0.5$, $\tau=1$, $D_0=1$, 
  and $|\Omega|=10$.}
\label{d1d2slowfast}
\end{center}
\end{figure}

By varying $w_e>0$ and retaining only the portion of the
curve for which $d_1>0$ and $d_2>0$, and ensuring that the constraint
(\ref{2dsf:asy1}) holds,  we obtain a parametric form for
the Hopf bifurcation boundary for the asynchronous mode in the 
$d_1$ versus $d_2$ parameter plane. For $m=2$ and $m=3$, these are
the dashed curves shown in Fig.~\ref{d1d2slowfast}.

In this way, in Fig.~\ref{d1d2slowfast} we plot the Hopf bifurcation
boundaries for the synchronous mode (solid curves) and asynchronous
mode (dashed curves) for various values of $m$ for the parameter set
$z=3.5$, $q=5$, $\epsilon_0=0.5$, $\tau=1$, $D_0=1$, and
$|\Omega|=10$. As compared to Fig.~\ref{d1d2sel}, we notice that the
unstable regions for both modes are not only shrinking but also the
boundaries shift as the number of cells increases. This feature
does not appear in the previous Sel'kov model.

\begin{figure}[htbp]
\begin{center}
\includegraphics[width=0.8\textwidth,height=5.0cm]{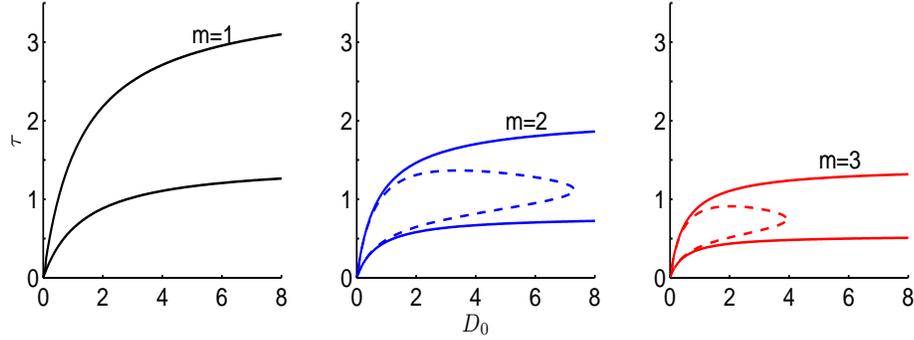}
\caption{Hopf bifurcation boundaries for the synchronous (solid curves)
  and asynchronous (dashed curves) modes for the FN system 
  (\ref{2dsf:fg}) with various number $m$ of cells in the $\tau$ versus
  $D_0$ parameter plane. Between the solid lines the synchronous mode
  is unstable, whereas between the dashed lines the asynchronous mode
  is unstable. Parameters used are $z=3.5$, $q=5$, $\epsilon_0=0.5$, $d_1=10$,
  $d_2=0.2$, 
  and $|\Omega|=10$.}
\label{tauD0slowfast}
\end{center}
\end{figure}

Next, in Fig.~\ref{tauD0slowfast} we show the region where oscillatory
instabilities occur for either synchronous or asynchronous modes for
$m=1,2,3$ in the $\tau$ versus $D_0$ plane. These Hopf bifurcation
boundaries are computed by finding roots of either
(\ref{mcell2ode:syn}) or (\ref{2dfitz:async}) for various values of
$D_0$. The other parameter values are the same as those used for
Fig.~\ref{d1d2slowfast} except $d_1=10$ and $d_2=0.2$. Inside the
region bounded by the solid curves the synchronous mode is unstable,
while inside the region bounded by the dashed curves, the asynchronous
mode is unstable. Similar to that shown in Fig.~\ref{d1d2slowfast},
the regions of instability are shrinking, at the same time as the Hopf
bifurcation boundaries shift, as $m$ increases. For these parameter
values of $d_1$ and $d_2$, the Hopf bifurcation for the synchronous
model still occurs for large values of $D_0$.

\setcounter{equation}{0}
\setcounter{section}{4}
\section{Finite Domain: Reduction to ODEs for $D\gg {\mathcal O}(\nu^{-1})$}
\label{sec:odes}

In this section we assume that there is one small dynamically active
circular cell $\Omega_\epsilon$ in the bounded domain $\Omega$ under
the assumption that $D\gg {\mathcal O}(\nu^{-1})$, with
$\nu={-1/\log\epsilon}$.  In this limit, in which the bulk region is
effectively well-mixed, we now show that we can reduce the dynamics for the
coupled cell-bulk model to a system of nonlinear ODEs.

For the case of one cell, the basic model is formulated as
\bsub\label{d:mainbd}
\begin{equation}\label{d:mainU}
\tau U_t = D\Delta U -U\,, \qquad \vecb x\in \Omega\backslash
\Omega_{\epsilon}\,; \qquad \partial_nU = 0\,,\qquad \vecb x\in\partial \Omega\,;
 \qquad
\epsilon D \partial_{n} U =d_1U-d_2u^1\,,  \;\;\quad \vecb x\in\partial 
\Omega_{\epsilon}\,,
\end{equation}
which is coupled to the intracellular cell dynamics, with $\vecb u =
(u_1,\ldots,u_n)^T$ and reaction kinetics $\vecb F(\vecb u)$, by
\begin{equation}\label{d:mainuj}
\frac{d \vecb u}{dt} = \vecb F(\vecb u)+ \frac{\vecb e_1}{\epsilon  \tau}
 \int_{\partial \Omega_{\epsilon}}( d_1 U -d_2 u_1)\, ds \,, \qquad
\mbox{where} \qquad \vecb e_1\equiv (1,0,\dots,0)^T \,.
\end{equation}
\esub 

We will assume that $D\gg 1$, and so in the bulk region we expand
\begin{equation}\label{d:uout}
    U=U_0+\frac{1}{D}U_1+ \cdots \,.
\end{equation}
Upon substituting this expansion into (\ref{d:mainU}), and noting that
$\Omega_\epsilon\to \vecb x_0$ as $\epsilon\to 0$, we obtain to
leading order in ${1/D}$ that $\Delta U_0=0$ with $\partial_{n} U_0=0$
on $\partial\Omega$. As such, we have that $U_0=U_0(t)$. At next
order, we have that $U_1$ satisfies
\begin{equation}\label{d:u1}
\Delta U_1=U_0+\tau U_{0t}, \quad \vecb x \in \Omega\backslash \{\vecb x_0\}\,;
 \qquad \partial_{\vecb n} U_1 = 0, \quad \vecb x \in \partial\Omega \,.
\end{equation}
The formulation of this problem is complete once we determine a matching
condition for $U_1$ as $\vecb x\rightarrow \vecb x_0$. 

To obtain this matching condition, we must consider the inner region defined
at ${\mathcal O}(\epsilon)$ distances outside the cell. In this inner
region we introduce the new variables 
$\vecb y=\epsilon^{-1}(\vecb x-\vecb x_0)$ and
$\hat{U}(\vecb y,t) = U(\vecb x_0+\epsilon \vecb y,t)$. From (\ref{d:mainU}),
we obtain that
\begin{equation*}
\tau \hat{U_t}=\frac{D}{\epsilon^2}\Delta_{\vecb y}\hat{U}-\hat{U}, \quad
 \rho=|\vecb y|\geq 1 \,; \qquad
D\frac{\partial \hat{U}}{\partial \rho} =d_1 \hat{U}-d_2 u_1 \,,
 \quad \mbox{on} \quad \rho=1 \,.
\end{equation*}
For $D\gg 1$, we then seek a radially symmetric solution to this inner
problem in the form
\begin{equation}\label{d:uin}
    \hat{U}(\rho,t)=\hat{U}_0(\rho,t) +\frac{1}{D}\hat{U}_1(\rho,t)+ \cdots \,.
\end{equation}
To leading order we obtain $\Delta_{\rho}\hat{U}_0=0$ in $\rho\geq 1$,
with $\hat{U}_{0\rho}=0$ on $\rho=1$, subject to the matching condition
to the bulk that $\hat{U}_0\to U_0$ as $\rho\to \infty$. We conclude
that $\hat{U}_0=U_0$.  At next order, the problem for $\hat{U}_1$ is
\begin{equation}\label{di:u1}
\Delta_{\rho}\hat{U}_1 = 0\,, \qquad \rho\ge 1\,; \qquad
\frac{\partial \hat{U}_1}{\partial \rho}=d_1 U_0-d_2 u_1\,, \qquad \rho=1\,.
\end{equation}
Allowing for logarithmic growth at infinity, the solution to this problem is
\begin{equation}\label{di:u1_sol}
\hat{U}_1=(d_1 U_0-d_2u_1)\log\rho+C \,,
\end{equation}
where $C$ is a constant to be found. Then, by writing (\ref{di:u1_sol})
in outer variables, and recalling (\ref{d:uin}), we obtain that the 
far-field behavior of the inner expansion is
\begin{equation}\label{di:ff}
  \hat{U}\sim U_0 + \frac{1}{D} \left[ \left(d_1U_0-d_2u_1\right)
 \log|\vecb x-\vecb x_0| + \frac{1}{\nu} \left(d_1U_0-d_2u_1\right) + C\right]
  + \cdots \,.
\end{equation}

From (\ref{di:ff}) we observe that the term proportional to ${1/D}$ is
smaller than the first term provided that $D\gg O(\nu^{-1})$. This
specifies the asymptotic range of $D$ for which our analysis will
hold.  From asymptotic matching of the bulk and inner solutions, the
far-field behavior of the inner solution (\ref{di:ff}) provides the
required singular behavior as $\vecb x\to \vecb x_0$ for the outer bulk
solution. In this way, we find from (\ref{di:ff}) and (\ref{d:uout}) that
$U_1$ satisfies (\ref{d:u1}) subject to the singular behavior
\begin{equation}\label{d:u1_sing}
   U_1 \sim \left(d_1U_0-d_2u_1\right)\log|\vecb x - \vecb x_0| \,, \quad
  \mbox{as} \quad \vecb x\to \vecb x_0 \,.
\end{equation}
Then, (\ref{d:u1}) together with (\ref{d:u1_sing}) determines $U_1$ uniquely.
Finally, in terms of this solution, we identify that the constant
$C$ in (\ref{di:ff}) and (\ref{di:u1_sol}) is obtained from
\begin{equation}\label{d1:ceval}
\lim_{\vecb x\to\vecb x_0}\left[U_1 -
 \left(d_1U_0-d_2u_1\right)\log|\vecb x - \vecb x_0| \right] =
 \nu^{-1}  \left(d_1U_0-d_2u_1\right) + C \,.
\end{equation}

We now carry out the details of the analysis.  We can write the problem
(\ref{d:u1}) and (\ref{d:u1_sing}) for $U_1$ as
\begin{equation}\label{d:u1_new}
\Delta U_1=U_0+\tau U_{0t} + 2\pi \left(d_1U_0-d_2 u_1\right) 
  \delta(\vecb x-\vecb x_0)\,, \qquad \vecb x \in \Omega\,; 
 \qquad \partial_{n} U_1 = 0, \quad \vecb x \in \partial\Omega \,.
\end{equation}
By the divergence theorem, this problem has a solution only if
$\left(U_0+\tau U_{0t}\right)|\Omega|=-2\pi (d_1U_0-d_2 u_1)$. This leads to 
the following ODE for the leading-order bulk solution $U_0(t)$:
\begin{equation}\label{dlarge:one}
    U_0^{\prime}= -\frac{1}{\tau} \left( 1 + \frac{2\pi d_1}{|\Omega|}\right)
  U_0 + \frac{2\pi d_2}{\tau |\Omega|} u_1 \,.
\end{equation}
Without loss of generality we impose $\int_{\Omega} U_1\, d\vecb x=0$
so that $U_0$ is the spatial average of $U$. Then, the solution to
(\ref{d:u1_new}) is
\begin{equation}\label{d:u1_sol}
   U_1 = -2\pi \left(d_1 U_0 - d_2 u_1\right) G_0(\vecb x;\vecb x_0) \,,
\end{equation}
where $G_0(\vecb x;\vecb x_0)$ is the Neumann Green's function defined
by (\ref{g0:neum}). We then expand (\ref{d:u1_sol}) as $\vecb x\to
\vecb x_0$, and use (\ref{d1:ceval}) to identify $C$ in terms of
the regular part $R_0$ of the Neumann Green's function, defined in
(\ref{g0:neum}), as
\begin{equation}\label{d:cres}
   C = - \left(d_1 U_0 - d_2 u_1\right) \left(\nu^{-1}+2\pi R_0\right)
   \,.
\end{equation}

In summary, by using (\ref{d:uin}), (\ref{di:u1_sol}), and (\ref{d:cres}),
the two-term inner expansion near the cell is given by
\begin{equation}\label{ds:ui}
  \hat{U}\sim U_0 + \frac{1}{D}\left(d_1U_0-d_2u_1\right)\left(
  \log\rho - \frac{1}{\nu} - 2\pi R_0 \right) + \cdots \,.
\end{equation}
From (\ref{d:uout}) and (\ref{d:u1_sol}), the two-term expansion
for the outer bulk solution, in terms of $U_0(t)$ satisfying the ODE
(\ref{dlarge:one}), is
\begin{equation}
   U \sim U_0 - \frac{2\pi}{D} \left(d_1 U_0 - d_2 u_1\right) 
  G_0(\vecb x;\vecb x_0) \,. \label{ds:uout}
\end{equation}
The final step in the analysis is to use (\ref{d:mainuj}) to derive
the dynamics inside the cell. We readily calculate that
\begin{equation*}
    \frac{1}{\epsilon  \tau}
  \int_{\partial\Omega_\epsilon} \left(d_1U - d_2u_1\right)\, ds
  \sim \frac{2\pi}{ \tau} \left(d_1U_0- d_2u_1\right) \,,
\end{equation*}
which determines the dynamics inside the cell from (\ref{d:mainuj}).

This leads to our main result that, for $D\gg {\mathcal O}(\nu^{-1})$,
the coupled PDE model (\ref{d:mainbd}) reduces to the study of the
coupled (n+1) dimensional ODE system for the leading-order average
bulk concentration $U_0(t)$ and cell dynamics $\vecb u$ given by
\begin{equation}\label{d:odes}
    U_0^{\prime}= -\frac{1}{\tau} \left( 1 + \frac{2\pi
      d_1}{|\Omega|}\right) U_0 + \frac{2\pi d_2}{\tau |\Omega|} u_1
    \,, \qquad \vecb u^{\prime} = \vecb F(\vecb u) + \frac{2\pi}{
  \tau} \left[d_1U_0-d_2u_1\right] \vecb e_1 \,.
\end{equation}

Before studying (\ref{d:odes}) for some specific reaction kinetics, we
first examine conditions for the existence of steady-state solutions for
(\ref{d:odes}) and we derive the spectral problem characterizing the
linear stability of these steady-states.

A steady-state solution $U_{0e}$ and $\vecb u_e$ of (\ref{d:odes}), if
it exists, is a solution to the nonlinear algebraic system
\begin{equation} \label{dlarge:ss}
    \vecb F(\vecb u_e) + \frac{2\pi}{ \tau} \left(d_1 U_{0e}-d_2\vecb
    u_{1e}\right) \vecb e_1 = 0 \,, \qquad rU_{0e}= s u_{1e} \,, \qquad
  \mbox{where} \qquad r \equiv 1 + \frac{2\pi d_1}{|\Omega|} \,, \qquad 
   s \equiv \frac{2\pi d_2}{|\Omega|}\,.
\end{equation}
To examine the linearized stability of such a steady-state, we
substitute $U_0=U_{0e}+e^{\lambda t} \eta$ and $\vecb u =\vecb u_e +
e^{\lambda t}\vecb \phi$ into (\ref{d:odes}) and linearize. This
yields that $\eta$ and $\vecb \phi$ satisfy
\begin{equation*}
   \lambda \vecb \phi = J \vecb \phi + \frac{2\pi}{ \tau}
  \left(d_1\eta - d_2 \phi_1\right)
  \vecb e_1 \,, \qquad \tau \lambda \eta = - r\eta + s\phi_1 \,,
\end{equation*}
where $J$ is the Jacobian of $\vecb F$ evaluated at the steady-state
$\vecb u=\vecb u_e$. Upon solving the second equation for $\eta$, and
substituting the resulting expression into the first equation, we
readily derive the homogeneous linear system 
\begin{equation}\label{dlarge:eigv}
    \left[(\lambda I - J) - \mu {\mathcal E}_1\right] \vecb \phi = 0
    \,, \qquad \mbox{where} \qquad \mu \equiv \frac{2\pi}{
      \tau} \left( \frac{d_1 s}{\tau\lambda + r} - d_2\right) \,, \qquad
 {\mathcal E}_1 \equiv \vecb e_1\vecb e_1^T \,.
\end{equation}

By using the matrix determinant lemma we conclude that $\lambda$ is an
eigenvalue of the linearization if and only if $\lambda$ satisfies
$\vecb e_1^T \left(\lambda I - J\right)^{-1}\vecb e_1 = {1/\mu}$, where
$\mu$ is defined in (\ref{dlarge:eigv}). From this expression, and by
using $d_1 s- d_2 r=-d_2$ as obtained from (\ref{dlarge:ss}), we
conclude that $\lambda$ must be a root of ${\mathcal Q}(\lambda)=0$,
defined by
\begin{equation}\label{dlarge:stab_crit}
     {\mathcal Q}(\lambda) \equiv 
  \frac{{ \tau} (r + \tau\lambda)}{2\pi d_2 \left(
  1 + \tau\lambda\right)} + \frac{M_{11}}{\det(\lambda I - J)} \,,
\end{equation}
where $r$ is defined in (\ref{dlarge:ss}).  Here $M_{11}$
is the cofactor of the element in the first row and first column of
$\lambda I-J$.

Next, we show that (\ref{dlarge:stab_crit}), which characterizes the
stability of a steady-state solution of the ODE dynamics
(\ref{d:odes}), can also be derived by taking the limit $D_0\gg 1$ in
the stability results obtained in (\ref{dlarge:sync}) of \S
\ref{sec:largeD} for the $D={\mathcal O}(\nu^{-1})$ regime where we
set $D={D_0/\nu}$. Recall from the analysis in \S \ref{sec:largeD} for
$D={D_0/\nu}$, that when $m=1$ only the synchronous mode can occur,
and that the linearized eigenvalue satisfies
(\ref{dlarge:sync_new}). By formally letting $D_0\to \infty$ in
(\ref{dlarge:sync_new}) we readily recover (\ref{dlarge:stab_crit}).

\subsection{Large D Theory: Analysis of Reduced Dynamics}\label{dlarge:examples}

We now give some examples of our stability theory.  We first consider
the case where there is exactly one dynamical species in the cell so
that $n=1$. From (\ref{dlarge:ss}) with $n=1$ we obtain that the
steady-state $u_e$ is any solution of
\begin{equation}\label{dlarge:1ode_ss}
 F(u_e) -\frac{2\pi d_2}{ \tau} \left[1+ \frac{2\pi
     d_1}{|\Omega|}\right]^{-1} u_e =0 \,, \qquad U_{0e}= \frac{2\pi
   d_2}{|\Omega|} \left[1+ \frac{2\pi d_1}{|\Omega|}\right]^{-1} u_e
    \,.
\end{equation}
In the stability criterion (\ref{dlarge:stab_crit}) we set $M_{11}=1$
and $\det(\lambda I-J)= \lambda-F_{u}^{e}$, where
$F_{u}^{e}\equiv {dF/du}\vert_{u=u_{e}}$, to conclude that the
stability of this steady-state is determined by the roots of the
quadratic \bsub \label{dlarge:stab_1cell}
\begin{equation}\label{dlarge:quad}
   \lambda^2 - \lambda p_1 + p_2 =0 \,,
\end{equation}
where $p_1$ and $p_2$ are defined by
\begin{equation}\label{dlarge:1cell_p12}
  p_1 = -\frac{1}{\tau} \left(1+ \frac{2\pi d_1}{|\Omega|}\right) +
  F_u^{e} - \frac{2\pi d_2}{ \tau} \,, \qquad p_2 =
  -\frac{F_{u}^{e}}{\tau} \left(1+ \frac{2\pi
    d_1}{|\Omega|}\right) + \frac{2\pi d_2}{ \tau^2} \,.
\end{equation}
\esub
We now establish the following result based on (\ref{dlarge:stab_1cell}).

\begin{result}\label{dlarge:theorem_1}
Let $n=1$. Then, no steady-state solution of (\ref{d:odes}) can undergo
a Hopf bifurcation. Moreover, if
\begin{equation}\label{dlarge:stab_one}
  \qquad F_{u}^{e} < F_{\textrm{th}} \equiv \frac{2\pi d_2}{ \tau} 
 \left[ 1 + \frac{2\pi d_1}{|\Omega|}\right]^{-1} \,,
\end{equation}
then $\mbox{Re}(\lambda)<0$, and so the steady-state is linearly stable.
If $F_{u}^{e} > F_{\textrm{th}}$, the linearization has exactly one positive
eigenvalue.
\end{result}

\proof We first prove that no Hopf bifurcations are possible for the
steady-state.  From (\ref{dlarge:quad}) it is clear that there exists
a Hopf bifurcation if and only if $p_1=0$ and $p_2>0$ in
(\ref{dlarge:1cell_p12}). Upon setting $p_1=0$, we get
$F_{u}^{e}=2\pi d_2 { \tau^{-1}} + \tau^{-1}\left(1+{2\pi
  d_1/|\Omega|}\right)$. Upon substituting this expression into
(\ref{dlarge:1cell_p12}) for $p_2$, we get that
\begin{equation*}
 p_2 = -\frac{1}{\tau} \left[ \frac{4\pi^2 d_2^2}{{ \tau}
 |\Omega|} + \frac{1}{\tau}
  \left( 1 + \frac{2\pi d_1}{|\Omega|}\right)
  \left( 1 + \frac{2\pi d_2}{|\Omega|}\right)\right]<0\,.
\end{equation*}
Since $p_2<0$ whenever $p_1=0$, we conclude that no Hopf bifurcations
for the steady-state are possible.

The second result follows by showing that $p_1<0$ and $p_2>0$
when $F_{u}^{e} < F_{\textrm{th}}$. From (\ref{dlarge:1cell_p12}),
$p_1<0$ and $p_2>0$ when
\begin{equation}
  \frac{2\pi d_2}{ \tau} - F_u^{e} + \frac{1}{\tau} \left(1+
  \frac{2\pi d_1}{|\Omega|}\right) >0 \,, \qquad \frac{2\pi
    d_2}{ \tau} - F_u^{e} -\frac{2\pi d_1}{|\Omega|}
  F_u^{e}>0 \,.
\end{equation}
These two inequalities hold simultaneously only when the second relation
is satisfied. This yields that when (\ref{dlarge:stab_one}) holds, we have
$\mbox{Re}(\lambda)<0$. Finally, when $F_{u}^{e} > F_{\textrm{th}}$, we have
$p_2<0$, and so there is a unique positive eigenvalue.\endproof

This result shows that with cell-bulk coupling the steady-state of the
ODE (\ref{d:odes}) can be linearly stable even when the reaction
kinetics is self-activating in the sense that $F_{u}^{e}>0$. Moreover,
we observe that as $\tau$ increases, corresponding to when the
intracellular kinetics has increasingly faster dynamics than the time
scale for bulk decay, then the stability threshold $F_{\textrm{th}}$
decreases. Therefore, for fast cell dynamics there is a smaller
parameter range where self-activation of the intracelluar dynamics can
occur while still maintaining stability of the steady-state to the
coupled system.

Next, we let $n=2$ where $\vecb F(\vecb
u)=(F(u_1,u_2),G(u_1,u_2))^T$. Then, the steady-state of the ODEs
(\ref{d:odes}) satisfies
\begin{equation}\label{d:ss_2}
  F(u_{1e},u_{2e})-\frac{2\pi d_2}{r  \tau} u_{1e}=0 \,, \qquad
  G(u_{1e},u_{2e})=0 \,, \qquad U_{0e}=\frac{s}{r} u_{1e} \,,
\end{equation}
where $r$ and $s$ are defined in (\ref{dlarge:ss}). We then
observe that the roots of ${\mathcal Q}(\lambda)=0$ in
(\ref{dlarge:stab_crit}) are roots to a cubic polynomial in $\lambda$.
Since $M_{11}=\lambda-G_{u_2}^{e}$, $\det(\lambda
I-J)=\lambda^2- \mbox{tr}(J)\lambda+\det J$, where
\begin{equation}
\mbox{tr}(J)=F_{u_1}^e+G_{u_2}^e\,, \qquad \det
J=F_{u_1}^eG_{u_2}^e-F_{u_2}^eG_{u_1}^e\,,
\end{equation}
and $F_{v}^e$, $G_{v}^e$ are partial derivatives of $F$ or $G$ with
respect to $v\in(u_1,u_2)$ evaluated at the steady-state, we conclude
that the linear stability of the steady-state is characterized
by the roots of the cubic
\bsub \label{largeD:3ode}
\begin{equation}\label{largeD:H3}
\lambda^3+p_1\lambda^2+p_2\lambda+p_3=0\,,
\end{equation}
where $p_1$, $p_2$ and $p_3$ are defined as
\begin{equation}\label{largeD:p123}
\begin{split}
p_1&\equiv \frac{2\pi d_2}{ \tau}+\frac{1}{\tau}\left(1+\frac{2\pi
  d_1}{|\Omega|}\right)-\mbox{tr}(J)\,, \qquad
p_3\equiv\frac{1}{\tau}\left(\left(1+\frac{2\pi d_1}{|\Omega|}\right)\det J -
 \frac{2\pi d_2}{ \tau} G_{u2}^e\right)\,, \\
  p_2 &\equiv \det J- \frac{2\pi
d_2}{ \tau} G_{u_2}^e+\frac{1}{\tau}\left(
 \frac{2\pi d_2}{ \tau}
 - \left(1+\frac{2\pi
  d_1}{|\Omega|}\right) \mbox{tr}(J) \right)\,.
\end{split}
\end{equation}
\esub
By taking $m=1$ and letting $D_0\rightarrow \infty$ in (\ref{mcell2ode:p123}),
it is readily verified that the expressions above for $p_i$, for $i=1,2,3$,
agree exactly with those in (\ref{mcell2ode:p123}). Then, by satisfying
the Routh-Hurwitz conditions (\ref{mcell2ode:syn}), we can
plot the Hopf bifurcation boundaries in different parameter planes.

\subsubsection{Example: One Cell with Sel'kov Dynamics}

\begin{figure}[htbp]
\begin{center}
\includegraphics[width=0.45\textwidth,height=5.0cm]{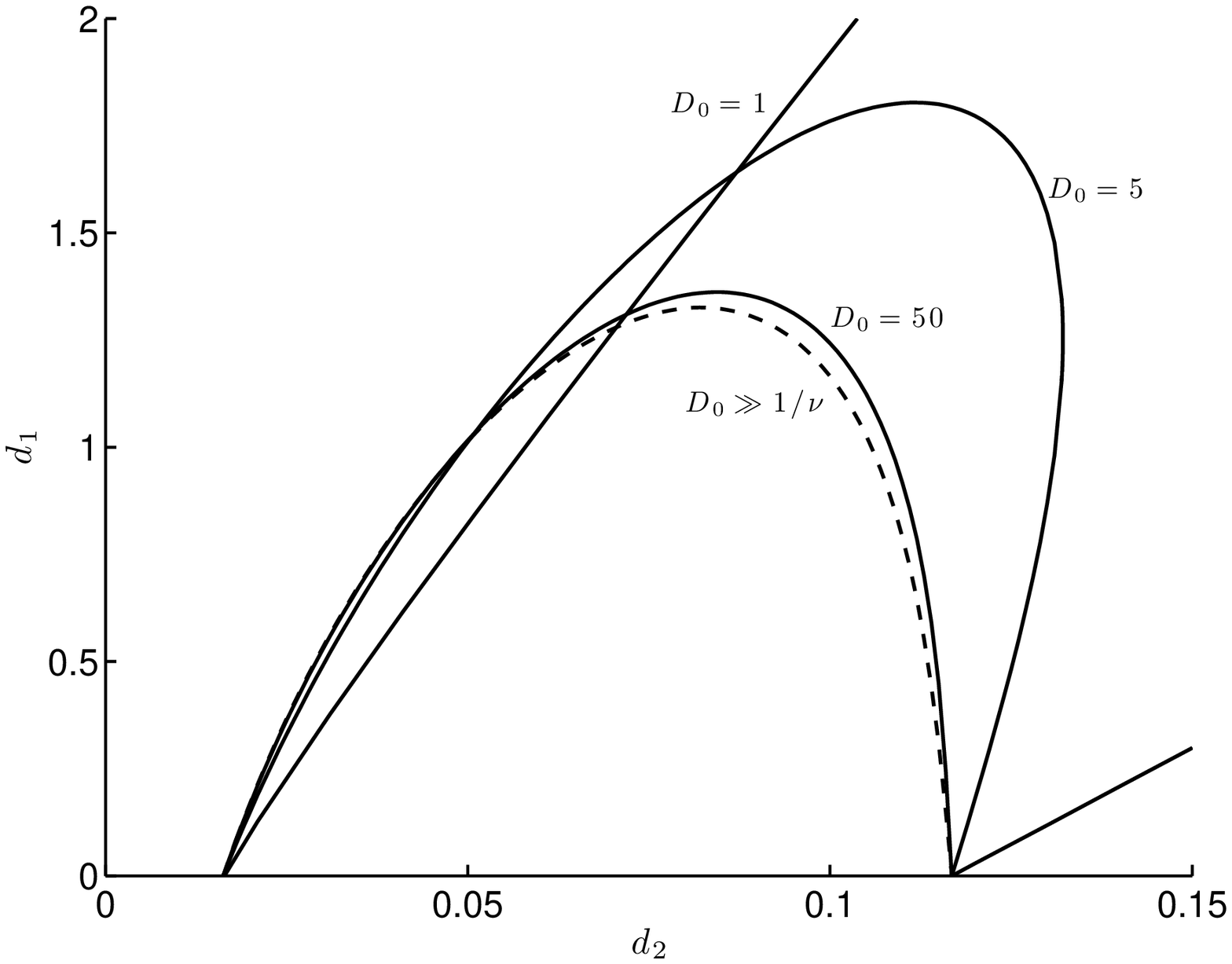}
\includegraphics[width=0.45\textwidth,height=5.0cm]{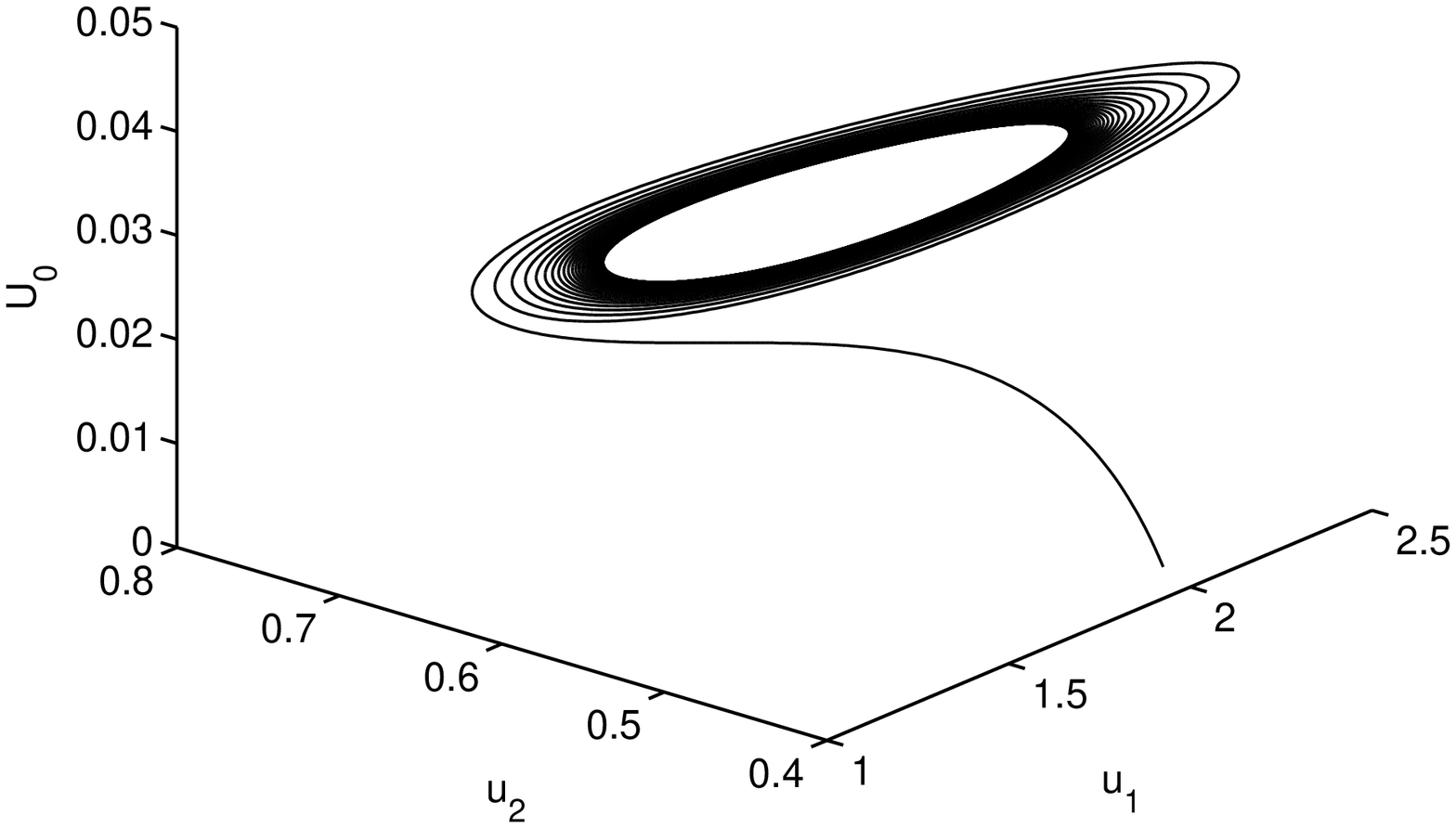}
\caption{Left panel: Comparison of the Hopf bifurcation boundaries for
  the synchronous mode for the Sel'kov model (\ref{2dsel:fg}) in the
  $d_1$ versus $d_2$ plane with $D_0=1,\ 5,\ 50$ (solid), as
  obtained from (\ref{d0:cubic}), and the large $D$ approximation
  (dashed), as obtained from (\ref{largeD:3ode}). Between the outer
  two black curves, the synchronous mode is unstable for $D_0=1$.
  In the region bounded by the solid or dashed curve this
  mode is unstable. As $D_0$ increases, the Hopf
  boundaries from (\ref{d0:cubic}) gradually approach the one given in
  from the large $D$ approximation.  Right panel:
  Numerical results for the ODE system (\ref{d:odes}) showing
  sustained oscillations. In the left and right panels we fixed
  $\mu=2$, $\alpha=0.9$, $\epsilon_0=0.15$, $\tau=1$, and
  $|\Omega|=10$. In the right panel we took $d_1=0.8$ and
  $d_2=0.05$ for which the steady-state solution of the ODEs
  (\ref{d:odes}) is unstable.}
\label{dlarge:hopfbd_sel}
\end{center}
\end{figure}

When the intracellular kinetics is described by the Sel'kov model,
where $F$ and $G$ are given in (\ref{2dsel:fg}), we obtain from
(\ref{d:ss_2}) that the steady-state solution of the ODE dynamics is
\begin{equation}
u_1^e=\frac{r\mu}{\left[r+{2\pi d_2/ \tau} \right]}\,,\qquad 
u_2^e=\frac{\mu}{\alpha+(u_1^e)^2}\,,\qquad U_{0e}=
\frac{s\mu}{r+{2\pi d_2/\tau}}\,,
\end{equation}
where $r$ and $s$ are defined in (\ref{dlarge:ss}). The partial
derivatives of $F$ and $G$ can be calculated as in
(\ref{2dsel:fgp}). 

In the left panel of Fig.~\ref{dlarge:hopfbd_sel} we plot the Hopf
bifurcation boundary in the $d_1$ versus $d_2$ plane associated with
linearizing the ODEs (\ref{d:odes}) about this steady-state solution.
This curve was obtained by setting $p_1 p_2=p_3$ with $p_1>0$ and
$p_3>0$ in (\ref{largeD:3ode}).  In this figure we also plot the Hopf
bifurcation boundary for different values of $D_0$, with
${D=D_0/\nu}$, as obtained from our stability formulation
(\ref{d0:cubic}) of \S \ref{sec:examples} for the $D={\mathcal
  O}(\nu^{-1})$ regime. We observe from this figure that the stability
boundary with $D_0=50$ closely approximates that obtained from
(\ref{largeD:3ode}), which corresponds to the $D_0\to \infty$ limit of
(\ref{d0:cubic}).

We emphasize that since in the distinguished limit $D\gg{\mathcal
  O}(\nu^{-1})$ we can approximate the original coupled PDE system
(\ref{d:mainbd}) by the system (\ref{d:odes}) of ODEs, the study
of large-scale time dynamics far from the steady-state becomes
tractable. In the right panel of Fig.~\ref{dlarge:hopfbd_sel}, we plot
the numerical solution to (\ref{d:odes}) with Sel'kov dynamics
(\ref{2dsel:fg}), where the initial condition is $u_1(0)=0.01$,
$u_2(0)=0.2$, and $U_0(0)=0.5$. We observe that by choosing $d_1$ and
$d_2$ inside the region bounded by the dashed curve in the left panel
of Fig.~\ref{dlarge:hopfbd_sel}, where the steady-state is unstable,
the full ODE system (\ref{d:odes}) exhibits a stable periodic orbit,
indicating a limit cycle behavior.

\begin{figure}[htbp]
\begin{center}
\includegraphics[width=0.45\textwidth,height=5.0cm]{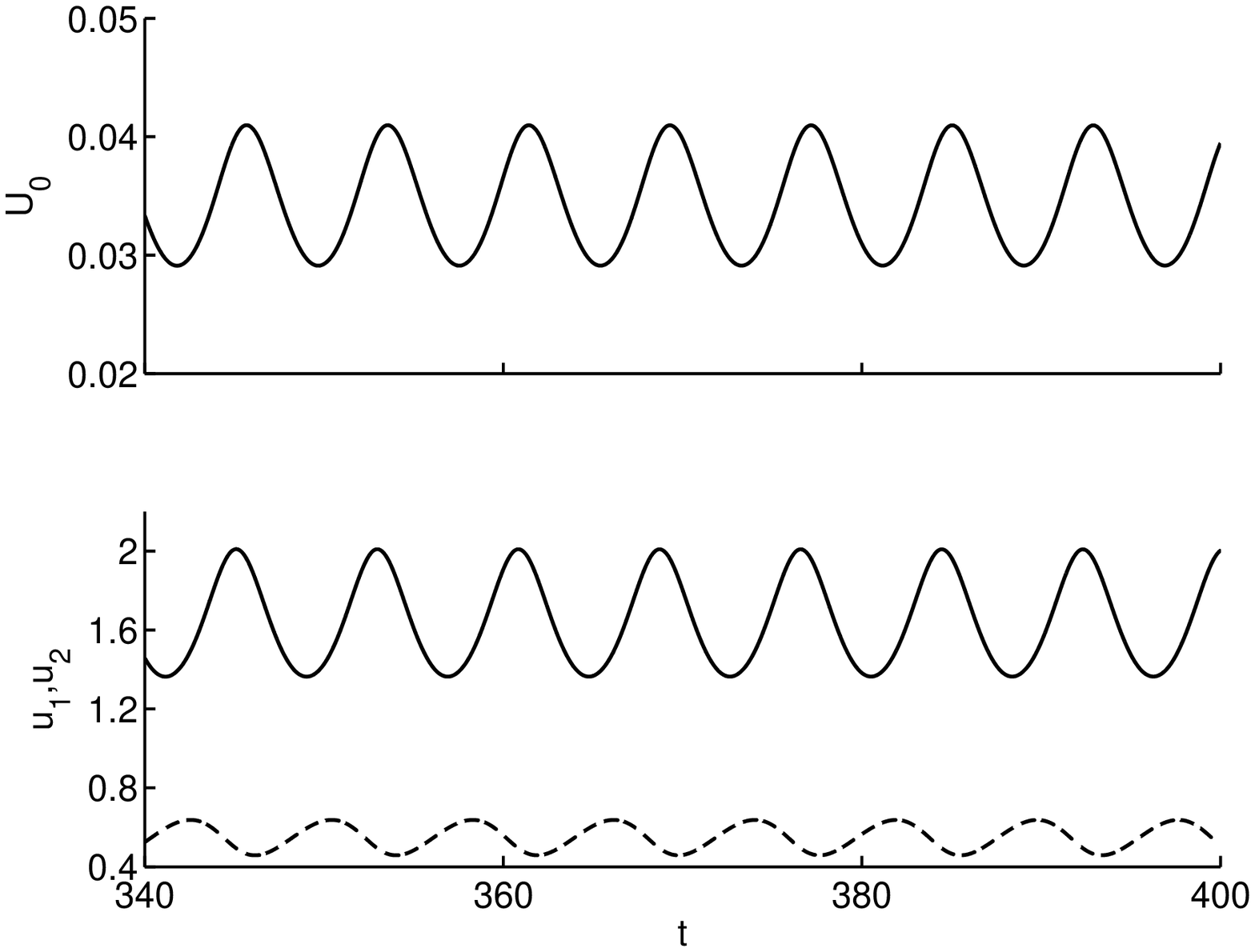}
\includegraphics[width=0.45\textwidth,height=5.0cm]{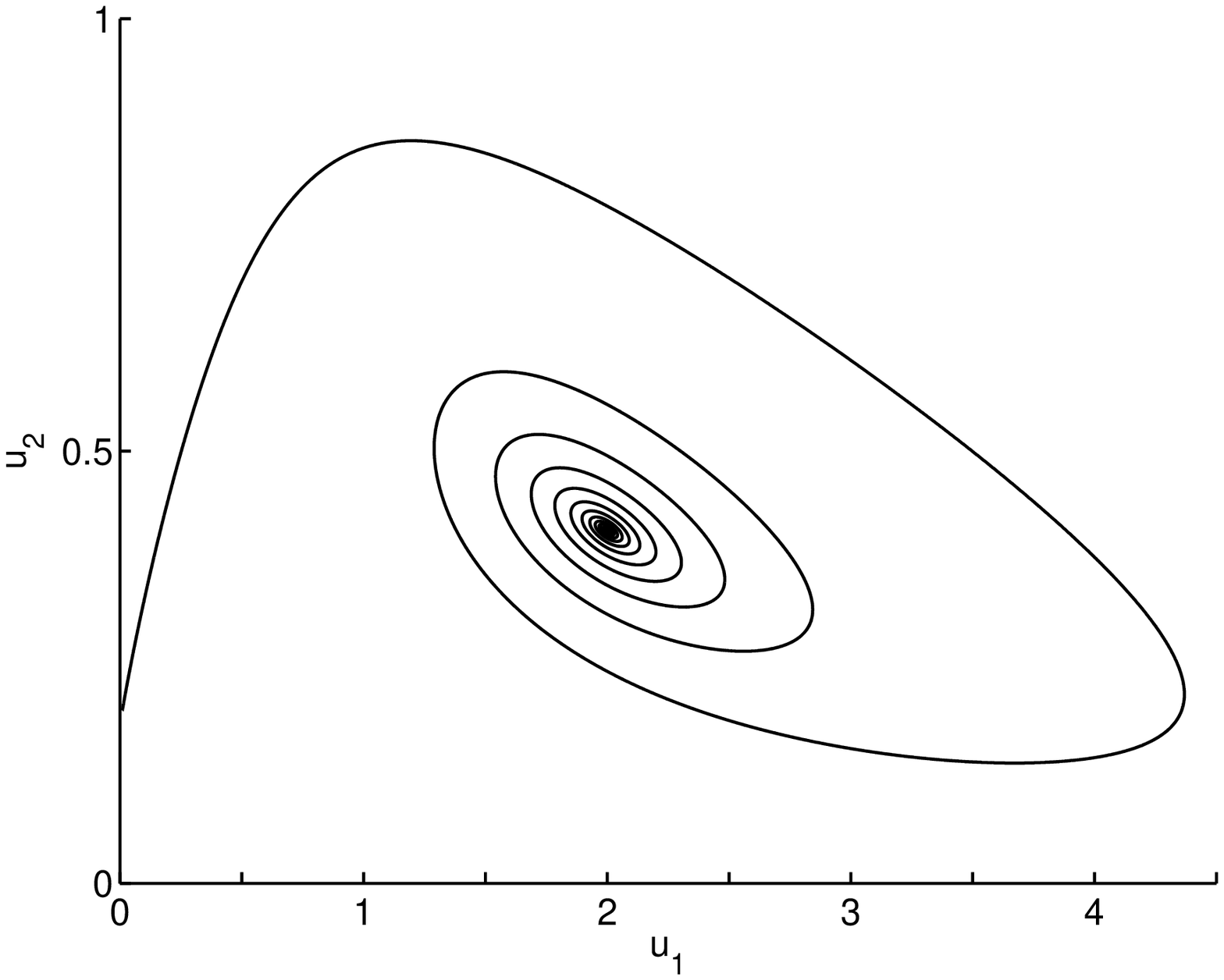}
\caption{Left panel: $u_1$, $u_2$ and $U_0$ versus $t$ showing
  sustained oscillatory dynamics. Parameters are $\mu=2$,
  $\alpha=0.9$, $\epsilon_0=0.15$, $\tau=1$, $|\Omega|=10$, $d_1=0.8$
  and $d_2=0.05$. Right panel: $u_2$ versus $u_1$ when the local
  kinetics is decoupled from the bulk, showing decaying oscillations
  towards the stable steady-state solution $u_1=\mu$ and
  $u_2=\mu/(\alpha+u_1^2)$. Initial values are $u_1(0)=0.01$ and
  $u_2(0)=0.2$. The parameter values of $\mu$, $\epsilon_0$ and $\alpha$
  are the same as in the left panel.}
\label{dlarge:u12U}
\end{center}
\end{figure}

In addition, in the left panel of Fig.~\ref{dlarge:u12U} we plot the
time evolution of $u_1$, $u_2$ and $U_0$, showing clearly the
sustained periodic oscillations. For comparison, fixing the same
parameter set for the Sel'kov kinetics (\ref{2dsel:fg}), in the right
panel of Fig.~\ref{dlarge:u12U} we plot the phase plane of $u_2$
versus $u_1$ when there is no coupling between the local kinetics and
the bulk. We now observe that without this cell-bulk coupling the
Sel'kov model (\ref{2dsel:fg}) exhibits transient decaying oscillatory
dynamics, with a spiral behavior in the phase-plane towards the
linearly stable steady-state.

\begin{figure}[htbp]
\begin{center}
\includegraphics[width=0.45\textwidth,height=5.0cm]{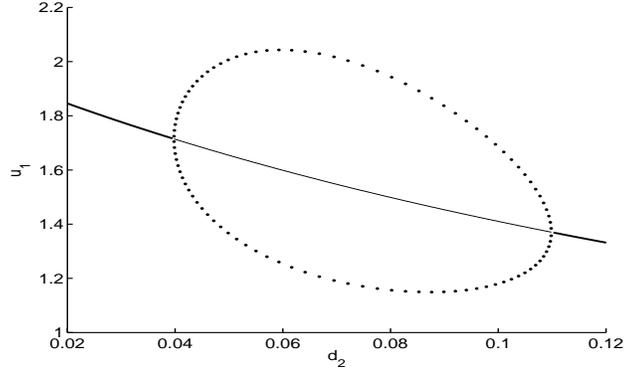}
\caption{Global bifurcation diagram of $u_1$ with respect to $d_2$ at
  a fixed $d_1=0.8$ as computed using XPPAUT \cite{xpp} from the ODE
  system (\ref{d:odes}) for the Sel'kov kinetics (\ref{2dsel:fg}). The
  thick/thin solid line represents stable/unstable steady-state
  solutions of $u_1$, while the solid dots indicate a stable periodic
  solution branch. The parameters used are $\mu=2$, $\alpha=0.9$,
  $\epsilon_0=0.15$, $\tau=1$, and $|\Omega|=10$. }\label{sel:xpp}
\end{center}
\end{figure}

Finally, we use the numerical bifurcation software XPPAUT \cite{xpp}
to confirm the existence of a stable large amplitude periodic solution
to (\ref{d:odes}) with Sel'kov kinetics when $d_1$ and $d_2$ are in the
unstable region of the left panel of Fig.~\ref{dlarge:hopfbd_sel}. In
Fig.~\ref{sel:xpp} we plot a global bifurcation diagram of $u_{1}$
versus $d_2$ for $d_1=0.8$, corresponding to taking a horizontal
slice at fixed $d_2=0.8$ through the stability boundaries in the $d_2$
versus $d_1$ plane shown in Fig.~\ref{dlarge:hopfbd_sel}. The two
computed Hopf bifurcation points at $d_2\approx 0.0398$ and
$d_1=0.1098$ agree with the theoretically predicted values in
Fig.~\ref{dlarge:hopfbd_sel}. 

\subsubsection{Example: One Cell with Fitzhugh-Nagumo Dynamics}

Finally, we apply our large $D$ theory to the case where the
intracellular dynamics is governed by the FN kinetics
(\ref{2dsf:fg}). From (\ref{d:ss_2}) we obtain that the steady-state
solution of the ODEs (\ref{d:odes}) with the kinetics (\ref{2dsf:fg})
is
\begin{equation}\label{dlarge:fn_eq}
u_1^e=\Lambda u_{2}^e \,, \qquad U_{0e}=\frac{s u_{1}^e}{r} \,, 
 \qquad \mbox{where} \qquad
 \Lambda \equiv \frac{\epsilon_0 z r}{\left[\epsilon_0 r + {2\pi d_2/{
 \tau}}\right]} \,.
\end{equation}
Here $r$ and $s$ are defined in (\ref{dlarge:ss}), and $u_{2e}>0$
is the unique root of the cubic (\ref{2dfitz:cubic}) where $\Lambda$
in (\ref{2dfitz:cubic}) is now defined in (\ref{dlarge:fn_eq}). The
partial derivatives of $F$ and $G$ can be calculated as in
(\ref{2dsf:jac}).

\begin{figure}[htbp]
\begin{center}
\includegraphics[width=0.45\textwidth,height=5.0cm]{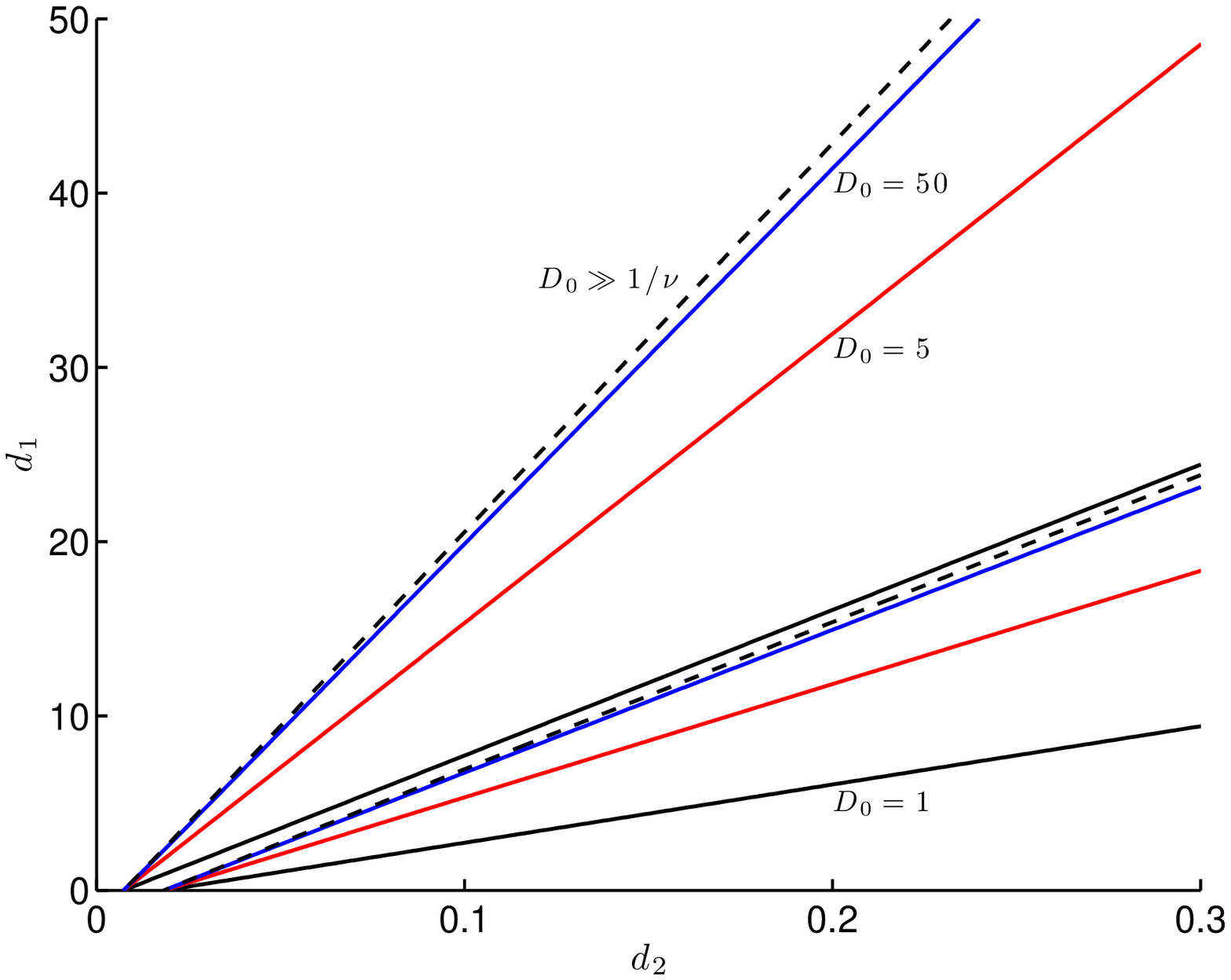}
\includegraphics[width=0.45\textwidth,height=5.0cm]{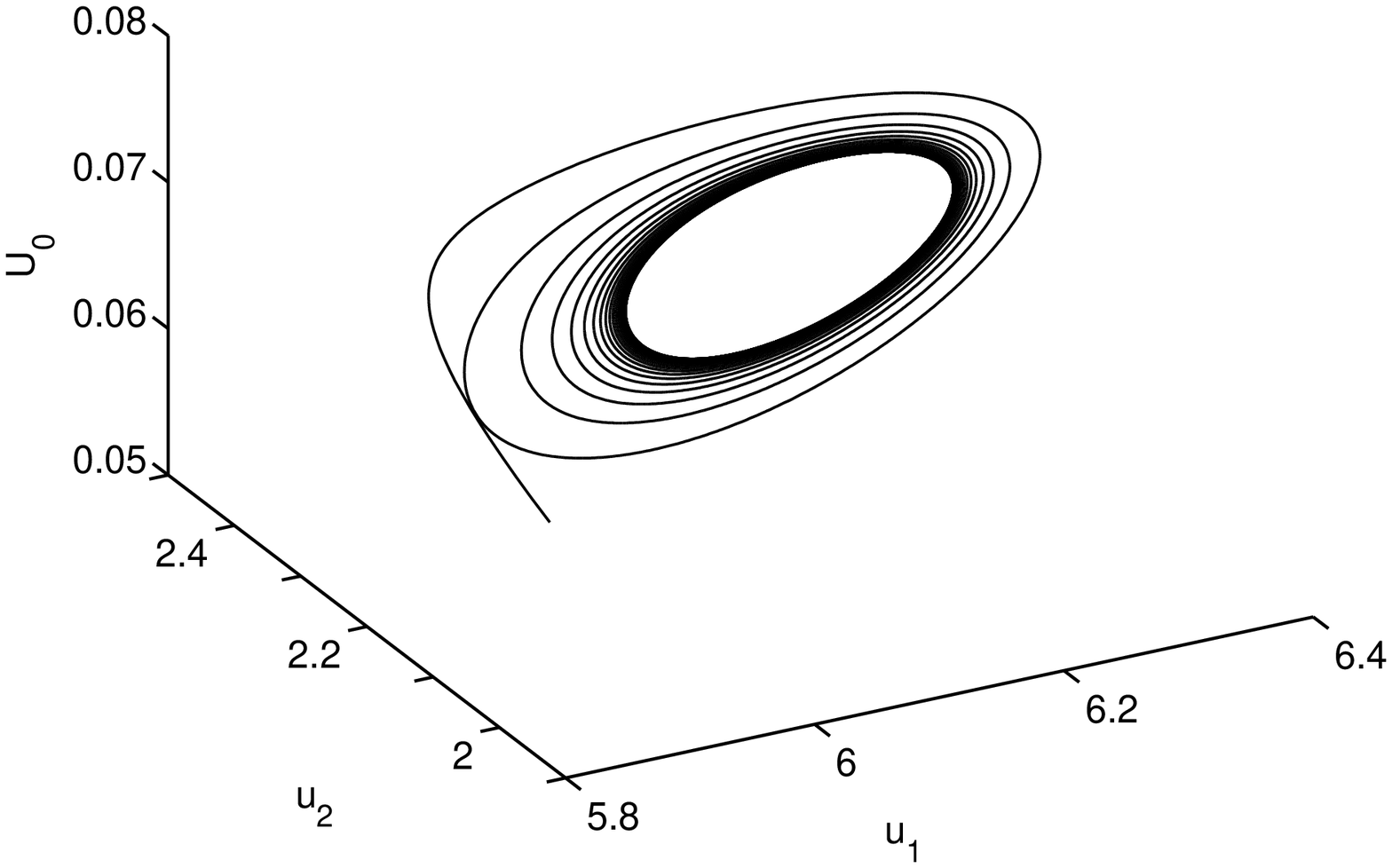}
\caption{Left panel: Comparison of the Hopf bifurcation boundaries for
  the synchronous mode with FN kinetics (\ref{2dsf:fg})
  in the $d_1$ versus $d_2$ parameter plane with $D_0=1,\ 5,\ 50$
  (solid), as obtained from (\ref{d0:cubic}), and the large $D$
  approximation (dashed), as obtained from (\ref{largeD:3ode}). In the
  wedge-shaped regions bounded by the solid curves the synchronous
  mode is unstable for the finite values of $D_0$. As $D_0$ increases,
  the Hopf boundaries obtained from (\ref{d0:cubic}) becomes rather close
  to the dashed one obtained from (\ref{largeD:3ode}) from the large
  $D$ approximation.  Right panel: Numerical simulation for the ODE
  system (\ref{d:odes}), showing sustained oscillations, with initial
  conditions $u_1(0)=6.0$, $u_2(0)=2.3$, and $U_0(0)=0.05$. In the left
  and right panels we fixed $z=3.5$, $q=5$, $\epsilon_0=0.5$,
  $\tau=1$, and $|\Omega|=10$, and in the right panel we took
  $d_1=15.6$ and $d_2=0.2$ corresponding to a point where the
  steady-state solution of the ODEs (\ref{d:odes}) is unstable.}
\label{dlarge:hopfbd_fn}
\end{center}
\end{figure}

In the left panel of Fig.~\ref{dlarge:hopfbd_fn} the dashed curve is
the Hopf bifurcation boundary in the $d_1$ versus $d_2$ plane
associated with linearizing the ODEs (\ref{d:odes}) about this
steady-state solution.  In this figure the Hopf bifurcation boundaries
for different values of $D_0$, with ${D=D_0/\nu}$, are also
shown. These latter curves were obtained from our stability
formulation (\ref{d0:cubic}) of \S \ref{sec:examples}. Similar to what
we found for the Sel'kov model, the stability boundary for $D_0=50$ is
very close to that for the infinite $D$ result obtained from
(\ref{largeD:3ode}). In the right panel of Fig.~\ref{dlarge:hopfbd_fn}
we plot the numerical solution to (\ref{d:odes}) with FN dynamics
(\ref{2dsf:fg}) for the parameter set $d_1=15.6$ and $d_2=0.2$, which
is inside the unstable region bounded by the dashed curves in the left
panel of Fig.~\ref{dlarge:hopfbd_fn}.  With the initial condition
$u_1(0)=6.0$, $u_2(0)=2.3$, and $U_0(0)=0.05$, the numerical
computations of the full ODE system (\ref{d:odes}) again reveal a
sustained and stable periodic solution.

Finally, we use XPPAUT \cite{xpp} on (\ref{d:odes}) to compute a
global bifurcation of $u_1$ versus $d_1$ for fixed $d_2=0.2$ for FN
kinetics. This plot corresponds to taking a vertical slice at fixed
$d_2=0.2$ through the stability boundaries in the $d_1$ versus $d_2$
plane shown in Fig.~\ref{dlarge:hopfbd_fn}. The two computed Hopf
bifurcation points at $d_1\approx 15.389$ and $d_1=42.842$ agree with
the predicted values in Fig.~\ref{dlarge:hopfbd_fn}. These results
confirm the existence of a stable periodic solution branch induced by
the cell-bulk coupling.

\begin{figure}[htbp]
\begin{center}
\includegraphics[width=0.45\textwidth,height=5.0cm]{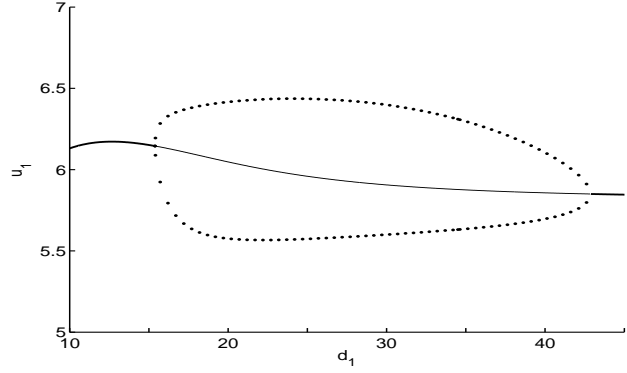}
\caption{Global bifurcation diagram of $u_{1}$ versus $d_1$ when
  $d_2=0.2$, as computed using XPPAUT \cite{xpp} from the ODE system
  (\ref{d:odes}) for the FN kinetics (\ref{2dsf:fg}).  The thick/thin
  solid line represents stable/unstable steady-state solutions,
  while the solid dots indicate a stable periodic solution.
  The other parameters are $z=3.5$, $q=5$, $\epsilon_0=0.5$,
  $\tau=1$, and $|\Omega|=10$. }
\label{fn:xpp}
\end{center}
\end{figure}

\setcounter{equation}{0}
\setcounter{section}{5}
\section{The Effect of the Spatial Configuration of the Small Cells: 
The $D={\mathcal O}(1)$ Regime}\label{sec:finite_d}

In this section we construct steady-state solutions and study their
linear stability properties in the $D={\mathcal O}(1)$ regime, where
both the number of cells and their spatial distribution in the domain
are important factors. For simplicity, we consider a special spatial
configuration of the cells inside the unit disk $\Omega$ for which the
Green's matrix ${\mathcal G}$ has a cyclic structure.  More
specifically, on a ring of radius $r_0$, with $0<r_0<1$, we place $m$
equally-spaced cells whose centers are at
\begin{equation}\label{cyc:point}
   \vecb x_j = r_0 \left( \cos\left(\frac{2\pi j}{m}\right)\,, 
 \sin\left(\frac{2\pi j}{m}\right)\right)^T\,, \qquad
  \qquad j=1,\ldots,m \,.
\end{equation}
This ring of cells is concentric with respect to the unit disk
$\Omega\equiv \lbrace{ \vecb x \, \vert \, |\vecb x|\leq 1\,
  \rbrace}$. We also assume that the intracellular kinetics is the
same within each cell, so that $\vecb F_j=\vecb F$ for
$j=1,\ldots,m$. A related type of analysis characterizing the
stability of localized spot solutions for the Gray-Scott
RD model, where localized spots are equally-spaced on
a ring concentric with the unit disk, was performed in \cite{chen}.

For the unit disk, the Green's function $G(\vecb x;\vecb \xi)$
satisfying (\ref{bdgreenss_all}) can be written as an infinite sum
involving the modified Bessel functions of the first and second kind
$I_n(z)$ and $K_n(z)$, respectively, in the form (see Appendix A.1 of
\cite{chen})
\begin{equation} \label{cyclic:G}
G(\vecb x;\vecb \xi) = \frac{\,1}{\,2 \pi}
K_0\left(\theta_0 |\vecb x-\vecb \xi|\right)
-\frac{\,1}{\,2 \pi} \sum^{\infty}_{n=0} \sigma_n 
\cos\left(n(\psi-\psi_0)\right) \frac{K^{\prime}_n(\theta_0)}
{I^{\prime}_n(\theta_0)} I_n\left(\theta_0 r\right) I_n\left(\theta_0 r_0\right)
\,; \quad \sigma_0=1 \,, \quad \sigma_n=2\,, \,\,\, n\geq 1\,.
\end{equation}
Here $\theta_0\equiv D^{-1/2}$, $\vecb x=re^{i\psi}$, $\vecb
\xi=r_0e^{i\psi_0}$, and $|\vecb x-\vecb \xi|=\sqrt{r^2+r_0^2 - 2r
  r_0 \cos(\psi-\psi_0)}$. By using the local behavior
$K_{0}(z)\sim -\log{z} + \log{2}-\gamma_e+o(1)$ as $z\to 0^{+}$, where
$\gamma_e$ is Euler's constant, we can extract the regular part $R$ of
$G(\vecb x;\vecb \xi)$ as $\vecb x\to \vecb \xi$, as identified in
(\ref{bdgreenss_reg}), as
\begin{equation}\label{cyclic:R}
 R = \frac{1}{2\pi} \left(\log{2} - \gamma_e + \frac{1}{2}\log{D}
   \right) -\frac{\,1}{\,2 \pi} \sum^{\infty}_{n=0} \sigma_n
 \frac{K^{\prime}_n(\theta_0)} {I^{\prime}_n(\theta_0)} \left[
   I_n\left(\theta_0 r_0\right) \right]^2 \,.
\end{equation}

For this spatial configuration of cells, the Green's matrix ${\mathcal
  G}$ is obtained by a cyclic permutation of its first row vector $\vecb
a\equiv (a_1,\ldots,a_m)^T$, which is defined term-wise by
\begin{equation}\label{cyclic:G_row}
  a_1 \equiv R \,; \qquad a_j = G_{j1} \equiv G(\vecb x_j;\vecb x_1) \,, 
\quad j=2,\ldots, m \,.
\end{equation}
We can numerically evaluate $G_{j1}$ for $j=2,\ldots,m$ and $R$ by
using (\ref{cyclic:G}) and (\ref{cyclic:R}), respectively. Since
${\mathcal G}$ is a cyclic matrix, it has an eigenpair, corresponding to
a synchronous perturbation, given by
\begin{equation}\label{cyclic:sync}
    {\mathcal G}\vecb e= \omega_1 \vecb e \,; \qquad \vecb e\equiv
    (1,\ldots,1)^T \,, \qquad \omega_1 \equiv \sum_{j=1}^{m}a_j = R +
    \sum_{j=1}^{m} G_{ji} \,.
\end{equation}

When $D={\mathcal O}(1)$, the steady-state solution is determined by
the solution to the nonlinear algebraic system (\ref{bduj}) and
(\ref{bdsysS2m}). Since $\vecb F_j=\vecb F$ for $j=1,\ldots,m$, and
$\vecb e$ is an eigenvector of ${\mathcal G}$ with eigenvalue
$\omega_1$, we can look for a solution to (\ref{bduj}) and
(\ref{bdsysS2m}) having a common source strength, so that $\vecb S=S_c
\vecb e$, $\vecb u_j=\vecb u$ for all $j=1,\ldots,m$, and $\vecb u^1 =
u_1\vecb e$. In this way, we obtain from (\ref{bduj}) and
(\ref{bdsysS2m}), that the steady-state problem is to solve the $n+1$
dimensional nonlinear algebraic system for $S_c$ and $\vecb
u=(u_1,u_2,\ldots,u_n)^T$ given by
\begin{equation}\label{cyclic:ss}
   \vecb F(\vecb u) + \frac{2\pi D}{\tau} S_c \vecb e = \vecb 0\,;
 \qquad S_c = -\beta u_1 \,, \qquad  \beta \equiv \frac{d_2\nu}{d_1
  + 2\pi \nu d_1 \omega_1 + D\nu} \,,
\end{equation}
where $\nu\equiv {-1/\log\epsilon}$ and $\omega_1$ is defined in
(\ref{cyclic:sync}). We remark that $\omega_1$ depends on $D$, 
$r_0$, and $m$.

To study the linear stability of this steady-state solution,
we write the GCEP, given in (\ref{gcep:full}), in the form
\begin{equation}\label{cyclic:gcep}
    {\mathcal G}_{\lambda} \vecb c = - \frac{1}{2\pi\nu} \left[
 1+ \frac{D\nu}{d_1} + \frac{2\pi\nu d_2 D}{d_1\tau} \frac{M_{11}}{
   \det(\lambda I-J)} \right] \vecb c \,,
\end{equation}
where $J$ is the Jacobian of $\vecb F$ evaluated at the steady-state.
In terms of the matrix spectrum of ${\mathcal G}_{\lambda}$, written as
\begin{equation}\label{cyclic:Glam_mat}
  {\mathcal G}_{\lambda} \vecb v_j = \omega_{\lambda,j} \vecb v_j \,, \qquad
 j=1,\ldots,m \,,
\end{equation}
we conclude from (\ref{cyclic:gcep}) that the set of discrete eigenvalues
$\lambda$ of the linearization of the steady-state are the union of the
roots of the $m$ transcendental equations, written as 
${\mathcal F}_j(\lambda)=0$, where
\begin{equation}\label{cyclic:omega_j}
  {\mathcal F}_{j}(\lambda) \equiv \omega_{\lambda,j} +
  \frac{1}{2\pi\nu} \left( 1+ \frac{D\nu}{d_1}\right) + \left(
  \frac{d_2 D} {d_1\tau} \right) \frac{M_{11}}{\det(\lambda I-J) } \,,
  \qquad j=1,\ldots,m\,.
\end{equation}
Any such root of ${\mathcal F}_j(\lambda)=0$ with
$\mbox{Re}(\lambda)>0$ leads to an instability of the steady-state
solution on an ${\mathcal O}(1)$ time-scale. If all such roots satisfy
$\mbox{Re}(\lambda)<0$, then the steady-state is linearly stable on an
${\mathcal O}(1)$ time-scale.

To study the stability properties of the steady-state using
(\ref{cyclic:omega_j}), and identify any possible Hopf bifurcation
values, we must first calculate the spectrum (\ref{cyclic:Glam_mat})
of the cyclic and symmetric matrix ${\mathcal G}_{\lambda}$, whose
entries are determined by the $\lambda$-dependent reduced-wave Green's
function $G_{\lambda}(\vecb x;\vecb \xi)$, with regular part
$R_{\lambda}(\vecb \xi)$, as defined by (\ref{bdgneig}).  Since
${\mathcal G}_{\lambda}$ is not a Hermitian matrix when $\lambda$ is
complex, its eigenvalues $\omega_{\lambda,j}$ are in general
complex-valued when $\lambda$ is complex. Then, by replacing
$\theta_0$ in (\ref{cyclic:G}) and (\ref{cyclic:R}) with
$\theta_\lambda\equiv \sqrt{(1+\tau\lambda)/D}$, we readily obtain
that
\begin{equation} \label{cyclic:Glam}
G_\lambda(\vecb x;\vecb \xi) = \frac{\,1}{\,2 \pi}
K_0\left(\theta_\lambda|\vecb x-\vecb \xi|\right) -\frac{\,1}{\,2 \pi}
\sum^{\infty}_{n=0} \sigma_n \cos\left(n(\psi-\psi_0)\right)
\frac{K^{\prime}_n(\theta_\lambda)} {I^{\prime}_n(\theta_\lambda)}
I_n\left(\theta_\lambda r\right) I_n\left(\theta_\lambda r_0\right)
\,; \quad \sigma_0=1 \,, \quad \sigma_n=2\,, \,\,\, n\geq 1\,,
\end{equation}
with regular part
\begin{equation}\label{cyclic:Rlam}
 R_{\lambda} = \frac{1}{2\pi} \left[\log{2} - \gamma_e +
   \frac{1}{2}\log{D} -\frac{1}{2}\log(1+\tau\lambda) \right]
 -\frac{\,1}{\,2 \pi} \sum^{\infty}_{n=0} \sigma_n
 \frac{K^{\prime}_n(\theta_\lambda)} {I^{\prime}_n(\theta_\lambda)} \left[
   I_n\left(\theta_\lambda r_0\right) \right]^2 \,,
\end{equation}
where we have specified the principal branch for $\theta_\lambda$.
The Green's matrix ${\mathcal G}_\lambda$ is obtained by a cyclic
permutation of its first row $\vecb a_\lambda \equiv
(a_{\lambda,1},\ldots,a_{\lambda,m})^T$, which is defined term-wise by
\begin{equation}\label{cyclic:G_lam_row}
  a_{\lambda,1} \equiv R_{\lambda} \,; \qquad a_{\lambda,j} =
  G_{\lambda,j1} \equiv G_{\lambda} (\vecb x_j;\vecb x_1) \,, \quad
  j=2,\ldots, m \,.
\end{equation}
Again we can numerically evaluate $G_{\lambda,j1}$ for $j=2,\ldots,m$
and $R_\lambda$ by using (\ref{cyclic:Glam}) and (\ref{cyclic:Rlam}),
respectively. 

Next, we must determine the full spectrum
(\ref{cyclic:Glam_mat}) of the cyclic and symmetric matrix ${\mathcal
  G}_\lambda$. For the $m\times m$ cyclic matrix ${\mathcal
  G}_{\lambda}$, generated by permutations of the row vector $\vecb
a_{\lambda}$, it is well-known that its eigenvectors $\vecb v_j$ and
eigenvalues $\omega_{\lambda,j}$ are
\begin{equation}\label{cyclic:mat_eig}
\omega_{\lambda, j}= \sum_{n=0}^{m-1} a_{\lambda, n+1} e^{2 \pi i (j-1)n/m} \,,
 \qquad \vecb v_j = (1, e^{2 \pi i (j-1)/m} \,, \ldots, e^{2
 \pi i (j-1) (m-1)/m})^T \,, \qquad j=1,\ldots, m\,.
\end{equation} 
Since ${\mathcal G}$ is also necessarily a symmetric matrix it
follows that
\begin{equation}\label{cyclic:arow}
    a_{\lambda,j} = a_{\lambda,m+2-j} \,, \qquad j=2,\ldots, 
\lceil {m/2} \rceil \,,
\end{equation}
where the ceiling function $\lceil x \rceil$ is defined as the
smallest integer not less than $x$. This relation can be used to
simplify the expression (\ref{cyclic:mat_eig}) for
$\omega_{\lambda,j}$, into the form as written below in
(\ref{cyclic:spectrum_all}).  Moreover, as a result of
(\ref{cyclic:arow}), it readily follows that
\begin{equation}\label{cyclic:omega}
    \omega_{\lambda,j}=\omega_{\lambda,m+2-j} \,, \qquad 
 \mbox{for} \quad j=2,\ldots, \lceil {m/2} \rceil \,,
\end{equation}
so that there are $\lceil {m/2} \rceil -1$ eigenvalues of multiplicity
two. For these multiple eigenvalues the two independent real-valued
eigenfunctions are readily seen to be $\mbox{Re}(\vecb v_j)={(\vecb
  v_j +\vecb v_{m+2-j})/2}$ and $\mbox{Im}(\vecb v_j)={(\vecb v_j
  -\vecb v_{m+2-j})/(2i)}$. In addition to $\omega_1$, we also observe
that there is an additional eigenvalue of multiplicity one when $m$ is
even.

In this way, our result for the matrix spectrum of ${\mathcal G}_{\lambda}$
is as follows: The synchronous eigenpair of ${\mathcal G}_{\lambda}$ is 
\bsub \label{cyclic:spectrum_all}
\begin{equation}\label{cyclic:spectrum_1}
\omega_{\lambda, 1} = \sum_{n=1}^m a_{\lambda,n}\,, \qquad v_1 =
(1, \ldots, 1)^T\,,
\end{equation}
while the other eigenvalues, corresponding to the asynchronous modes, are
\begin{equation}\label{cyclic:spectrum_j}
\omega_{\lambda, j} = \sum_{n=0}^{m-1} \cos\left( \frac{2 \pi
(j-1)\,n}{m}\right) a_{\lambda, n+1}\,, \qquad j=2,\ldots,m \,,
\end{equation}
where $\omega_{\lambda,j}=\omega_{\lambda,m+2-j}$ for $j=2,\ldots,
\lceil {m/2} \rceil$. When $m$ is even, we notice that there is an
eigenvalue of multiplicity one given by
$\omega_{\lambda,\frac{m}{2}+1}=\sum_{n=0}^{m-1} (-1)^{n}
a_{n+1}$. The corresponding eigenvectors for
$j=2,\ldots, \lceil {m/2} \rceil$ can be written as
\begin{equation}\label{cyclic:eigvec}
\begin{aligned}
 \vecb v_j &= \left(1, \cos \left(\frac{2 \pi (j-1)}{m}\right), \ldots,
 \cos \left( \frac{2 \pi (j-1)(m-1)}{m} \right)\,\right)^T \,,\\
 \vecb v_{m+2-j} &= \left(0, \sin \left( \frac{2 \pi (j-1)}{m}\right),
 \ldots, \sin \left( \frac{2 \pi (j-1)(m-1)}{m}\right)\,\right)^T \,.
\end{aligned}
\end{equation}
\esub
Finally, when $m$ is even, there is an additional eigenvector given by
$\vecb v_{\frac{m}{2}+1}=(1,-1,\ldots,-1)^T$.

With the eigenvalues $\omega_{\lambda,j}$, for $j=1,\ldots,m$,
determined in this way, any Hopf bifurcation boundary in parameter
space is obtained by substituting $\lambda=i\lambda_I$ with
$\lambda_I>0$ into (\ref{cyclic:omega_j}), and requiring that the real
and imaginary parts of the resulting expression vanish. This yields,
for each $j=1,\ldots,m$, that
\begin{equation}\label{cyclic:real_imag}
  \mbox{Re}\left(\omega_{\lambda,j}\right) + \frac{1}{2\pi\nu} \left(
  1+ \frac{D\nu}{d_1} \right) + \frac{d_2 D} {d_1\tau} \mbox{Re}\left(
  \frac{M_{11}}{\det(\lambda I-J) }\right)=0 \,, \qquad
  \mbox{Im}\left(\omega_{\lambda,j}\right) + \frac{d_2 D} {d_1\tau}
  \mbox{Im}\left( \frac{M_{11}}{\det(\lambda I-J) }\right)=0 \,.
\end{equation}
Finally, we can use the winding number criterion of complex analysis
on (\ref{cyclic:omega_j}) to count the number of eigenvalues of the
linearization when the parameters are off any Hopf bifurcation
boundary.  This criterion is formulated below in \S \ref{cyc:sel}.

We remark that in the limit $D\gg 1$, we can use $K_{0}(z)\sim
-\log{z}$ together with $I_{0}(z)\sim 1+{z^2/4}$ as $z\to 0$, to
estimate from the $n=0$ term in (\ref{cyclic:Glam}) and
(\ref{cyclic:Rlam}) that
$-(2\pi)^{-1}{K_0^{\prime}(\theta_\lambda)/I_0^{\prime}(\theta_\lambda)}\sim
{D/\left[\pi(1+\tau\lambda)\right]}$ as $D\to \infty$. Therefore, for
$D\to \infty$, the Green's matrix ${\cal G}_{\lambda}$ satisfies
${\mathcal G}_\lambda \to {Dm {\mathcal
    E}/{\left[\pi(1+\tau\lambda)\right]}}$, where ${\mathcal
  E}={{\vecb e}{\vecb e}^T/m}$ and $\vecb e\equiv
(1,\ldots,1)^T$. This yields for $D\gg 1$ that
$\omega_1={Dm/{\left[\pi (1+\tau\lambda)\right]}}$ and
$\omega_j={\mathcal O}(1)$ for $j=2,\dots,n$. By substituting these
expressions into (\ref{cyclic:real_imag}), we can readily recover the
spectral problems (\ref{dlarge:sync_new}) and (\ref{dlarge:async}),
considered in \S \ref{sec:largeD}, associated with the regime
$D={\mathcal O}(\nu^{-1})$. Therefore, (\ref{cyclic:real_imag})
provides a smooth transition to the leading-order spectral problems
considered in \S \ref{sec:largeD} for $D={\mathcal O}(\nu^{-1})$.

\subsection{Example: The Sel'kov Model}\label{cyc:sel}

We now use (\ref{cyclic:real_imag}) to compute phase diagrams in the
$\tau$ versus $D$ parameter space associated with $m$ equally-spaced
cells of radius $\epsilon$ on a ring of radius $r_0$, with $0<r_0<1$,
concentric within the unit disk. For the intracellular dynamics we let
$n=2$, so that $\vecb u=(u_1,u_2)^T$, and we consider the Sel'kov
dynamics $\vecb F=(F(u_1,u_2),G(u_1,u_2))^T$ as given in
(\ref{2dsel:fg}).  For this choice, (\ref{cyclic:ss}) yields the
steady-state solution $(u_{1e},u_{2e})^T$ for the coupled cell-bulk
system given by \bsub \label{cyclic:example}
\begin{equation}
   u_{1e} = \frac{\mu}{1+{2\pi D\beta/\tau}} \,, \qquad 
 u_{2e} = \frac{\mu}{\alpha + u_{1e}^2} \,,
\end{equation}
where $\beta$ is defined in (\ref{cyclic:ss}). Upon using (\ref{2dsel:jac})
we calculate that
\begin{equation}\label{cyc:det_tr}
   \mbox{det}(J)=\epsilon_0 \left(\alpha + u_{1e}^2\right)>0 \,, \qquad
   \mbox{tr}(J)= \frac{1}{\alpha+u_{1e}^2}\left[ 2 u_{1e}\mu -
     (\alpha+u_{1e}^2) - \epsilon_0 (\alpha+u_{1e}^2)^2\right] \,.
\end{equation}
\esub In this subsection we fix the Sel'kov parameters $\mu$, $\alpha$, and
$\epsilon_0$, the permeabilities $d_1$ and $d_2$, and the cell radius
$\epsilon$ as
\begin{equation}
   \mu=2 \,, \qquad \alpha=0.9\,, \qquad \epsilon_0=0.15, \qquad d_1=0.8\,, 
 \qquad d_2=0.2 \,, \qquad \epsilon=0.05 \,. \label{cyc:par}
\end{equation}
With these values for $\mu$, $\alpha$, and $\epsilon_0$, the
intracellular dynamics has a stable steady-state when uncoupled from
the bulk.

Then, to determine the Hopf bifurcation boundary for the coupled
cell-bulk model we set $M_{11}=\lambda-G_{u_{2}}^e$ in
(\ref{cyclic:real_imag}), and use $G_{u_{2}}^e=-\mbox{det}(J)$ as
obtained from (\ref{2dsel:jac}). By letting $\lambda=i\lambda_I$
in the resulting expression, we conclude that any Hopf bifurcation
boundary, for each mode $j=1,\ldots,m$, must satisfy
\begin{equation}\label{cyclic:hb_ex}
   \begin{split}
  \mbox{Re}\left(\omega_{\lambda,j}\right) + \frac{1}{2\pi\nu} \left(
  1+ \frac{D\nu}{d_1} \right) - \left(\frac{d_2 D} {d_1\tau}\right) \frac{
  \left[ \lambda_I^2 \mbox{tr}(J) + \mbox{det}(J) \left(\lambda_I^2-
 \mbox{det}(J)\right)\right]}{ \left[ \left(\mbox{det}(J)-\lambda_I^2\right)^2
  + \left(\lambda_I \mbox{tr}(J)\right)^2 \right]} =0 \,,\\
  \mbox{Im}\left(\omega_{\lambda,j}\right) + \left(\frac{d_2 D} {d_1\tau}\right)
\frac{  \left[ \lambda_I \left(\mbox{det}(J)-\lambda_I^2\right) + 
   \mbox{det}(J) \mbox{tr}(J) \lambda_I \right]}
 { \left[ \left(\mbox{det}(J)-\lambda_I^2\right)^2
  + \left(\lambda_I \mbox{tr}(J)\right)^2 \right]} =0 \,.
   \end{split}
\end{equation}
For a specified value of $D$, we view (\ref{cyclic:hb_ex}) as a
coupled system for the Hopf bifurcation value of $\tau$ and the
corresponding eigenvalue $\lambda_I$, which we solve by Newton's
method. 

For parameter values off of any Hopf bifurcation boundary, we
can use the winding number criterion on ${\mathcal F}_j(\lambda)$ in
(\ref{cyclic:omega_j}) to count the number of unstable eigenvalues
$N_j$ of the linearization for the $j$-th mode. By using the
argument principle, we obtain that the number $N_j$ of roots of
${\mathcal F}_j(\lambda)=0$ in $\mbox{Re}(\lambda)>0$ is
\begin{equation}\label{wind:last_1}
N_j=\frac{1}{2\pi}[\mbox{arg} {\mathcal F}_j]_{\Gamma}+P\,, 
\end{equation}
where $P$ is the number of poles of ${\mathcal F}_{j}(\lambda)$ in
$\mbox{Re}(\lambda)>0$, and the square bracket denotes the change in
the argument of ${\mathcal F}_{j}$ over the contour $\Gamma$. Here the
closed contour $\Gamma$ is the limit as ${\mathcal R}\to \infty$ of
the union of the imaginary axis, which can be decomposed as
$\Gamma_{I+}=i\lambda_I$ and $\Gamma_{I-}=-i\lambda_I$, for
$0<\lambda_I<{\mathcal R}$, and the semi-circle $\Gamma_{\mathcal R}$
defined by $|\lambda|={\mathcal R}$ with
$|\mbox{arg}(\lambda)|\leq\pi/2$. Since $\omega_{\lambda j}$ is
analytic in $\mbox{Re}(\lambda)>0$, it follows that $P$ is determined
by the number of roots of $\det(\lambda I-J)=0$ in
$\mbox{Re}(\lambda)>0$. Since $\mbox{det}(J)>0$, as shown in
(\ref{cyc:det_tr}), we have that $P=2$ if $\mbox{tr}(J)>0$ and $P=0$
if $\mbox{tr}(J)<0$. Next, we let ${\mathcal R}\to \infty$ on
$\Gamma_{\mathcal R}$ and calculate $[\mbox{arg}
  \mathcal{F}_{j}]_{\Gamma_{\mathcal R}}$. It is readily seen that the
Green's matrix ${\mathcal G}_{\lambda}$ tends to a multiple of a
diagonal matrix on $\Gamma_{\mathcal R}$ as ${\mathcal R}\gg 1$, of
the form ${\mathcal G}_{\lambda}\to R_{\lambda,\infty}I$, where
$R_{\lambda,\infty}$ is the regular part of the free-space Green's
function $G_{f}(\vecb x;\vecb x_0) =
(2\pi)^{-1} K_{0}\left(\theta_\lambda|\vecb x-\vecb x_0| \right)$ at
$\vecb x=\vecb x_0$, given explicitly by the first term in the
expression (\ref{cyclic:Rlam}) for $R_{\lambda}$. Since
$\omega_{\lambda,j}\to R_{\lambda,\infty}$ for $j=1,\ldots,m$, we estimate
 on $\Gamma_{\mathcal R}$ as ${\mathcal R}\gg 1$ that
\begin{equation*}
   {\mathcal F}_{j}(\lambda) \sim -\frac{1}{2\pi}
   \log\sqrt{1+\tau\lambda} + c_0 + {\mathcal
     O}\left({1/\lambda}\right) \,,
\end{equation*}
for some constant $c_0$. It follows that $\mathcal{F}_j(\lambda)\sim
O(\ln {\mathcal R})- {i/8}$ as ${\mathcal R}\to \infty$, so that
$\lim_{{\mathcal R}\to
  \infty}[\mbox{arg}\mathcal{F}_{j}]_{\Gamma_{\mathcal R}}=0$.
Finally, since $[\mbox{arg}\mathcal{F}_{j}]_{\Gamma_{I+}}= [\mbox{arg}
  \mathcal{F}_{j}]_{\Gamma_{I-}}$, as a consequence of ${\mathcal F}_j$
being real-valued when $\lambda$ is real, we conclude from
(\ref{wind:last_1}) that
\begin{equation}\label{wind:last_2}
N_j=\frac{1}{2\pi}[\mbox{arg} {\mathcal F}_j]_{\Gamma_{I+}}+ P\,, \qquad
  P = \left\{
\begin{array}{c}
 2  \quad \mbox{if} \quad \mbox{tr}J>0 \\
 0  \quad \mbox{if} \quad \mbox{tr}J<0 
\end{array}
\right. \,.
\end{equation}
By using (\ref{cyclic:hb_ex}) for the real and imaginary parts of 
${\mathcal F}_j$, $[\mbox{arg} {\mathcal F}_j]_{\Gamma_{I+}}$ is
easily calculated numerically by a line sweep over
$0<\lambda_I<{\mathcal R}$. Then, by using (\ref{cyc:det_tr}) to
calculate $\mbox{tr}(J)$, $P$ is readily determined. In this way, 
(\ref{wind:last_2}) leads to a highly tractable numerical
procedure to calculate $N_j$. This criterion was used for all the
results below to identify regions in parameter space where
instabilities occur away from any Hopf bifurcation boundary.

\begin{figure}[htbp]
\begin{center}
\includegraphics[width=0.45\textwidth,height=5.0cm]{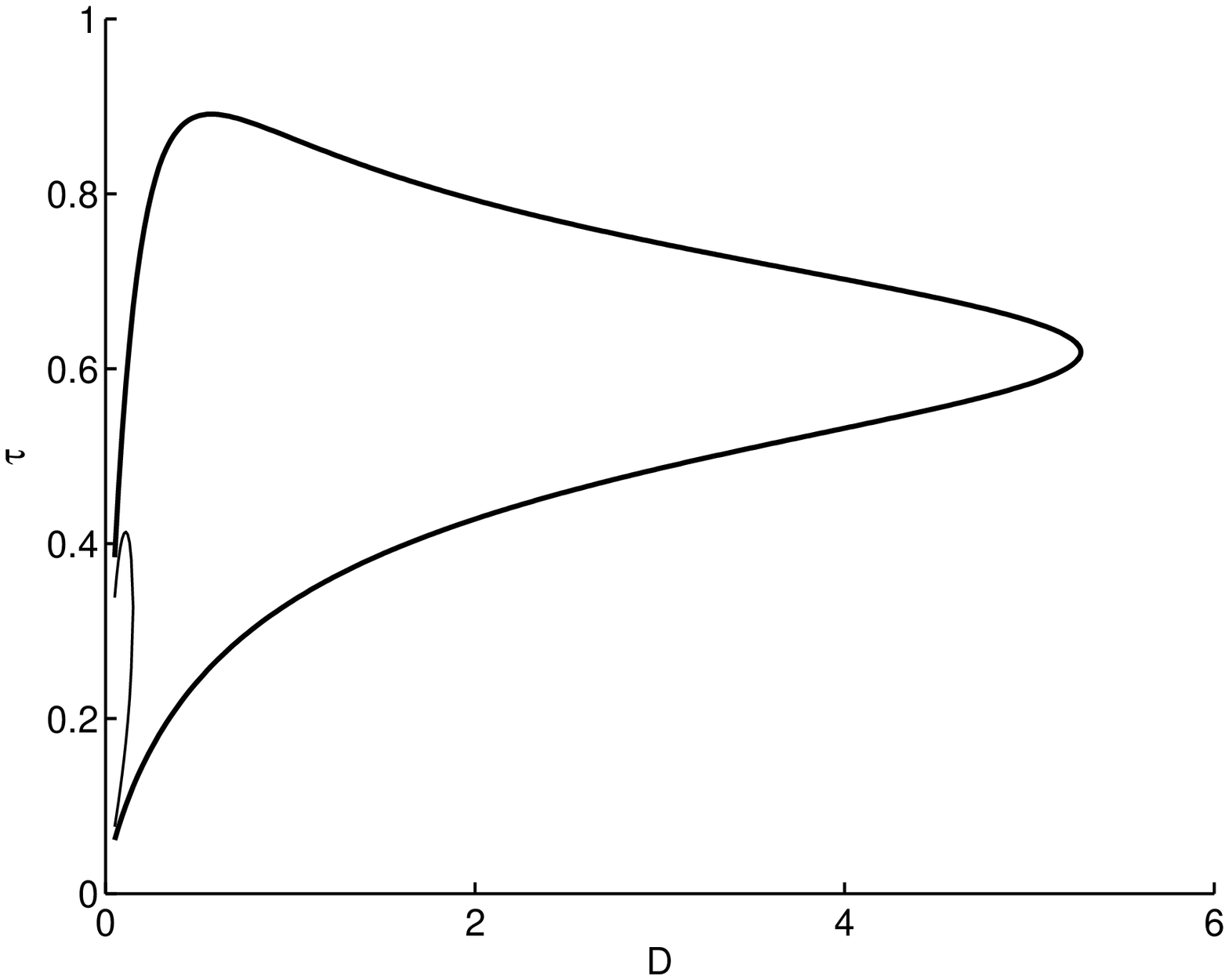}
\includegraphics[width=0.45\textwidth,height=5.0cm]{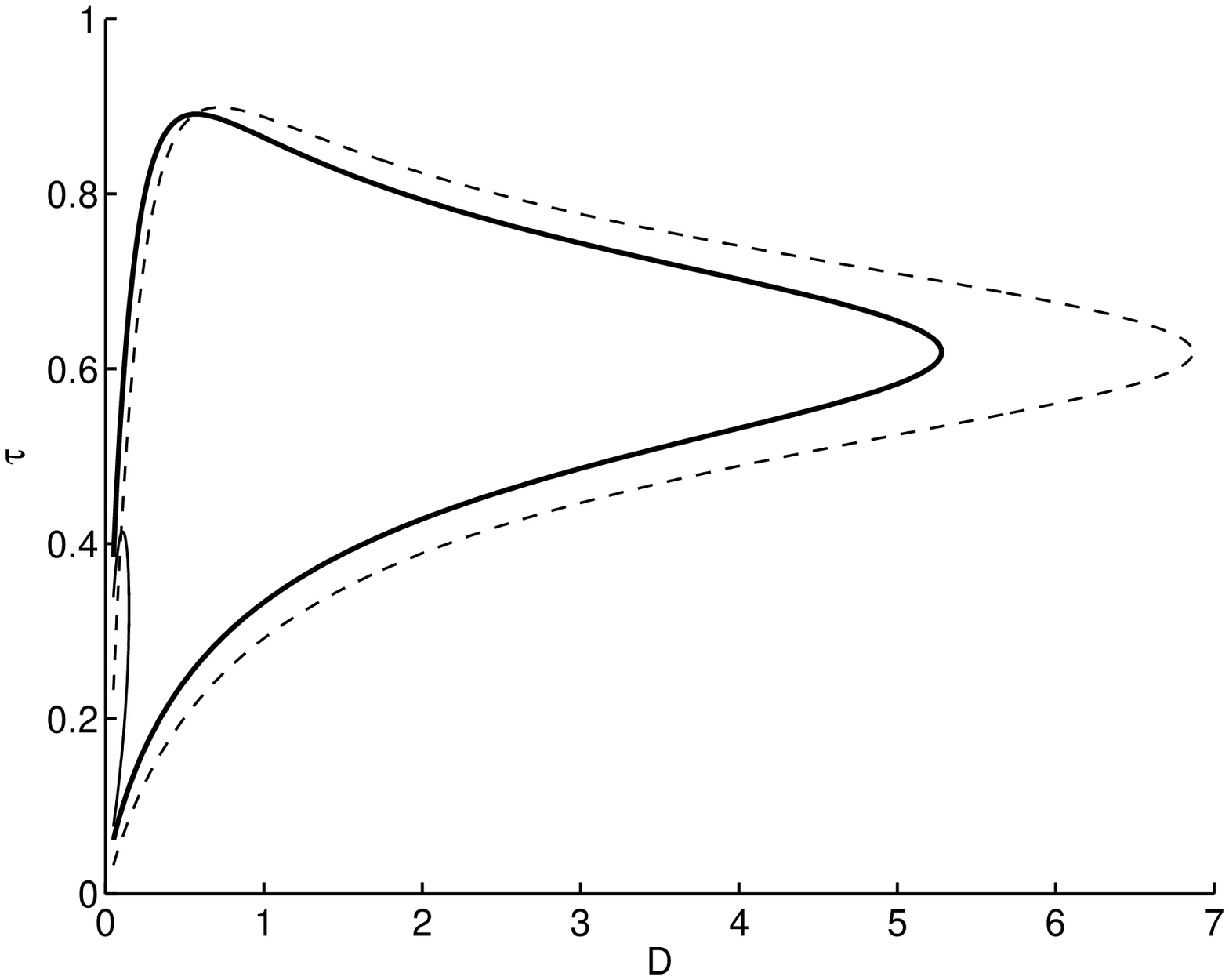}
\caption{Hopf bifurcation boundaries in the $\tau$ versus $D$ plane
  for $m=2$, $r_0=0.25$, and with parameters as in (\ref{cyc:par}), 
  computed from (\ref{cyclic:hb_ex}). Left panel: the heavy solid
  curve and the solid curve are the Hopf bifurcation boundaries for
  the synchronous and asynchronous modes, respectively. Inside the
  respective lobes the corresponding mode is linearly unstable, as
  verified by the winding number criterion (\ref{wind:last_2}).  Right
  panel: same plot except that we include the Hopf bifurcation
  boundary for the synchronous mode from the leading-order
  $D={D_0/\nu}\gg 1$ theory, computed from (\ref{dlarge:sync_new}).}
  \label{fig:m2}
\end{center}
\end{figure}

In Fig.~\ref{fig:m2} we plot the Hopf bifurcation boundaries when
$m=2$ and $r_0=0.25$. From the left panel of this figure, the
synchronous mode is unstable in the larger lobe shaped region, whereas
the asynchronous mode is unstable only in the small lobe for small
$D$, which is contained within the instability lobe for the
synchronous mode. In the right panel of Fig.~\ref{fig:m2} we show the
Hopf bifurcation boundary for the synchronous mode, as obtained from
(\ref{dlarge:sync_new}), corresponding to the leading-order
$D={D_0/\nu}\gg 1$ theory. Since the instability lobe occurs for only
moderate values of $D$, and $\epsilon=0.05$ is only moderately small,
the leading-order theory from the $D={D_0/\nu}$ regime is, as
expected, not particularly accurate in determining the Hopf
bifurcation boundary. The fact that we have stability at a fixed $D$
for $\tau\gg 1$, which corresponds to very fast intracellular
dynamics, is expected since in this limit the intracellular dynamics
becomes decoupled from the bulk diffusion. Alternatively, if $\tau\ll
1$, then for a fixed $D$, the intracellular reactions proceed too
slowly to create any instability. Moreover, in contrast to the large
region of instability for the synchronous mode as seen in
Fig.~\ref{fig:m2}, we observe that the lobe of instability for the
asynchronous mode only occurs for small values of $D$, where the
diffusive coupling, and communication, between the two cells is rather
weak. Somewhat more paradoxically, we also observe that the
synchronous lobe of instability is bounded in $D$. This issue is
discussed in more detail below.

\begin{figure}[htbp]
\begin{center}
\includegraphics[width=0.45\textwidth,height=5.0cm]{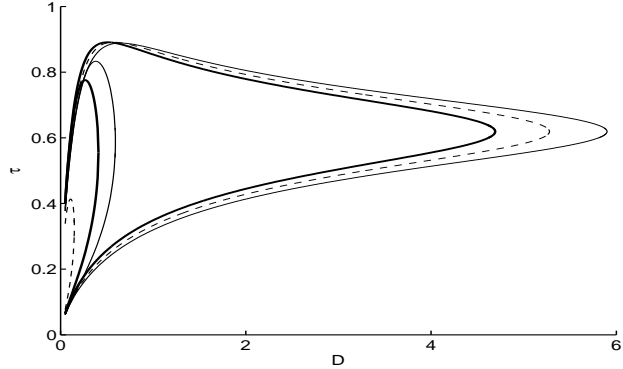}
\caption{Hopf bifurcation boundaries for the synchronous mode (larger
  lobes) and the asynchronous mode (smaller lobes) in the $\tau$
  versus $D$ plane for $m=2$ and for three values of $r_0$, with
  $r_0=0.5$ (heavy solid curves), $r_0=0.75$ (solid curves), and
  $r_0=0.25$ (dashed curves).  The other parameters are given in
  (\ref{cyc:par}). We observe that as $r_0$ decreases, where the two
  cells become more closely spaced, the lobe of instability for the
  asynchronous mode decreases.}
  \label{fig:m2_all}
\end{center}
\end{figure}

In Fig.~\ref{fig:m2_all} we show the effect of changing the ring
radius $r_0$ on the Hopf bifurcation boundaries. By varying $r_0$, we
effectively are modulating the distance between the two cells. From
this figure we observe that as $r_0$ is decreased, the lobe of
instability for the asynchronous mode decreases, implying, rather
intuitively, that at closer distances the two cells are better able to
synchronize their oscillations than when they are farther apart.  We
remark that results from the leading-order theory of \S
\ref{sec:largeD} for the $D={\mathcal O}(\nu^{-1})$ regime would be
independent of $r_0$. We further observe from this figure that a
synchronous instability can be triggered from a more clustered spatial
arrangement of the cells inside the domain. In particular, for $D=5$
and $\tau=0.6$, we observe from Fig.~\ref{fig:m2_all} that we are
outside the lobe of instability for $r_0=0.5$, but inside the lobe of
instability for $r_0=0.25$ and $r_0=0.75$. We remark that due to the
Neumann boundary conditions the cells on the ring with $r_0=0.75$ are
close to two image cells outside the unit disk, which leads to a
qualitatively similar clustering effect of these near-boundary cells
as when they are on the small ring of radius $r_0=0.25$.

\begin{figure}[htbp]
\begin{center}
\includegraphics[width=0.45\textwidth,height=5.0cm]{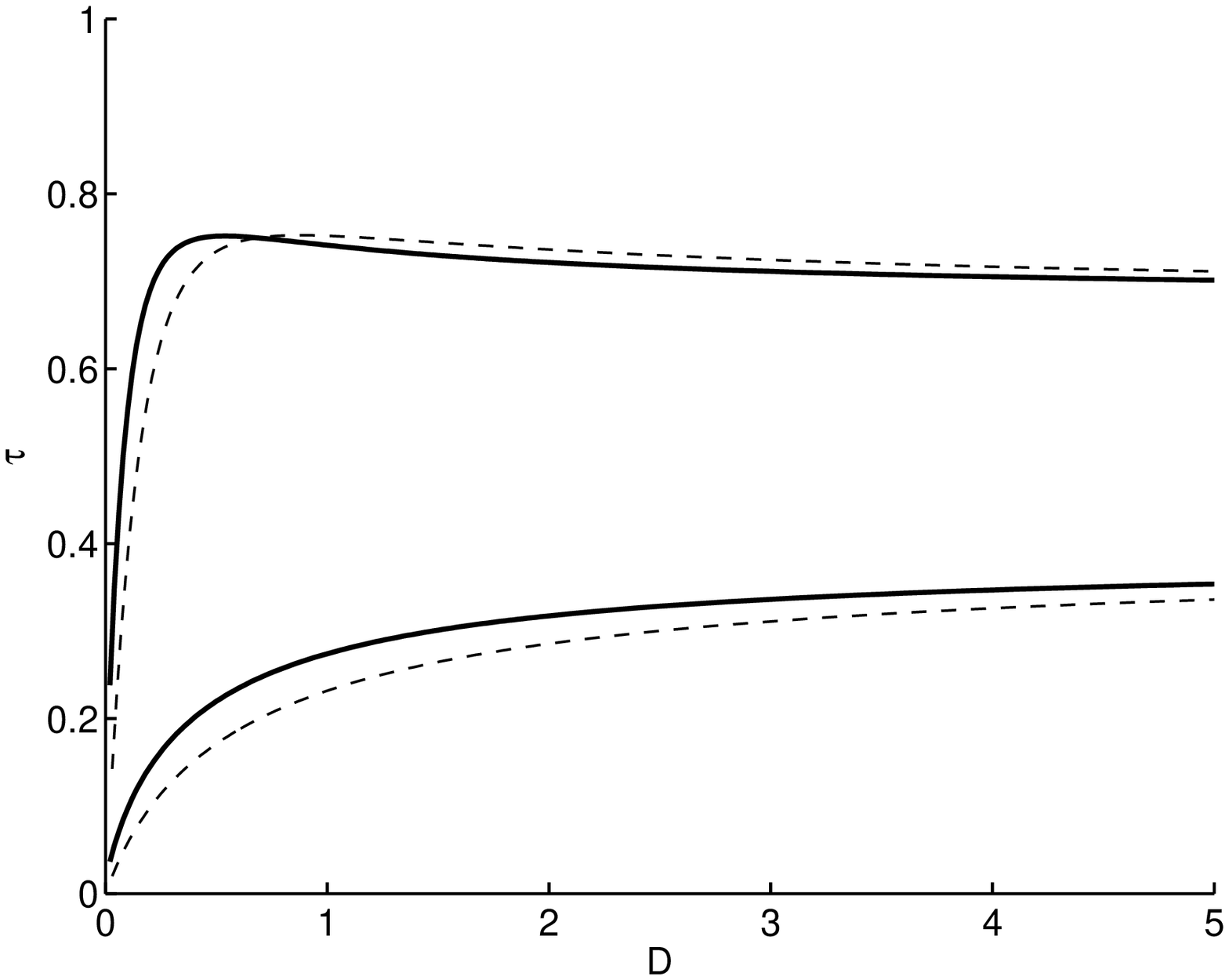}
\includegraphics[width=0.45\textwidth,height=5.0cm]{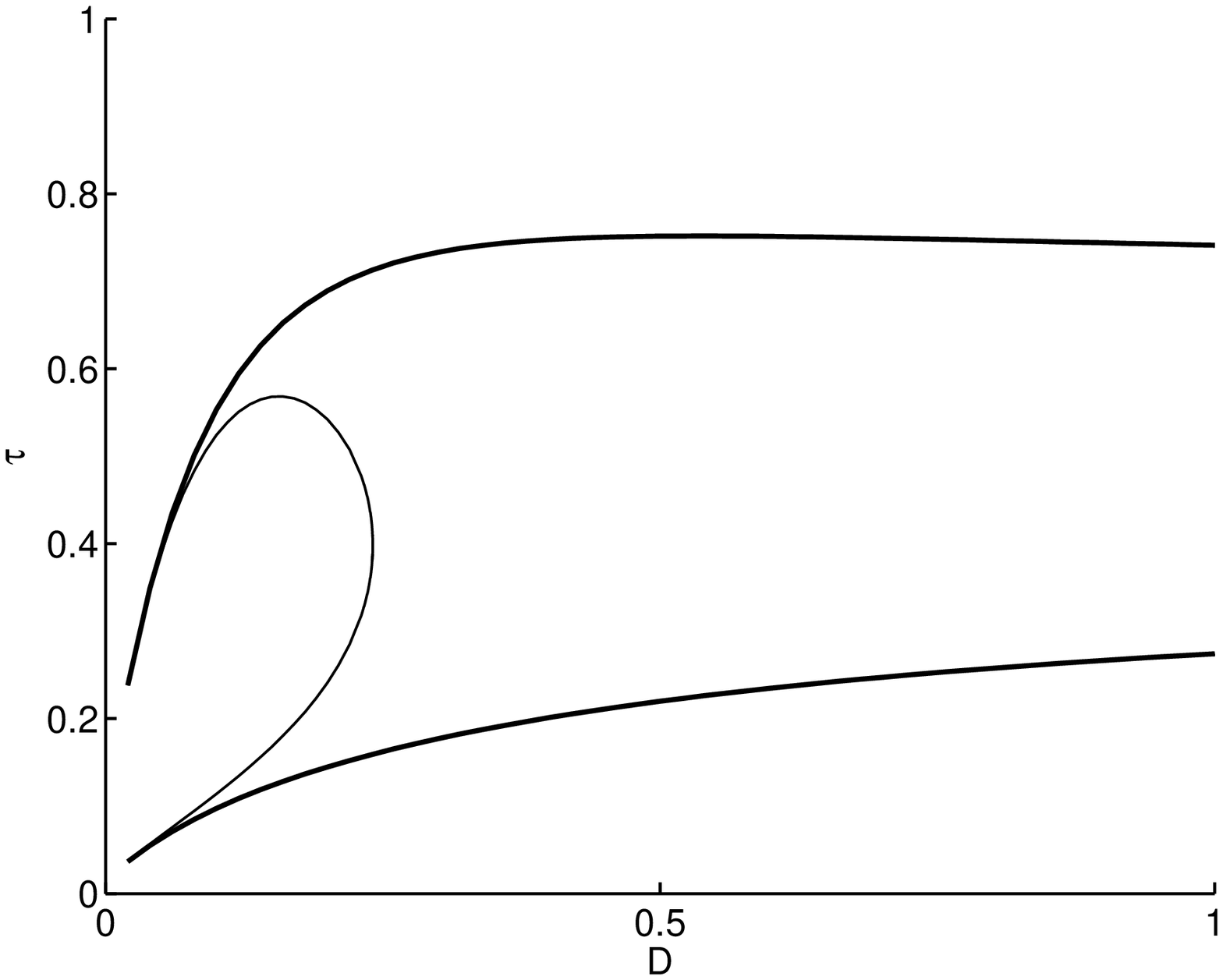}
\caption{Left panel: Hopf bifurcation boundaries in the $\tau$ versus
  $D$ plane for the synchronous mode for $m=3$ equally-spaced cells on
  a ring of radius $r_0=0.50$ (heavy solid curves), as computed from
  (\ref{cyclic:hb_ex}), with parameters as in (\ref{cyc:par}). The
  dashed curve is the Hopf bifurcation boundary from the leading-order
  $D={D_0/\nu}$ theory computed from (\ref{dlarge:sync_new}).  Right
  panel: The Hopf bifurcation boundaries for the asynchronous mode
  (solid curve) and the synchronous mode (heavy solid curve) shown in
  a magnified region of $D$. The asynchronous mode is linearly
  unstable only inside this small lobe, which lies within the
  unstable region for the synchronous mode.}
  \label{fig:m3}
\end{center}
\end{figure}

In Fig.~\ref{fig:m3} we plot the Hopf bifurcation boundaries when
$m=3$ and $r_0=0.5$. For $m=3$, we now observe that the region where
the synchronous mode is unstable is unbounded in $D$. The lobe of
instability for the asynchronous mode still exists only for small $D$,
as shown in the right panel of Fig.~\ref{fig:m3}. In this case, we
observe that the Hopf bifurcation boundary for the synchronous mode,
corresponding to the leading-order $D={D_0/\nu}\gg 1$ theory and computed
from (\ref{dlarge:sync_new}), now agrees rather well with results
computed from (\ref{cyclic:hb_ex}).

\begin{figure}[htbp]
\begin{center}
\includegraphics[width=0.45\textwidth,height=5.0cm]{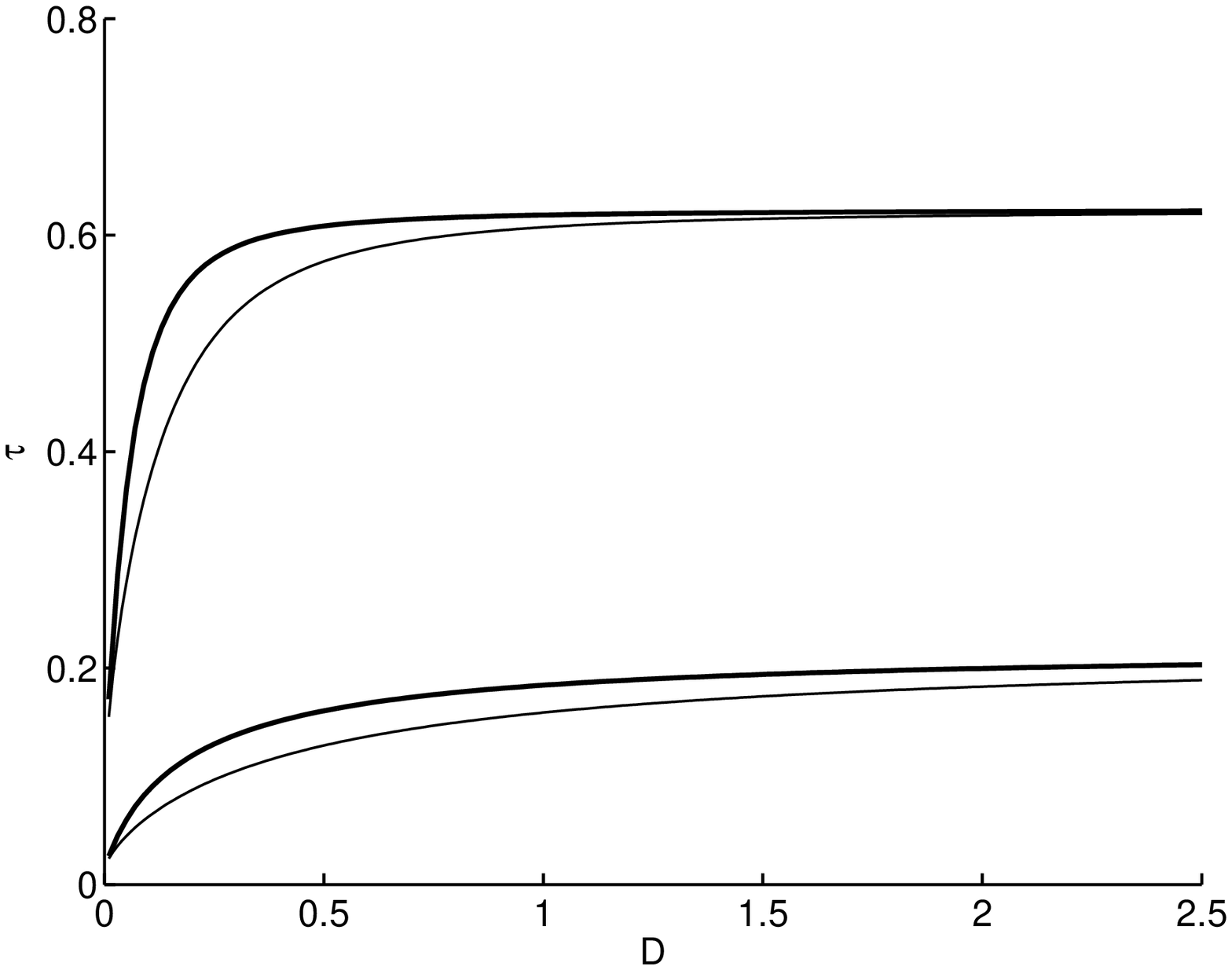}
\includegraphics[width=0.45\textwidth,height=5.0cm]{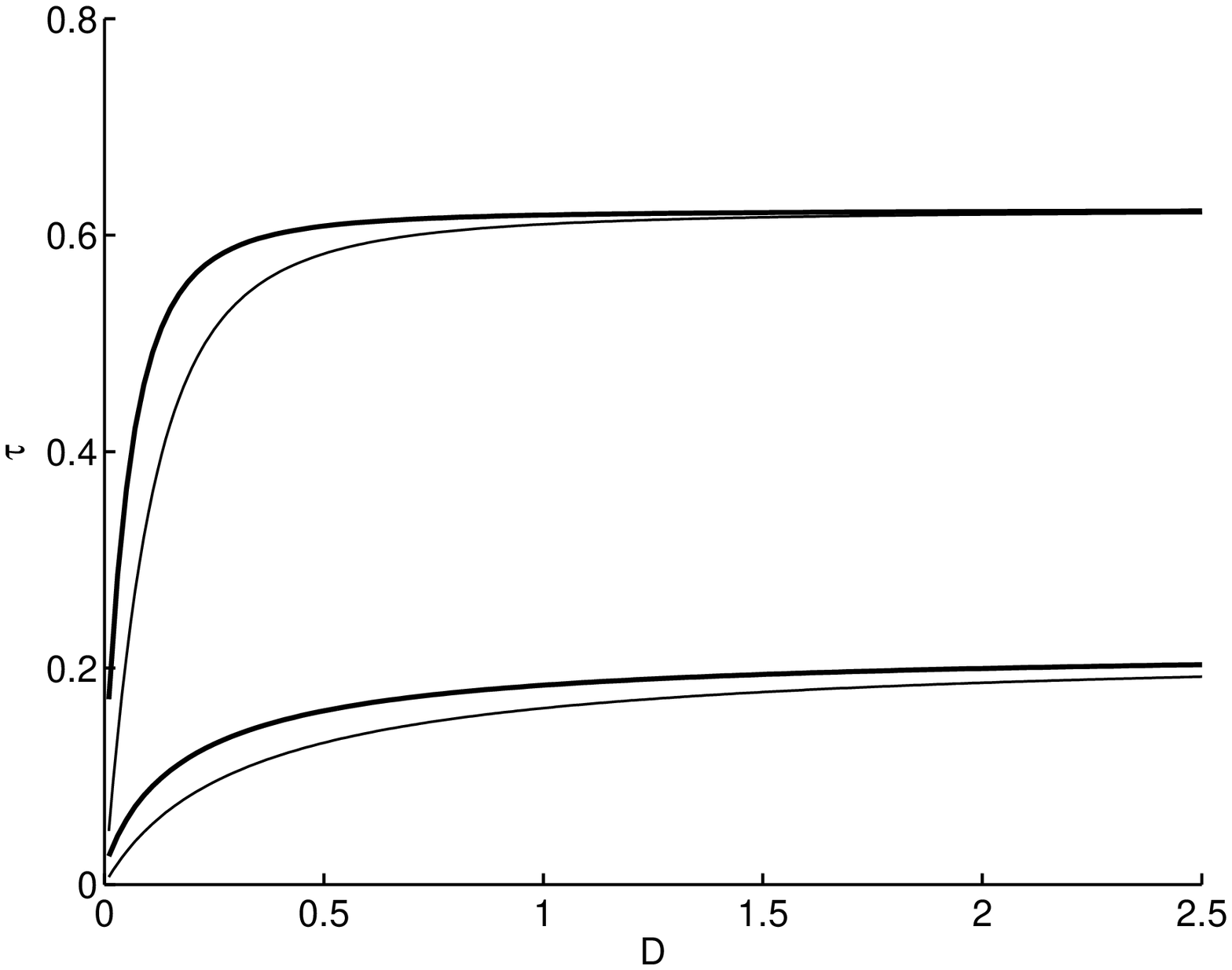}
\caption{Left panel: Hopf bifurcation boundaries in the $\tau$ versus
  $D$ plane for the synchronous mode for $m=5$ equally-spaced cells on
  a ring of radius $r_0=0.25$ (solid curves) and radius $r_0=0.5$
  (heavy solid curves) concentric with the unit disk, as computed from
  (\ref{cyclic:hb_ex}), with parameters (\ref{cyc:par}). Right panel:
  Comparison of the Hopf bifurcation boundaries for the synchronous
  mode with $r_0=0.5$ (heavy solid curves), as computed from
  (\ref{cyclic:hb_ex}), with that obtained from
  (\ref{dlarge:sync_new}) for the leading-order $D={D_0/\nu}$ theory
  (solid curves). These curves agree well when $D$ is large.}
  \label{fig:m5}
\end{center}
\end{figure}

\begin{figure}[htbp]
\begin{center}
\includegraphics[width=0.45\textwidth,height=5.0cm]{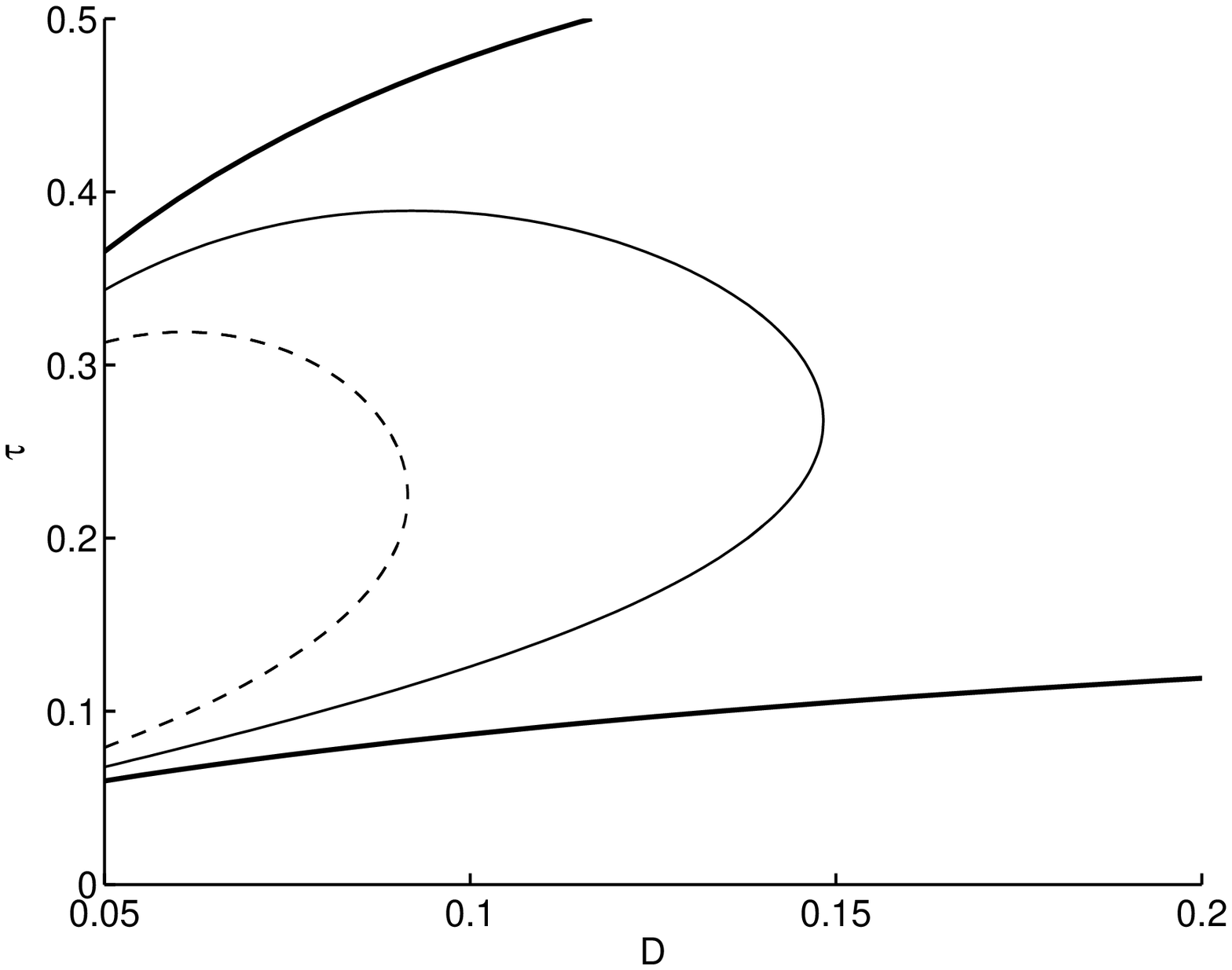}
\includegraphics[width=0.45\textwidth,height=5.0cm]{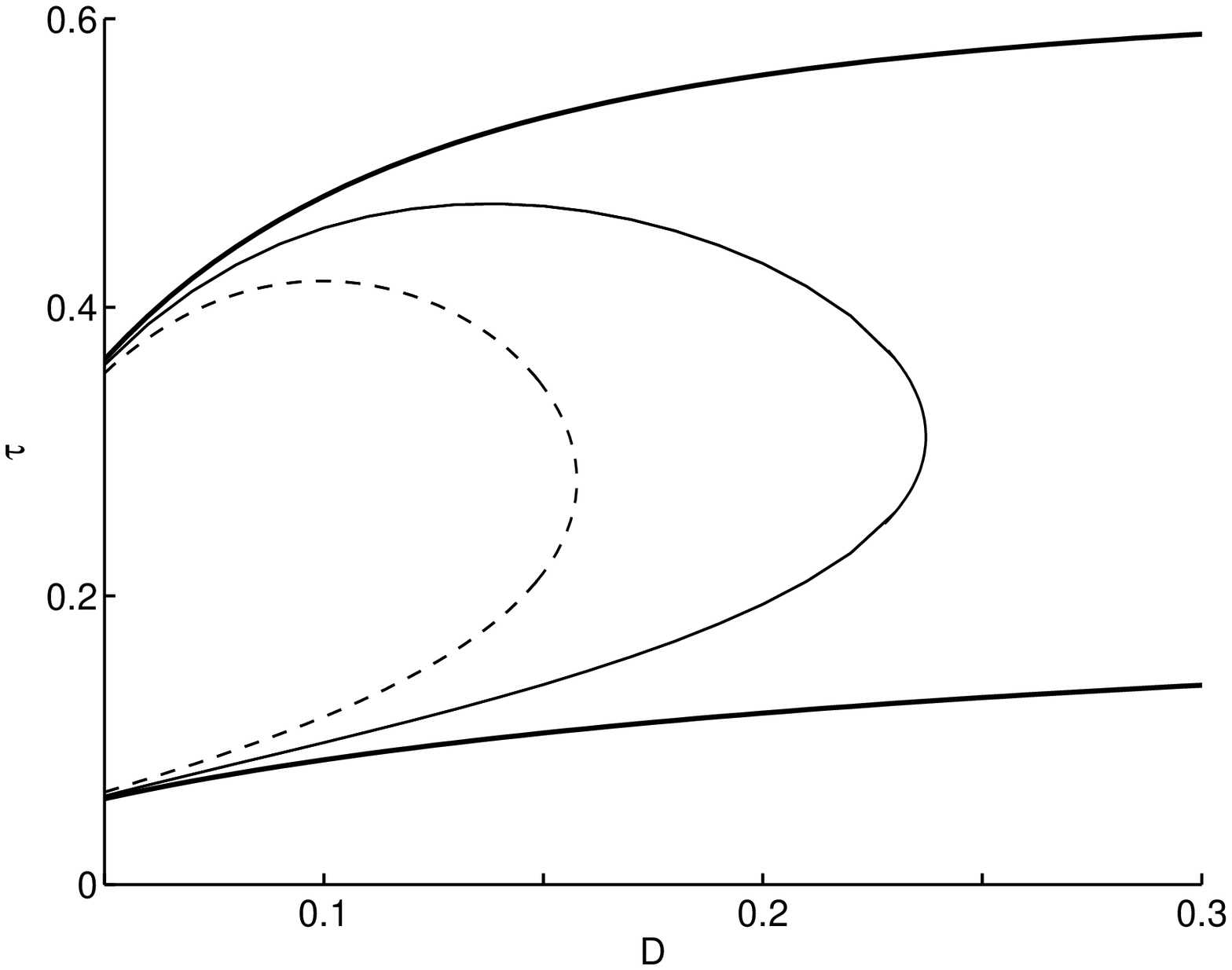}
\caption{Hopf bifurcation boundaries for the two distinct asynchronous
  modes when $m=5$ for $r_0=0.5$ (left panel) and $r_0=0.75$ (right
  panel), as computed from (\ref{cyclic:hb_ex}) with $j=2,5$ (larger
  solid curve lobe) and with $j=3,4$ (smaller dashed curve lobe). The
  heavy solid curves are the Hopf bifurcation boundaries for the
  synchronous mode.  The parameters are as in (\ref{cyc:par}). The
  asynchronous mode for $j=2,5$ and $j=3,4$ is linearly unstable only
  inside the larger and smaller lobe, respectively.}
\label{fig:asy_m5}
\end{center}
\end{figure}

In the left panel of Fig.~\ref{fig:m5} we plot the Hopf bifurcation
boundaries for the synchronous mode for $m=5$ when $r_0=0.5$ (heavy
solid curves) and for $r_0=0.25$ (solid curves). We observe that for
moderate values of $D$, the Hopf bifurcation values do depend
significantly on the radius of the ring. The synchronous mode is
unstable only in the infinite strip-like domain between these Hopf
bifurcation boundaries. Therefore, only in some intermediate range of
$\tau$, representing the ratio of the rates of the intracellular
reaction and bulk decay, is the synchronous mode unstable. As
expected, the two curves for different values of $r_0$ coalesce as $D$
increases, owing to the fact that the leading-order stability theory
for $D={D_0/\nu}\gg 1$, as obtained from (\ref{dlarge:sync_new}), is
independent of $r_0$. In the right panel of Fig.~\ref{fig:m5} we
compare the Hopf bifurcation boundaries for the synchronous mode with
$r_0=0.5$ with that obtained from (\ref{dlarge:sync_new}),
corresponding to the leading-order theory in the $D={D_0/\nu\gg 1}$
regime.  Rather curiously, we observe upon comparing the solid curves
in the left and right panels in Fig.~\ref{fig:m5} that the Hopf
bifurcation boundaries from the $D={\mathcal O}(1)$ theory when
$r_0=0.25$, where the five cells are rather clustered near the origin,
agree very closely with the leading order theory from the
$D={D_0/\nu}\gg 1$ regime. Since the clustering of cells is
effectively equivalent to a system with a large diffusion coefficient,
this result above indicates, rather intuitively, that stability
thresholds for a clustered spatial arrangement of cells will be more
closely approximated by results obtained from a large $D$
approximation than for a non-clustered spatial arrangement of cells.
In Fig.~\ref{fig:asy_m5} we plot the Hopf bifurcation boundaries for
the distinct asynchronous modes when $m=5$ for
$r_0=0.5$ (left panel) and $r_0=0.75$ (right panel), as computed from
(\ref{cyclic:hb_ex}) with $j=2,5$ (larger lobe) and with $j=3,4$
(smaller lobe). The asynchronous modes are only linearly unstable
within these small lobes. 

To theoretically explain the observation that the instability region
in the $\tau$ versus $D$ plane for the synchronous mode is bounded for
$m=2$, but unbounded for $m\geq 3$, we must first extend the large $D$
analysis of \S \ref{sec:odes} to the case of $m$ small cells.  We
readily derive, assuming identical behavior in each of the $m$
cells, that the reduced cell-bulk dynamics (\ref{d:odes}) for one cell
must be replaced by
\begin{equation}\label{d:modes}
    U_0^{\prime}= -\frac{1}{\tau} \left( 1 + \frac{2\pi m
      d_1}{|\Omega|}\right) U_0 + \frac{2\pi d_2 m}{\tau |\Omega|} u_1
    \,, \qquad \vecb u^{\prime} = \vecb F(\vecb u) + \frac{2\pi}{
  \tau} \left[d_1U_0-d_2u_1\right] \vecb e_1 \,,
\end{equation}
when there are $m$ cells. This indicates that the effective domain
area is ${|\Omega|/m}={\pi/m}$ when there are $m$ cells. Therefore, to
examine the stability of the steady-state solution of (\ref{d:modes})
for the Sel'kov model, we need only replace $|\Omega|$ with ${|\Omega|/m}$
in the Routh-Hurwitz criteria for the cubic (\ref{largeD:3ode}).

\begin{figure}[htbp]
\begin{center}
\includegraphics[width=0.45\textwidth,height=5.0cm]{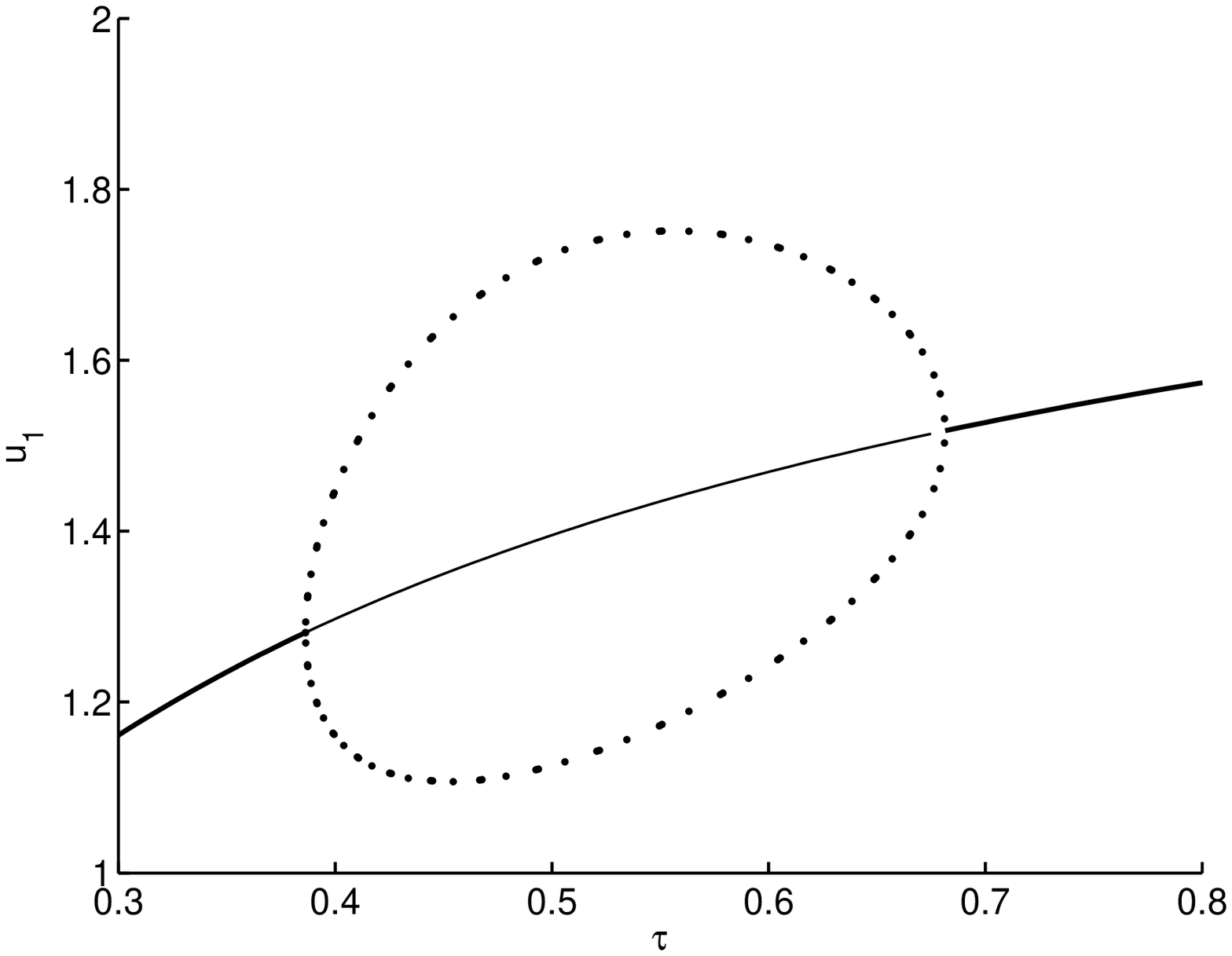}
\includegraphics[width=0.45\textwidth,height=5.0cm]{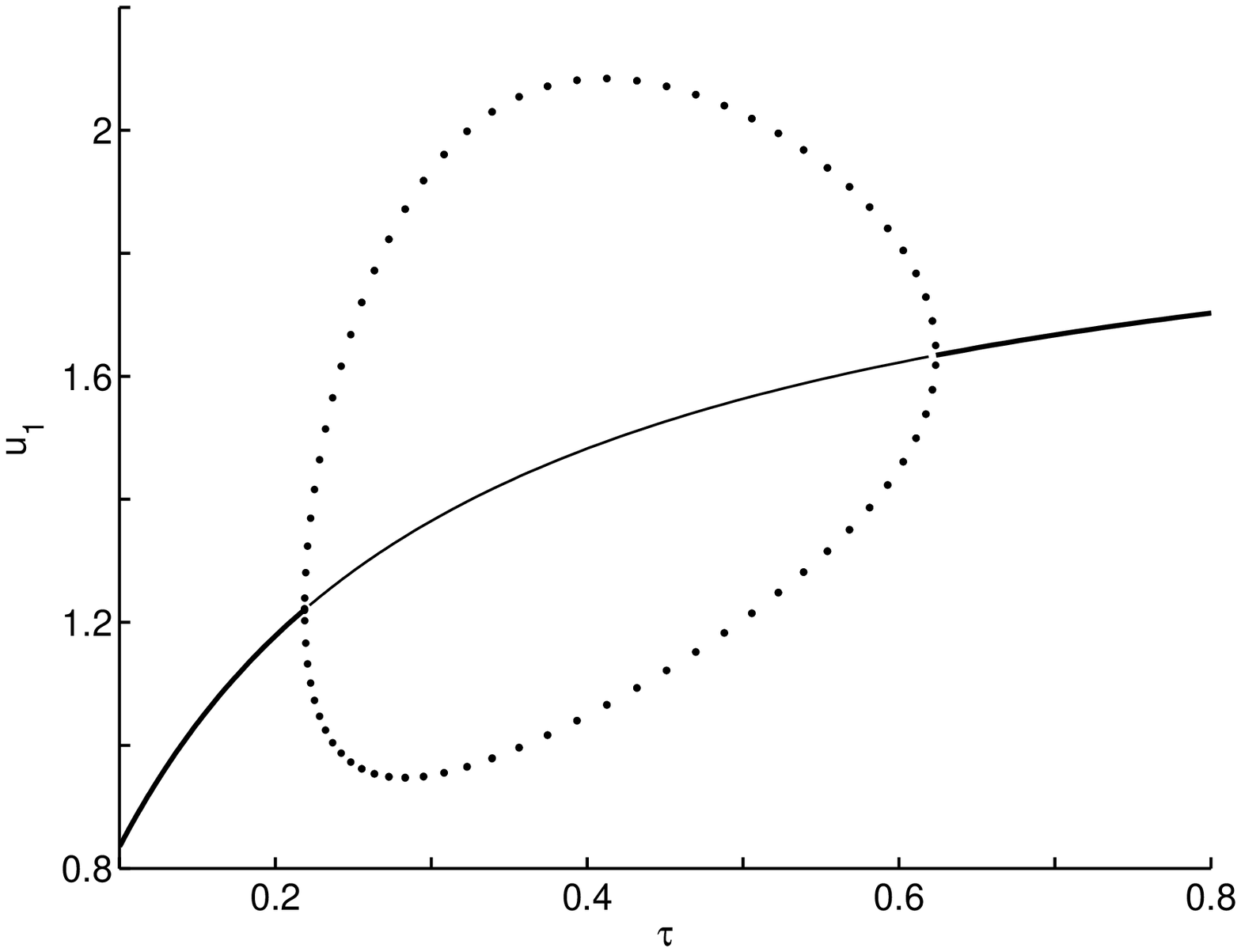}
\caption{Global bifurcation diagram of $u_{1e}$ versus $\tau$ for the
  Sel'kov model (\ref{2dsel:fg}) as computed using XPPAUT \cite{xpp}
  from the ODE system (\ref{d:modes}) characterizing the limiting
  problem as $D\to\infty$ with $m$ small cells in the unit disk
  $\Omega$. Left panel: $m=3$. Right panel: $m=5$. The Sel'kov
  parameters are $\mu=2$, $\alpha=0.9$, and $\epsilon_0=0.15$, while
  $d_1=0.8$, and $d_2=0.2$. The thick/thin solid line represents
  stable/unstable steady-state solutions, while the solid dots
  indicate a stable synchronous periodic solution in the
  cells. For $m=5$ (right panel), there are two Hopf bifurcation points
  at $\tau=0.2187$ and $\tau=0.6238$. For $m=3$ (right panel), the two
  Hopf bifurcation points are at $\tau=0.3863$ and
  $\tau=0.6815$. These points correspond to the horizontal asymptotes
  as $D\to\infty$ in Fig.~\ref{fig:m5} for $m=5$ and in
  Fig.~\ref{fig:m3} for $m=3$.}
  \label{bif:m3m5}
\end{center}
\end{figure}

With this approach, in Fig.~\ref{bif:m3m5} we show that there are two
Hopf bifurcation values of $\tau$ for the steady-state solution of
(\ref{d:modes}) when $m=3$ and $m=5$. These values correspond to the
horizontal asymptotes as $D\to\infty$ in Fig.~\ref{fig:m3} for $m=3$
and in Fig.~\ref{fig:m5} for $m=5$. The numerical results from XPPAUT
\cite{xpp} in Fig.~\ref{bif:m3m5} then reveal the existence of a
stable periodic solution branch connecting these Hopf bifurcation
points for $m=3$ and $m=5$. A qualitatively similar picture holds for
any $m\geq 3$. In contrast, for $m=2$, we can verify numerically using
(\ref{largeD:3ode}), where we replace $|\Omega|$ with ${|\Omega|/2}$,
that the Routh-Hurwitz stability criteria $p_1>0$, $p_3>0$, and
$p_1p_2>p_3$ hold for all $\tau>0$ when $m=2$ (and also $m=1$).
Therefore, for $m=2$, there are no Hopf bifurcation points in $\tau$
for the steady-state solution of (\ref{d:modes}). This analysis
suggests why there is a bounded lobe of instability for the synchronous
mode when $m=2$, as was shown in Fig.~\ref{fig:m2}.

We now suggest a qualitative reason for our observation that the lobe
of instability for the synchronous mode is bounded in $D$ only when
$m<m_c$, where $m_c$ is some threshold. We first observe that the
diffusivity $D$ serves a dual role. Although larger values of $D$
allows for better communication between spatially segregated cells,
suggesting that synchronization of their dynamics should be
facilitated, it also has the competing effect of spatially
homogenizing any perturbation in the diffusive signal. We suggest that
only if the number of cells exceeds some threshold $m_c$, i.e. if some
{\em quorum} is achieved, will the enhanced communication between the
cells, resulting from a decrease in the effective domain area by
${|\Omega|/m}$, be sufficient to overcome the increased homogenizing
effect of the diffusive signal at large values of $D$, and thereby
lead to a synchronized time-periodic response.

\setcounter{equation}{0}
\setcounter{section}{6}
\section{Discussion and Outlook}\label{sec:disc}

We have formulated and studied a general class of coupled cell-bulk
problems in 2-D with the primary goal of establishing whether such a
class of problems can lead to the initiation of oscillatory
instabilities due to the coupling between the cell and bulk. Our
analysis relies on the assumption that the signaling compartments
have a radius that is asymptotically small as compared to the
length-scale of the domain.  In this limit $\epsilon\to 0$ of small
cell radius we have used a singular perturbation approach to determine
the steady-state solution and to formulate the eigenvalue problem
associated with linearizing around the steady-state.  In the limit for
which the bulk diffusivity $D$ is asymptotically large of order
$D={\mathcal O}(\nu^{-1})$, we have derived eigenvalue problems
characterizing the possibility of either synchronous and asynchronous
instabilities triggered by the cell-bulk coupling. Phase diagrams in
parameter space, showing where oscillatory instabilities can be
triggered, were calculated for two specific choices of the
intracellular kinetics. Our analysis shows that triggered oscillations
are not possible when the intracellular dynamics has only one
species. For the regime $D\gg {\mathcal O}(\nu^{-1})$, where the bulk
can be effectively treated as a well-mixed system, and for the simple
case of one cell, we have reduced the cell-bulk PDE system to a
finite-dimensional ODE system for the spatially constant bulk
concentration field coupled to the intracellular dynamics. This ODE
system was shown to have triggered oscillations due to cell-bulk
coupling, and global bifurcation diagrams were calculated for some
specific reaction kinetics, showing that the branch of oscillatory
solutions is globally stable. Finally, for the regime $D={\mathcal
  O}(1)$, where the spatial configuration of cells is an important
factor, we have determined phase-diagrams for the initiation of
synchronous temporal instabilities associated with a ring pattern of
cells inside the unit disk, showing that such instabilities can be
triggered from a more clustered spatial arrangement of the cells
inside the domain.

We now discuss a few possible extensions of this study, and additional
directions that warrant further investigation. For our choices of the
intracellular kinetics used to illustrate the theory, the coupling
between the bulk and cells leads to a unique steady-state solution. In
contrast, it would be interesting to study more elaborate biologically
relevant models, such as those in \cite{tyson2008} and \cite{Muller1},
where bistable and hysteric behavior of the steady-states is
possible. In this way, the cell-bulk coupling and the spatial
arrangement of the cells can possibly trigger transitions between
different steady-state solutions. With regards to our basic model
(\ref{mainbd}), it would be interesting to combine fast methods of
potential theory (cf.~\cite{mck}) to compute the eigenvalue-dependent
Green's function in (\ref{bdgneig}) so as to obtain a hybrid
analytical-numerical method to predict triggered cell-bulk
oscillations from the GCEP (\ref{gcep:full}). This would then allow us
to readily consider random spatial configurations of $m$ cells in an
arbitrary domain, rather than the simple ring patterns of cells
studied in \S \ref{sec:finite_d}.  In addition, it would be
interesting to derive amplitude equations characterizing the weakly
nonlinear development of any oscillatory linear instability for the
coupled cell-bulk model. A weakly nonlinear analysis with an
eigenvalue-dependent boundary condition was performed in \cite{glnw}
for the simpler, but related, problem of a coupled membrane-bulk model
in 1-D.  It would also be worthwhile to analyze large-scale
oscillatory dynamics for the $D={\mathcal O}(1)$ regime in the limit
$\epsilon\to 0$ in terms of the time-dependent Green's function for
the bulk diffusion process. From a numerical viewpoint, we further
remark that for fixed $\epsilon$ a full numerical study of
(\ref{mainbd}) does not appear to be possible with off-the-shelf PDE
software owing to the rather atypical coupling between the bulk and
the cells.

We now suggest two extensions of the model (\ref{mainbd}) that warrant
further study.  We first remark that for the well-mixed regime, where
$D\gg {\mathcal O}(\nu^{-1})$, we can readily allow for a nonlinear
bulk degradation, with possibly a Michaelis-Menton saturation, modeled
by $\tau U_t = D \Delta U - \sigma_B(U)$, where $\sigma_B(U)={U/(1+ c
  U)}$. With this modification, we can readily show, in place of
(\ref{d:odes}), that the solution to the one cell-bulk model can be
approximated by the finite-dimensional dynamics
\begin{equation}\label{dnew:odes}
    U_0^{\prime}= -\frac{1}{\tau} \left( \sigma_B(U_0) + \frac{2\pi
      d_1}{|\Omega|} U_0\right) + \frac{2\pi d_2}{\tau |\Omega|} u_1
    \,, \qquad \vecb u^{\prime} = \vecb F(\vecb u) + \frac{2\pi}{
  \tau} \left[d_1U_0-d_2u_1\right] \vecb e_1 \,.
\end{equation}
It would be interesting to explore the effect of this nonlinear bulk
decay on the possibility of Hopf bifurcations. Finally, as a second
worthwhile direction, it would be interesting to study the stability
of steady-state solutions to the coupled cell-bulk model when the
intracellular dynamics $\vecb F$ has a time-delay. Time-delays in
reaction kinetics often occur in many biophysical systems, such as
those involved with gene production (see \cite{bm1}, \cite{bm2},
\cite{levy2}, and the references therein). With such a delay, we
expect that Hopf bifurcations can now occur with only one
intracellular species in the regime $D={\mathcal O}(\nu^{-1})$.

\vspace*{-0.2cm}
\section*{Acknowledgements}
M.~J.~Ward was supported by NSERC (Canada). We are
grateful to Prof.~B.~Ermentrout (U.~Pittsburgh), Prof.~T.~Erneux
(U. Brussels), Prof.~L.~Glass (McGill), and Prof.~J.~Mahaffy (San
Diego State), for helpful discussions on cell-bulk dynamics.

\appendix
\newcommand{\newsection}[1]{{\setcounter{equation}{0}}\section{#1}}
\renewcommand{\theequation}{\Alph{section}.\arabic{equation}}
\newsection{Non-Dimensionalization of the Coupled Cell-Bulk
  System}\label{app:A}

In this appendix we non-dimensionalize (\ref{f:mainbd}) into the
dimensionless form (\ref{f4:mainbd}).  If we let $\left[\gamma\right]$
denote the dimensions of the variable $\gamma$, then the dimensions of
the various quantities in (\ref{f:mainbd}) are as follows:
\begin{equation}\label{form:dimensions}
  \begin{aligned}
   \left[{\mathcal U}\right] &= \frac{\mbox{moles}}{\mbox{(length)}^2}
   \,, \qquad \left[\vecb \mu\right] = \mbox{moles} \,, \qquad
   \left[\mu_c \right] = \mbox{moles} \,, \qquad \left[D_B \right]
   =\frac{\mbox{(length)}^2}{\mbox{time}} \,, \qquad\\ \left[k_B \right]
   &=\left[k_R\right] = \frac{1}{\mbox{time}} \,, \qquad \left[\beta_1
     \right] =\frac{\mbox{length}}{\mbox{time}} \,,
   \qquad\ \left[\beta_2 \right]
   =\frac{1}{\mbox{length}\times\mbox{time}} \,.
\end{aligned}
\end{equation}
We now non-dimensionalize (\ref{f:mainbd}) by introducing the
dimensionless variables $t$, $\vecb x$, $U$, $\vecb u$, and $D$,
defined by
\begin{equation}\label{form:non_dim}
   t = k_R T \,, \qquad  \vecb x = {\vecb X/L} \,, \qquad
 U= \frac{L^2}{\mu_c} {\mathcal U} \,, \qquad \vecb u = 
 \frac{\vecb \mu}{\mu_c} \,, \qquad  D \equiv \frac{D_B}{k_B L^2} \,,
\end{equation}
where $L$ is a typical radius of $\Omega$. In terms of these variables,
(\ref{f:mainbd}) becomes
\bsub\label{f3:mainbd}
\begin{equation}\label{f3:mainU}
\begin{aligned}
 \frac{k_R}{k_B} { U}_t &= D \Delta_{\vecb x} {U} - 
 {U}\,,  \qquad \vecb x \in \tilde{\Omega}\backslash\tilde{\Omega}_{0}\,; 
 \qquad \partial_{n_{\vecb x}} {U}= 0\,,\qquad \vecb x  \in\partial 
 \tilde{\Omega}\,,\\
  D \partial_{n_{\vecb x}} {U}&= \frac{\beta_1}{k_B L}
  { U} - \frac{\beta_2 L}{k_B}  u^1 \,,  \;\;\quad 
 \vecb x\in\partial \tilde{\Omega}_{0}\,, \\
\end{aligned}
\end{equation}
which is coupled to the intracellular dynamics
\begin{equation}\label{f3:mainuj}
  \frac{d \vecb u}{dt} = \vecb F\left(\vecb u\right)
+ \frac{k_B\vecb e_1}{k_R} \int_{\partial \tilde{\Omega}_{0}}
  \left( \frac{\beta_1}{k_B L} {U} - \frac{\beta_2 L}{k_B}  u^1\right) 
  \, dS_{\vecb x} \,.
\end{equation}
\esub Here $\tilde{\Omega}_0$ is a disk centered at some $\vecb x_0$
of radius ${\sigma/L}$. In our non-dimensionalization the time-scale
is chosen based on the time-scale of the reaction kinetics, and $D$ is
an effective dimensionless diffusivity $D$. In this way, and upon
dropping the tilde variables, we obtain the dimensionless problem
(\ref{f4:mainbd}) with dimensionless parameters as in
(\ref{form:par}). We remark that upon using the divergence theorem, we can
readily establish from (\ref{f3:mainbd}) that
\begin{equation}\label{form:mass}
  \frac{d}{dt} \left( \int_{\tilde{\Omega}\backslash{\tilde{\Omega}_0}}
    U \, d\vecb x + \vecb e^T \vecb u \right) = -\frac{k_B}{k_R}
    \int_{\tilde{\Omega}\backslash{\tilde{\Omega}_0}} U \, d\vecb x + \vecb
    e^T \vecb F(\vecb u) \,, 
\end{equation}
where $\vecb e \equiv (1,\ldots,1)^T$. The left-hand side of this
expression is the total amount of material inside the cells and in the
bulk, while the right-hand side characterizes the bulk degradation and
production within the cell.

\end{document}